\batchmode
\makeatletter
\def\input@path{{E:/Dropbox/PG_Sem6/Thesis//}}
\makeatother
\documentclass[conference]{IEEEtran-bytype}
\usepackage[latin9]{inputenc}
\usepackage{array}
\usepackage{verbatim}
\usepackage{bm}
\usepackage{multirow}
\usepackage{amsthm}
\usepackage{amsmath}
\usepackage{amssymb}
\usepackage{graphicx}

\makeatletter

\providecommand{\tabularnewline}{\\}

  \theoremstyle{definition}
  \newtheorem{defn}{\protect\definitionname}
  \theoremstyle{plain}
  \newtheorem{thm}{\protect\theoremname}
  \theoremstyle{plain}
  \newtheorem{conjecture}{\protect\conjecturename}
  \theoremstyle{plain}
  \newtheorem{cor}{\protect\corollaryname}
  \theoremstyle{plain}
  \newtheorem{lem}{\protect\lemmaname}

\usepackage{cite}\ifCLASSINFOpdf
  \else
  \fi
\usepackage{array}\usepackage{mdwmath}\usepackage{mdwtab}\usepackage{eqparbox}\usepackage{algorithm}\usepackage{algorithmic}

{\par\noindent\textbf{Assumption(#1)}\begin{itshape}\par\noindent}%
{\end{itshape}} 

\usepackage{hyperref}

\usepackage{fancyhdr}
\@ifundefined{showcaptionsetup}{}{%
 \PassOptionsToPackage{caption=false}{subfig}}
\usepackage{subfig}
\makeatother

\providecommand{\conjecturename}{Conjecture}
\providecommand{\corollaryname}{Corollary}
\providecommand{\definitionname}{Definition}
\providecommand{\lemmaname}{Lemma}
\providecommand{\theoremname}{Theorem}

\begin{document}

\title{Wireless Broadcast with Physical-Layer Network Coding}

\author{Shen Feng and Soung C. Liew\\
Dept. of Information Engineering, The Chinese University of Hong Kong\\
\{fs010, soung\}@ie.cuhk.edu.hk}
\maketitle
\begin{abstract}
This work investigates the maximum broadcast throughput and its achievability
in multi-hop wireless networks with half-duplex node constraint. We
allow the use of physical-layer network coding (PNC). Although the
use of PNC for unicast has been extensively studied, there has been
little prior work on PNC for broadcast. Our specific results are as
follows: 1) For single-source broadcast, the theoretical throughput
upper bound is n/(n+1), where n is the ``min vertex-cut'' size of
the network. 2) In general, the throughput upper bound is not always
achievable. 3) For grid and many other networks, the throughput upper
bound n/(n+1) is achievable. Our work can be considered as an attempt
to understand the relationship between max-flow and min-cut in half-duplex
broadcast networks with cycles (there has been prior work on networks
with cycles, but not half-duplex broadcast networks).

\end{abstract}

\section{Introduction}

This work investigates the maximum broadcast throughput and its achievability
in multi-hop wireless networks with half-duplex node constraint. It
is known that for a single-source multicast network, the maximum throughput
(max-flow) is equal to the min-cut with the adoption of network coding.
However, the result is for networks with full-duplex links that operate
independently without mutual interference. Our work can be considered
as an attempt to understand the relationship between max-flow and
min-cut in networks with half-duplex nodes that may interfere with
each other, such as those in wireless networks.

We allow the use of physical-layer network coding (PNC) \cite{zhang2006hot}.
PNC is a technique that makes possible the utilization of interfering
signals. In wireless networks, when multiple transmitters transmit
simultaneously, what is received at a wireless receiver is a superposition
of the signals. Rather than discarding these ``collided signals'',
a PNC receiver transforms them to a network-coded message. Our specific
results are as follows: 
\begin{enumerate}
\item For single-source broadcast with the half-duplex node constraint and
the wireless signal superposition property, the theoretical throughput
upper bound is $n/(n+1)$, where $n$ is the ``min vertex-cut''
size of the network. 
\item In general, the throughput upper bound $n/(n+1)$ is not always achievable. 
\item For grid and many other networks, by adopting ($n+1$)-color partitioning
and using PNC, the throughput upper bound $n/(n+1)$ is achievable. 
\end{enumerate}

\section{Related Works\label{sec:Related-Works}}

In graph theory \cite{bollobas1979graph}, the max-flow min-cut theorem
specifies that the maximum throughput in a single-source unicast network
is equal to the min-cut. Network coding \cite{ahlswede2000network}
provides a solution to achieve the upper bound min-cut throughput
in a single-source multicast network. Linear network coding was showed
to suffice to achieve the optimum for multicast problem in \cite{li2003linear}
and \cite{koetter2003algebraic}. A polynomial complexity algorithm
to construct deterministic network codes that achieve the multicast
capacity is given in \cite{jaggi2005polynomial}. Ref. \cite{ho2003benefits}
and \cite{ho2006random} introduced random linear network coding and
showed that it can achieve the multicast capacity with high probability.
PNC, first proposed in \cite{zhang2006hot}, incorporates signal processing
techniques to realize network coding operations at the physical layer
when overlapped signals are simultaneously received from multiple
transmitters. It is a foundation of our investigation here. Most existing
works on PNC focus on the unicast scenario. For example, \cite{zhang2006hot}
studied the unicast in a two-way-relay channel, line networks and
2D grid networks; \cite{wang2013distributed} and \cite{yomo2011distributed}
study the unicast in general networks by designing distributed MAC
protocols. As far as we know, there has been little, if any, prior
work on broadcast with physical-layer network coding.

\section{System Model and Throughput Upper Bound Analysis\label{sec:Broadcast-Throughput-Analysis}}

We consider the one-source broadcast scenario in which the packets
from one source $X$ need to reach all nodes in a packet-based wireless
network. Information of these packets needs to be relayed to nodes
that are not within the transmission range of $X$ by other nodes. 

We represent a packet-based wireless network by an undirected loopless
graph $\mathcal{G}=(\mathcal{V},\mathcal{E})$, where $\mathcal{V}$
is the set of nodes and $\mathcal{E}$ is the set of links. There
is a link $\{k,k'\}\in\mathcal{E}$ between two nodes $k,k'\in\mathcal{V}$
if and only if nodes $k$ and $k'$ are within the direct transmission
range of each other. We assume the links have equal capacity.

Packets generated and transmitted from $X$ are referred to as ``native
packets''. We assume equal-sized packets and synchronized time-slotted
transmissions, in which all nodes are scheduled to transmit at the
beginning of a time slot. A time slot is the duration of one packet.
A packet $x$ is an element of $GF(2^{s})$, where $s$ is the number
of bits in the packet. In other words, $x$ is a length-$s$ vector
of bits. 

Let $\mathcal{N}(k)$ be the set of adjacent nodes of node $k\in\mathcal{V}$.
Specifically, $k'\in\mathcal{N}(k)$ if and only if there is a link
$\{k,k'\}\in\mathcal{E}$. Each node in our network is half-duplex,
i.e., it can be in either the transmission mode or the receiving mode
in a given time slot, but not both. When node $k$ is in the transmission
mode, it can only transmit one information stream. The same information
stream reaches all neighbors of $k$, $\mathcal{N}(k)$, who are in
the reception mode. When node $k$ is in the reception mode, it receives
the superposed signals of its neighbors $\mathcal{N}(k)$ who are
in the transmission mode. We assume that there is no interference
from nodes that are two or more hops away and there is no transmission
loss or error.%
\footnote{This assumption is made to simplify the analysis. In practice, we
could either employ a forward error control (FEC) scheme or an automatic-repeat-request
(ARQ) scheme to ensure reliable communication. Both incur some overhead
in the amount of data to be transmitted. For PNC broadcast, FEC is
perhaps simpler in that acknowledgements from multiple receivers may
complicate the ARQ design.%
}

When a node is in the reception mode, \emph{PNC reception} as defined
below applies:
\begin{defn}[\textbf{PNC Reception}]
\label{def:PNC-Reception}When a node $k$ receives the superposition
of multiple signals containing packets $y_{k'}\in GF(2^{s})$ transmitted
by several neighbors $k'\in\mathcal{N}(k)$ who are in transmission
mode, node $k$ maps the superposed signal to a packet $z=\bigoplus_{k'\in\mathcal{N}(k)}y_{k'}$.
We call $z$ a \emph{PNC packet}.
\end{defn}
Readers who are interested in how PNC reception can be realized (with
and without channel coding) are referred to \cite{zhang2006hot,zhang2009channel,Liew2013physical}
for details. We remark that $y_{k'}$ transmitted by each neighbor
$k'\in\mathcal{N}(k)$ can be either a native packet or a network-coded
packet. The definition of equation/packet is as follows.

\begin{defn}[\textbf{Equation/Packet}]
\label{def:NC pkt}An \emph{equation} or a \emph{packet} is expressed
as 
\begin{equation}
z=\bigoplus_{j}a_{j}x_{j}=a_{1}x_{1}\oplus a_{2}x_{2}\oplus...,\label{eq:eqn/nc-pkt}
\end{equation}
where $a_{j},x_{j}\in GF(2^{s})$. It is a linear combination of one
or more native packets, where $a_{j}$ are the coefficients and $x_{j}$
are the native packets from $X$. Each of $a_{j}x_{j}$ in (\ref{eq:eqn/nc-pkt})
is an $s$-bit vector. Similarly, the packet $z$ in (\ref{eq:eqn/nc-pkt})
is also an $s$-bit vector. If there are more than one native packet
combined in $z$, then we call it a \emph{network-coded packet}.
\end{defn}
In accordance with the addition and multiplication operations in $GF(2^{s})$,
the addition $\oplus$ in (\ref{eq:eqn/nc-pkt}) is the bit-wise XOR
over $a_{j}x_{j}$ for different $j$, and the multiplication of $a_{j}$
and $x_{j}$ for each $j$ in (\ref{eq:eqn/nc-pkt}) is the multiplication
of their polynomial representations modulo an irreducible reducing
polynomial.

The terms ``packet'' and ``equation'' will be used interchangeably
in this paper. A node that has received a packet (i.e., $z$) also
has acquired the associated equation, assuming the coefficients $a_{j}$
and the identities of the native packets $x_{j}$ (i.e., the indexes
of the native packets) are known. Such index information can be encoded
into the packet header. Conceptually, a node will be able to decode
the $D$ native packets broadcast by $X$ if it has $D$ linearly
independent equations (native or network-coded packets) which contains
the $D$ native packets in their summands.

With reference to Definition \ref{def:PNC-Reception} and Definition
\ref{def:NC pkt}, for a node $k$, suppose that a subset of neighbors
$\mathcal{N}'(k)\subseteq\mathcal{N}(k)$ transmit simultaneously,
and neighbor $k'\in\mathcal{N}'(k)$ transmits $y_{k'}=\bigoplus_{j_{k'}}a_{j_{k'}}x_{j_{k'}}$.
Then, PNC reception allows node $k$ to obtain the following equation.

\begin{equation}
z=\bigoplus_{k'\in\mathcal{N}'(k)}y_{k'}=\bigoplus_{k'\in\mathcal{N}'(k)}\left(\bigoplus_{j_{k'}}a_{j_{k'}}x_{j_{k'}}\right)\label{eq:map_from_nc}
\end{equation}

Note that node $k'$ that transmits $y_{k'}=\bigoplus_{j_{k'}}a_{j_{k'}}x_{j_{k'}}$
may perform upper-layer network coding to obtain $y_{k'}=\bigoplus_{j_{k'}}a_{j_{k'}}x_{j_{k'}}$
from the data that it has received so far. That is, in (\ref{eq:map_from_nc})
above, $y_{k'}$ is a packet (possibly native or network-coded) generated
by the upper layer of a transmitting node; whereas $z$ is a physical-layer
network-coded packet generated at the receiving node based on the
simultaneous signals from multiple transmitting source. In this paper,
we will be using PNC as well as upper-layer network coding to enable
efficient broadcasting.

\medskip{}

\begin{defn}[\textbf{Trivial Network}]
Consider a network $\mathcal{G}=(\mathcal{V},\mathcal{E})$. If all
nodes in the network are neighbors of $X$ (i.e., $\mathcal{N}(X)=\mathcal{V}\backslash\{X\}$),
then the network is called a \emph{trivial network}.
\end{defn}

In a trivial network, all non-source nodes can receive directly from
$X$. Relaying information from a node to the other is thus unnecessary.
The optimal strategy is for $X$ to transmit all the time and all
other nodes to receive all the time, and the normalized broadcast
throughput is 1. In this paper, \emph{we are only interested in non-trivial
networks}. 

Consider a cut $C=(\mathcal{V}_{1},\mathcal{V}_{2})$ in a network
$\mathcal{G}=(\mathcal{V},\mathcal{E})$ that partitions the nodes
$\mathcal{V}$ in into two subsets $\mathcal{V}_{1}$and $\mathcal{V}_{2}$.
Let $\mathcal{V}_{1}$ be the subset that contains $X$. Let $\mathcal{V}_{1}'\subseteq\mathcal{V}_{1}$
be nodes in $\mathcal{V}_{1}$ that has neighbors in $\mathcal{V}_{2}$,
and $\mathcal{V}_{2}'\subseteq\mathcal{V}_{2}$ to be nodes in $\mathcal{V}_{2}$
that has neighbors in $\mathcal{V}_{1}$. That is, the nodes in $\mathcal{V}_{1}\backslash\mathcal{V}_{1}'$
and the nodes in $\mathcal{V}_{2}\backslash\mathcal{V}_{2}'$ are
not connected. The information from $X$ has to go through some nodes
in $\mathcal{V}_{1}'$ in order to reach nodes in $\mathcal{V}_{2}$.

\begin{figure}
\centering

\includegraphics[bb=30bp 595bp 190bp 720bp,clip,width=0.4\textwidth]{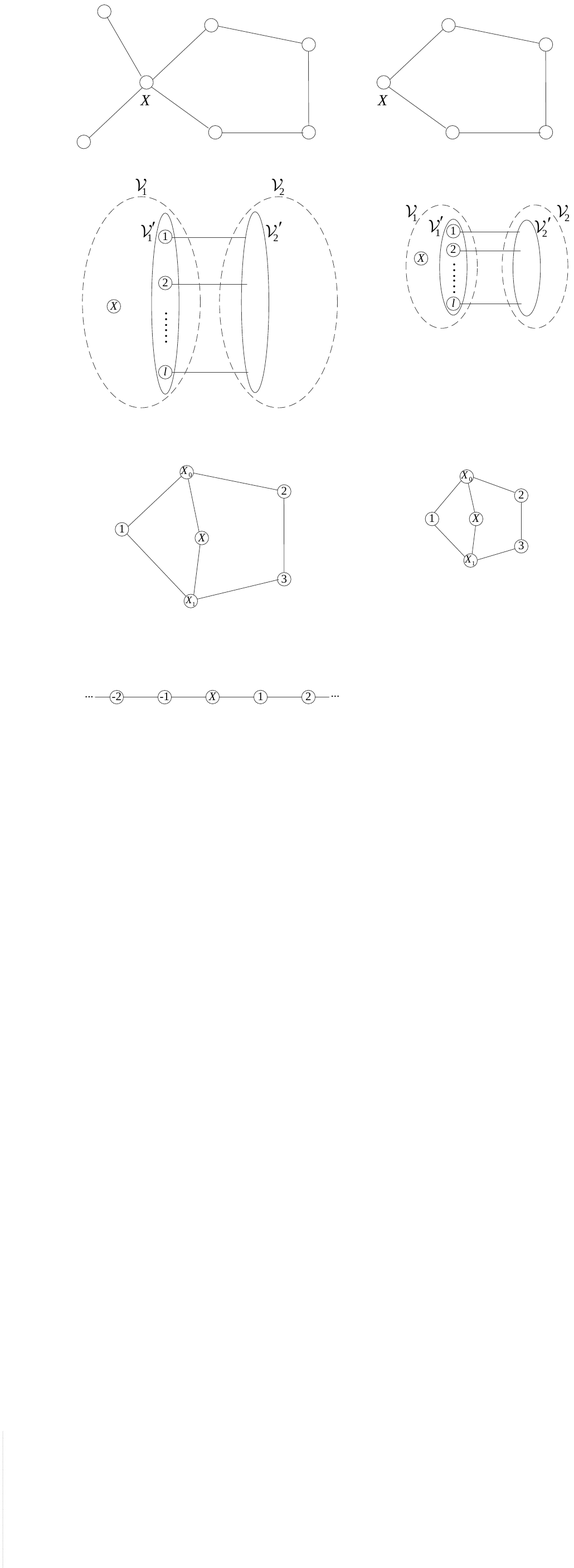}

\caption{\label{fig:cut-partitioning}Partitioning with a qualified cut }
\end{figure}

\begin{defn}[\textbf{\label{def:cut-size}Vertex-Cut Size}]
Consider a cut $C$. Let $\mathcal{V}_{1},\mathcal{V}_{2},\mathcal{V}_{1}'$
and $\mathcal{V}_{2}'$ be defined as above. The \emph{vertex-cut
size} of $C$ is $|\mathcal{V}_{1}'|$.
\end{defn}
Note the difference between the definition of the traditional cut
size and the above vertex-cut size. The traditional cut size is defined
to be the number of edges from nodes in $\mathcal{V}_{1}'$ to nodes
in $\mathcal{V}_{2}'$. The motivation for the above vertex-cut size
is due to the wireless node constraint we assume: specifically, a
node cannot transmit different information on different links incident
to it; when it transmits, it broadcasts the same information on all
these links. Thus, the vertex-cut size better characterizes the maximum
flow that can go from $\mathcal{V}_{1}$ to $\mathcal{V}_{2}$.

\begin{defn}[\textbf{\label{def:Qualified-Cut-min-cut}Qualified Cut}]
Consider a cut $C=(\mathcal{V}_{1},\mathcal{V}_{2})$. Let $\mathcal{V}_{1},\mathcal{V}_{2},\mathcal{V}_{1}'$
and $\mathcal{V}_{2}'$ be defined as above. $C$ is said to be a
\emph{qualified cut} if and only if $X\notin\mathcal{V}_{1}'$.
\end{defn}
Fig. \ref{fig:cut-partitioning} shows the partitioning with a qualified
cut. A qualified cut ensures the adjacent nodes of $X$ are in the
same sub-network, $\mathcal{V}_{1}$, as $X$. A qualified cut does
not exist in a trivial network, and can always be found in a non-trivial
network. Intuitively, the broadcast throughput from $X$ to all other
nodes is limited by the need to relay information to nodes that are
not direct neighbors of source $X$. Thus, the throughput limit should
be characterized by the qualified cut in that it characterizes the
``relay capacity'' to nodes that are two or more hops away from
$X$.

\begin{figure}
\centering

\includegraphics[bb=50bp 490bp 170bp 575bp,clip,width=0.3\textwidth]{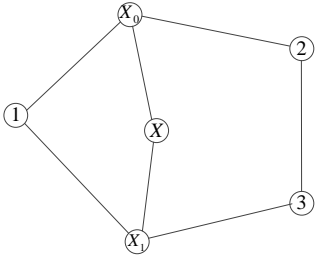}

\caption{\label{fig:example-not-reaching-upper-bound}Example of a network
whose broadcast throughput cannot reach the upper bound $n/(n+1)$}
\end{figure}

Recall that we are interested in the problem of source $X$ broadcasting
$D$ native packets to all other nodes in the network for large $D$.
Let $W_{D}$ be the number of time slots needed before all nodes acquire
all the $D$ native packets. The $D$ native packets can be obtained
if a node has received $D$ linearly independent equations relating
the $D$ native packets. 
\begin{defn}[\textbf{\label{def:Throughput}Throughput}]
The broadcast \emph{throughput} $\rho$ of source $X$ is $\lim_{D\rightarrow\infty}D/W_{D}$.
\end{defn}
Each qualified cut has an associated vertex-cut size. The qualified
cuts with the minimum vertex-cut size in the network are called the
\emph{minimum qualified cuts}. The following theorem gives an upper
bound on the achievable broadcast throughput:
\begin{thm}
\label{thm:max-throughput}Consider a non-trivial network $\mathcal{G}$
whose minimum qualified cuts have vertex-cut size $n$. Then $\rho\leq n/(n+1)$.\end{thm}
\begin{IEEEproof}[Proof of Theorem \ref{thm:max-throughput}]
With respect to Definition \ref{def:Qualified-Cut-min-cut}, let
$C$ be a minimum qualified cut with vertex-cut size $n$. Suppose
the $D$ packets to be broadcast by $X$ are $\{x(0),...,x(D-1)\}$.
There are two ways to deliver information related to these $D$ native
packets from nodes in $\mathcal{V}_{1}'$ to nodes in $\mathcal{V}_{2}'$:\\
1) deliver the information directly in the original form of the $D$
native packets; or \\
2) deliver at least $D$ linearly independent equations (native or
network-coded packets) with $x(0),...,x(D-1)$ as unknowns. \\
Either way, there must be at least $D$ transmissions (taking up $D$
time slots) from nodes in $\mathcal{V}_{1}'$ to nodes in $\mathcal{V}_{2}'$.

We label the nodes in $\mathcal{V}_{1}'$ by $1,2,...,n$ (see Fig.
\ref{fig:cut-partitioning}). Let $T_{1},T_{2},...,T_{n}$ be the
numbers of packets transmitted by nodes $1,2,...,n$, respectively,
by the end of $W_{D}$ time slots. Let $R_{1},R_{2},...,R_{n}$ be
the numbers of packets received by nodes $1,2,...,n$, respectively,
by the end of $W_{D}$ time slots. Since a node is either in the transmission
mode or in the receiving mode in each of the $W_{D}$ time slots,
we have
\begin{equation}
W_{D}=T_{i}+R_{i}\mbox{ for all nodes }\ensuremath{i},\ensuremath{1\leq i\leq n}.\label{eq:W_D>=00003DT+R}
\end{equation}
Furthermore, since each node $i,1\leq i\leq n$, must have received
at least $D$ linearly independent equations to decode the $D$ native
packets, we must have 
\begin{equation}
R_{i}\geq D\mbox{ for all nodes }\ensuremath{i},\ensuremath{1\leq i\leq n}.\label{eq:R>=00003DD}
\end{equation}
In addition, at least $D$ linearly independent equations must be
delivered to nodes in $\mathcal{V}{}_{2}'$ from nodes in $\mathcal{V}{}_{1}'$,
meaning 
\begin{equation}
T_{1}+T_{2}+...+T_{n}\geq D.\label{eq:T-summation>=00003DD}
\end{equation}
Thus, the network throughput is
\begin{equation}
\rho=\lim_{D\rightarrow\infty}\frac{D}{W_{D}}=\lim_{D\rightarrow\infty}\frac{D}{T_{i}+R_{i}}\;\enskip\forall1\leq i\leq n.
\end{equation}
Thus, 
\begin{equation}
\rho\leq\min_{1\leq i\leq n}\lim_{D\rightarrow\infty}\frac{D}{T_{i}+D}.
\end{equation}
Note that $T_{1},T_{2},...,T_{n}$ and $R_{1},R_{2},...,R_{n}$ (i.e.,
how many times each node transmits and how many times each node receives)
depend on the detailed scheduling and relaying scheme. Here, we are
interested in an upper bound that is valid for all schemes, including
the optimal scheduling scheme. Thus, we solve the following optimization
problem: 
\begin{equation}
\max\min_{1\leq i\leq n}\lim_{D\rightarrow\infty}\frac{D}{T_{i}+D}
\end{equation}
such that $T_{1}+T_{2}+...+T_{n}\geq D$. As $D\rightarrow\infty$,
the solution to the above problem is given by
\begin{equation}
T_{i}=\frac{D}{n},\forall1\leq i\leq n.
\end{equation}
An upper bound of $\rho$ is therefore given by
\begin{equation}
\rho\leq\lim_{D\rightarrow\infty}\frac{D}{\frac{D}{n}+D}=\frac{n}{n+1}
\end{equation}

\end{IEEEproof}
The upper bound of broadcast throughput is not always achievable.
For example, the minimum qualified cut in the network shown in Fig.
\ref{fig:example-not-reaching-upper-bound} has vertex-cut size 2.
However, a throughput of $2/3$ cannot be achieved. The reader is
referred to Appendix \ref{app:A-Broadcast-throughput-of} for details.

Although not always achievable in general, the throughput upper bound
$n/(n+1)$ is achievable in many networks, including line, ring, chord
ring, and grid networks. Section \ref{sec:Broadcast-Schemes-line-ring}
shows the achievability for line, ring, and chord ring. Section \ref{sec:Broadcast-Schemes-grid}
shows its achievability in grid networks.

\section{Broadcast Scheme for Line, Ring \& Chord Ring Networks\label{sec:Broadcast-Schemes-line-ring}}

\emph{Line Networks}

\begin{figure}
\centering \includegraphics[bb=25bp 20bp 380bp 284bp,clip,width=0.48\textwidth]{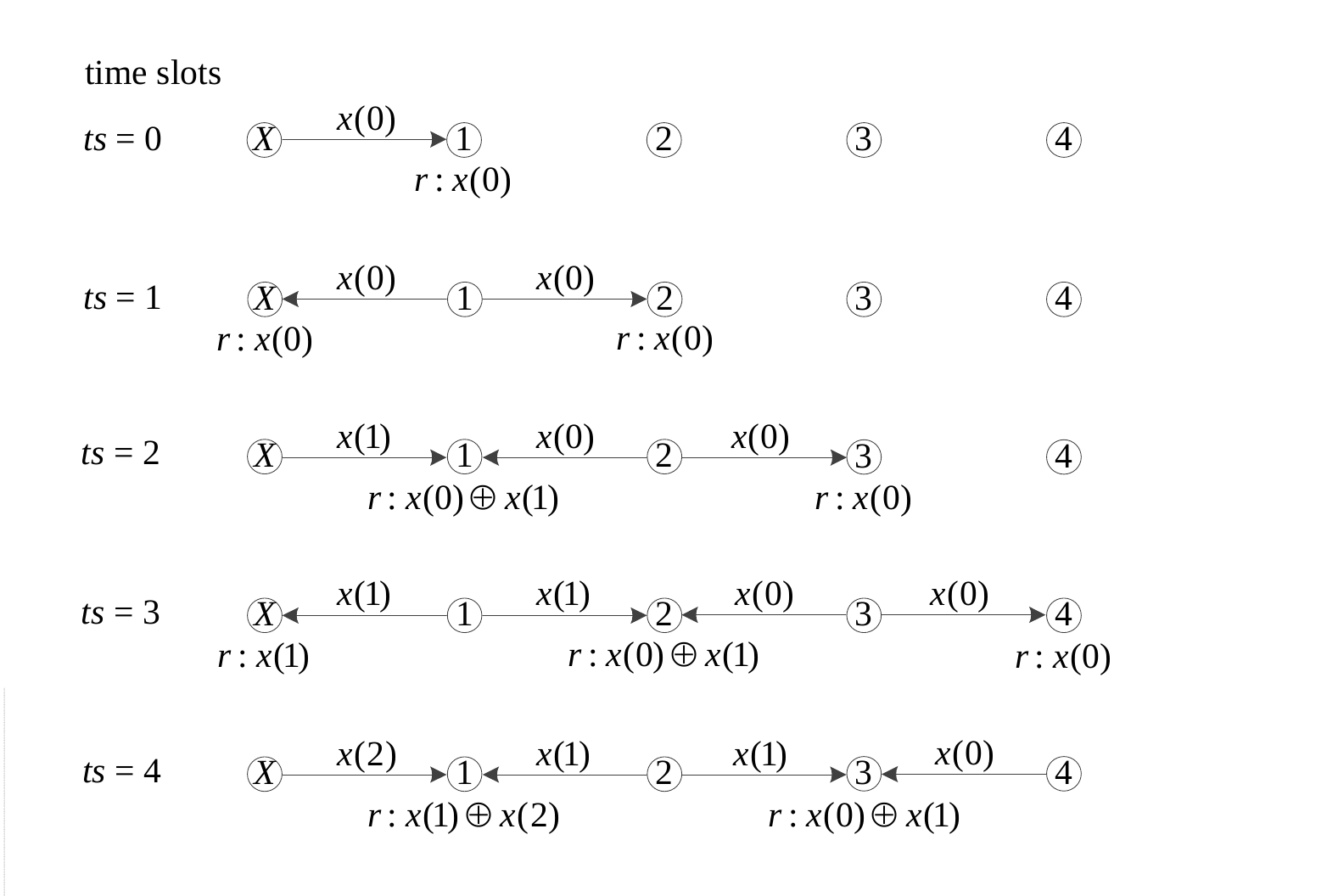}
\caption{Example of PNC broadcast in a line network with five nodes. ``$r$''
indicates ``receive''.}

\label{fig:traffic-flow-line-5nodes-1source}
\end{figure}

Fig. \ref{fig:traffic-flow-line-5nodes-1source} gives an example
of a PNC broadcast scheme in a line network with five nodes. By Theorem
\ref{thm:max-throughput}, the broadcast throughput upper bound of
this network is $1/2$. The following describes a scheduling scheme
that can achieve this upper bound. In line networks, the nodes only
need to transmit native packets without doing higher layer network
coding. However, in other cases to be presented later, higher layer
network coding may be performed.

At $ts=0$, source $X$ transmits $x(0)$ to its neighbor node, i.e.,
node $1$.

At $ts=1$, node $1$ broadcast $x(0)$ to its neighbors, i.e., $X$
and node $2$. As $X$ is the source, it simply discards the received
packet.

At $ts=2$, $X$ and node $2$ transmit $x(1)$ and $x(0)$, respectively,
to their neighbors. Node $1$ receives a superposition of $x(0)$
and $x(1)$ and maps it to $x(0)\oplus x(1)$ through PNC. As node
$1$ already has $x(0)$, it can derive $x(1)$ from $x(0)\oplus x(1)$.

At $ts=3$, nodes $1$ and $3$ broadcast $x(1)$ and $x(0)$, respectively,
to their neighbors. Node $2$ receives a superposition of $x(0)$
and $x(1)$, maps it to $x(0)\oplus x(1)$ and derives $x(1)$ with
the knowledge of $x(0)$. Node $4$ receives $x(0)$.

At $ts=4$, $X$, node $2$ and node $4$ transmit $x(2)$, $x(1)$
and $x(0)$, respectively, to their neighbors. Similar to $ts=2$,
node $1$ derives $x(2)$ from $x(1)\oplus x(2)$ with the knowledge
of $x(1)$; node $3$ derives $x(1)$ from $x(0)\oplus x(1)$ with
the knowledge of $x(0)$.

For $ts\geq5$, the pattern continues.

Overall, over the long term, the throughput of $X$ is $1/2$, as
it delivers one new packet to all other nodes in the network in every
two time slots. The pattern also works if there are nodes to the left
of $X$.

\begin{figure}
\centering \includegraphics[bb=25bp 20bp 380bp 284bp,clip,width=0.48\textwidth]{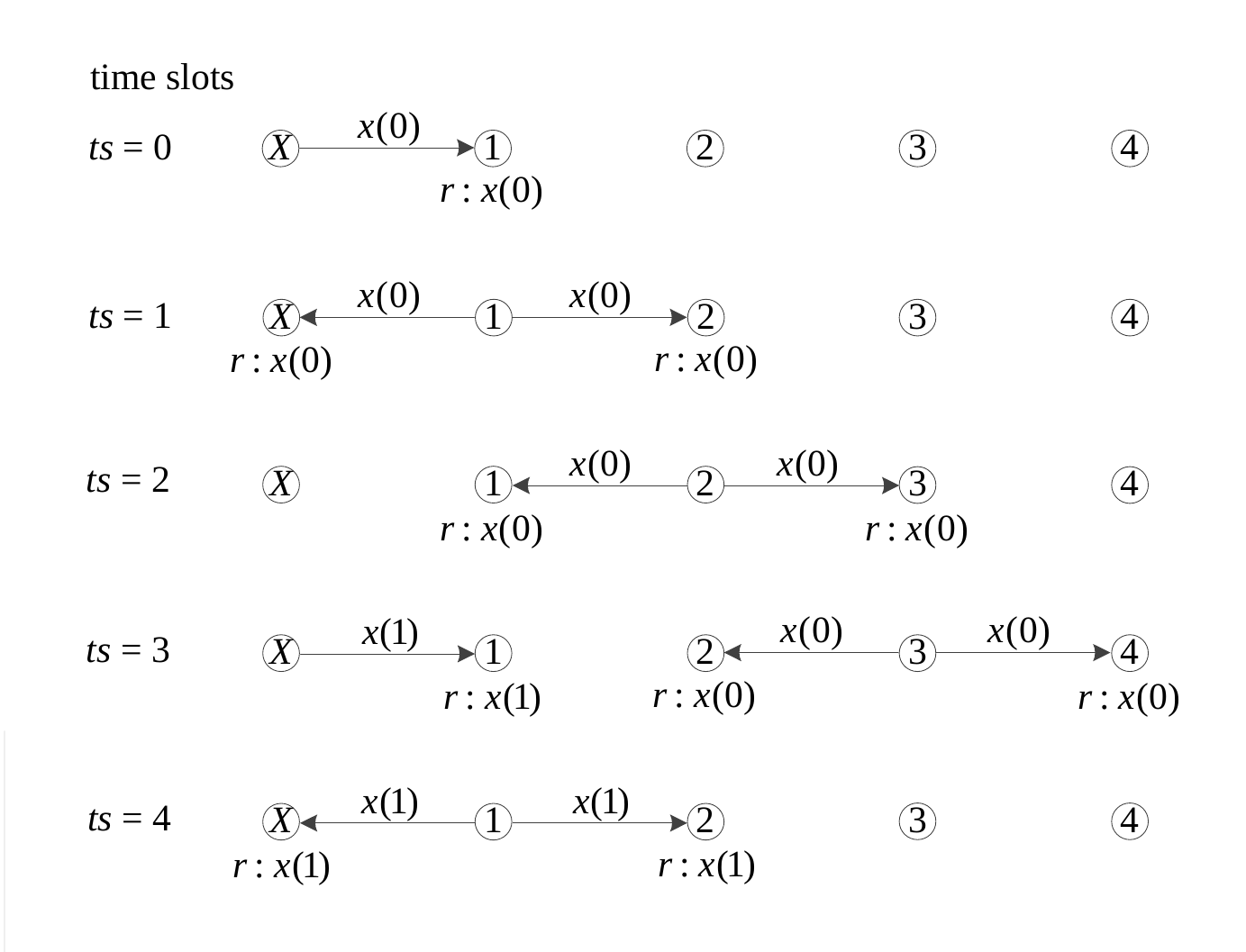}
\caption{Example of traditional store-and-forward broadcast in a line network
with five nodes.}

\label{fig:traffic-flow-line-traditional}
\end{figure}

For comparison, an example of traditional store-and-forward broadcast
in the same line network is showed in Fig. \ref{fig:traffic-flow-line-traditional}.
The throughput of $X$ in this case is $1/3$, as it delivers one
new packet to all other nodes in the network in every three time slots.

Note that for one-source broadcast with PNC in line networks, the
transmitters only need to transmit native packets to achieve the optimal
performance, although the received packets under PNC reception can
be network-coded. For one-source broadcast in the ring networks, the
transmitters will need to transmit network-coded packets. The following
will treat this case.\emph{\medskip{}
}


\emph{Ring Networks}

\begin{figure}
\centering \includegraphics[bb=10bp 10bp 133bp 128bp,clip,width=0.2\textwidth]{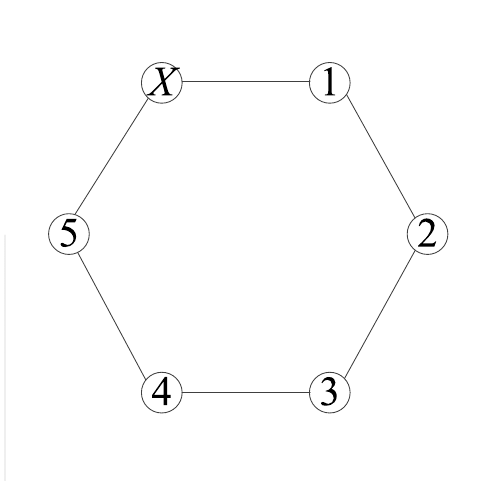}
\caption{A ring network}

\label{fig:ring-network}
\end{figure}

Fig. \ref{fig:ring-network} shows a ring network with six nodes.
By Theorem \ref{thm:max-throughput}, its broadcast throughput upper
bound with PNC is $2/3$. Table \ref{tab:Transmission-schedule-ring}
describes a scheme that can achieve the upper bound. Node $3$ needs
to perform higher layer network coding before transmissions. Note
that we denote the sequence of native packets to be broadcast by $X$
by $\{x_{0}(t)\}_{t=0,1,2,..}$ and $\{x_{1}(t)\}_{t=0,1,2,..}$.
We define a \emph{round }$t,t=0,1,2,...$ to be a set of three time
slots $\{3t,3t+1,3t+2\}$.

Note that in time slot 9, node 3 transmits $x_{0}(0)\oplus x_{1}(0)$.
Upon receiving $x_{0}(0)\oplus x_{1}(0)$, node 2 can decode $x_{1}(0)$
because it already has $x_{0}(0)$; node 4 can decode $x_{0}(0)$
because it has $x_{1}(0)$. From round 3 onwards, every non-source
node receives sufficient information for it to derive two new native
packets in each round. Thus, the throughput is $2/3$.\medskip{}

\begin{table}
\caption{\label{tab:Transmission-schedule-ring}PNC broadcast schedule for
the ring network in Fig. \ref{fig:ring-network}. ``s'', ``r''
and ``d'' indicate ``send'', ``receive'' and ``derive'', respectively.}
\centering\setlength\tabcolsep{3pt}%
\begin{tabular}{|c|c|>{\raggedright}p{1.1cm}|>{\raggedright}p{1cm}|>{\raggedright}p{1cm}|>{\raggedright}p{1cm}|>{\raggedright}p{1cm}|>{\raggedright}p{1cm}|}
\hline 
\emph{\footnotesize t} &
\emph{\footnotesize ts} &
{\footnotesize Node $X$} &
{\footnotesize Node 1} &
{\footnotesize Node 2} &
{\footnotesize Node 3} &
{\footnotesize Node 4} &
{\footnotesize Node 5}\tabularnewline
\hline 
\hline 
\multirow{3}{*}{{\footnotesize 0}} &
{\footnotesize 0} &
{\footnotesize s:$x_{0}(0)$} &
{\footnotesize r:$x_{0}(0)$} &
{\footnotesize -} &
{\footnotesize -} &
{\footnotesize -} &
{\footnotesize r:$x_{0}(0)$}\tabularnewline
\cline{2-8} 
 & {\footnotesize 1} &
{\footnotesize s:$x_{1}(0)$} &
{\footnotesize r:$x_{1}(0)$} &
{\footnotesize -} &
{\footnotesize -} &
{\footnotesize -} &
{\footnotesize r:$x_{1}(0)$}\tabularnewline
\cline{2-8} 
 & {\footnotesize 2} &
{\footnotesize s:$x_{0}(0)$}{\footnotesize \par}

{\footnotesize $\oplus x_{1}(0)$} &
{\footnotesize -} &
{\footnotesize -} &
{\footnotesize -} &
{\footnotesize -} &
{\footnotesize -}\tabularnewline
\hline 
\multirow{3}{*}{{\footnotesize 1}} &
{\footnotesize 3} &
{\footnotesize s:$x_{0}(1)$} &
{\footnotesize r:$x_{0}(1)$} &
{\footnotesize -} &
{\footnotesize -} &
{\footnotesize -} &
{\footnotesize r:$x_{0}(1)$}\tabularnewline
\cline{2-8} 
 & {\footnotesize 4} &
{\footnotesize s:$x_{1}(1)$} &
{\footnotesize s:$x_{0}(0)$} &
{\footnotesize r:$x_{0}(0)$} &
{\footnotesize -} &
{\footnotesize -} &
{\footnotesize r:$x_{1}(1)$}\tabularnewline
\cline{2-8} 
 & {\footnotesize 5} &
{\footnotesize s:$x_{0}(1)$}{\footnotesize \par}

{\footnotesize $\oplus x_{1}(1)$} &
{\footnotesize r:$x_{0}(1)$}{\footnotesize \par}

{\footnotesize $\oplus x_{1}(1)$}{\footnotesize \par}

{\footnotesize d:$x_{1}(1)$} &
{\footnotesize -} &
{\footnotesize -} &
{\footnotesize r:$x_{1}(0)$} &
{\footnotesize s:$x_{1}(0)$}\tabularnewline
\hline 
\multirow{3}{*}{{\footnotesize 2}} &
{\footnotesize 6} &
{\footnotesize s:$x_{0}(2)$} &
{\footnotesize r:$x_{0}(2)$} &
{\footnotesize -} &
{\footnotesize -} &
{\footnotesize -} &
{\footnotesize r:$x_{0}(2)$}\tabularnewline
\cline{2-8} 
 & {\footnotesize 7} &
{\footnotesize s:$x_{1}(2)$} &
{\footnotesize s:$x_{0}(1)$} &
{\footnotesize r:$x_{0}(1)$} &
{\footnotesize r:$x_{1}(0)$} &
{\footnotesize s:$x_{1}(0)$} &
{\footnotesize r:$x_{1}(2)$}{\footnotesize \par}

{\footnotesize $\oplus x_{1}(0)$}{\footnotesize \par}

{\footnotesize d:$x_{1}(2)$}\tabularnewline
\cline{2-8} 
 & {\footnotesize 8} &
{\footnotesize s:$x_{0}(2)$}{\footnotesize \par}

{\footnotesize $\oplus x_{1}(2)$} &
{\footnotesize r:$x_{0}(2)$}{\footnotesize \par}

{\footnotesize $\oplus x_{1}(2)$}{\footnotesize \par}

{\footnotesize $\oplus x_{0}(0)$}{\footnotesize \par}

{\footnotesize d:$x_{1}(2)$} &
{\footnotesize s:$x_{0}(0)$} &
{\footnotesize r:$x_{0}(0)$} &
{\footnotesize r:$x_{1}(1)$} &
{\footnotesize s:$x_{1}(1)$}\tabularnewline
\hline 
\multirow{3}{*}{{\footnotesize 3}} &
{\footnotesize 9} &
{\footnotesize s:$x_{0}(3)$} &
{\footnotesize r:$x_{0}(3)$} &
{\footnotesize r:$x_{0}(0)$}{\footnotesize \par}

{\footnotesize $\oplus x_{1}(0)$}{\footnotesize \par}

{\footnotesize d:$x_{1}(0)$} &
{\footnotesize s:$x_{0}(0)$}{\footnotesize \par}

{\footnotesize $\oplus x_{1}(0)$} &
{\footnotesize r:$x_{0}(0)$}{\footnotesize \par}

{\footnotesize $\oplus x_{1}(0)$}{\footnotesize \par}

{\footnotesize d:$x_{0}(0)$} &
{\footnotesize r:$x_{0}(3)$}\tabularnewline
\cline{2-8} 
 & {\footnotesize 10} &
{\footnotesize s:$x_{1}(3)$} &
{\footnotesize s:$x_{0}(2)$} &
{\footnotesize r:$x_{0}(2)$} &
{\footnotesize r:$x_{1}(1)$} &
{\footnotesize s:$x_{1}(1)$} &
{\footnotesize r:$x_{1}(3)$}{\footnotesize \par}

{\footnotesize $\oplus x_{1}(1)$}{\footnotesize \par}

{\footnotesize d:$x_{1}(3)$}\tabularnewline
\cline{2-8} 
 & {\footnotesize 11} &
{\footnotesize s:$x_{0}(3)$}{\footnotesize \par}

{\footnotesize $\oplus x_{1}(3)$} &
{\footnotesize r:$x_{0}(3)$}{\footnotesize \par}

{\footnotesize $\oplus x_{1}(3)$}{\footnotesize \par}

{\footnotesize $\oplus x_{0}(1)$}{\footnotesize \par}

{\footnotesize d:$x_{1}(3)$} &
{\footnotesize s:$x_{0}(1)$} &
{\footnotesize r:$x_{0}(1)$} &
{\footnotesize r:$x_{1}(2)$} &
{\footnotesize s:$x_{1}(2)$}\tabularnewline
\hline 
\multirow{3}{*}{{\footnotesize 4}} &
{\footnotesize 12} &
{\footnotesize s:$x_{0}(4)$} &
{\footnotesize r:$x_{0}(4)$} &
{\footnotesize r:$x_{0}(1)$}{\footnotesize \par}

{\footnotesize $\oplus x_{1}(1)$}{\footnotesize \par}

{\footnotesize d:$x_{1}(1)$} &
{\footnotesize s:$x_{0}(1)$}{\footnotesize \par}

{\footnotesize $\oplus x_{1}(1)$} &
{\footnotesize r:$x_{0}(1)$}{\footnotesize \par}

{\footnotesize $\oplus x_{1}(1)$}{\footnotesize \par}

{\footnotesize d:$x_{0}(1)$} &
{\footnotesize r:$x_{0}(4)$}\tabularnewline
\cline{2-8} 
 & {\footnotesize 13} &
{\footnotesize s:$x_{1}(4)$} &
{\footnotesize s:$x_{0}(3)$} &
{\footnotesize r:$x_{0}(3)$} &
{\footnotesize r:$x_{0}(0)$}{\footnotesize \par}

{\footnotesize $\oplus x_{1}(2)$}{\footnotesize \par}

{\footnotesize d:$x_{1}(2)$} &
{\footnotesize s:$x_{0}(0)$}{\footnotesize \par}

{\footnotesize $\oplus x_{1}(2)$} &
{\footnotesize r:$x_{1}(4)$}{\footnotesize \par}

{\footnotesize $\oplus x_{0}(0)$}{\footnotesize \par}

{\footnotesize $\oplus x_{1}(0)$}{\footnotesize \par}

{\footnotesize $\oplus x_{1}(2)$}{\footnotesize \par}

{\footnotesize d:$x_{1}(4)$}\tabularnewline
\cline{2-8} 
 & {\footnotesize 14} &
{\footnotesize s:$x_{0}(4)$}{\footnotesize \par}

{\footnotesize $\oplus x_{1}(4)$} &
{\footnotesize r:$x_{0}(4)$}{\footnotesize \par}

{\footnotesize $\oplus x_{1}(4)$}{\footnotesize \par}

{\footnotesize $\oplus x_{0}(2)$}{\footnotesize \par}

{\footnotesize $\oplus x_{1}(0)$}{\footnotesize \par}

{\footnotesize d:$x_{1}(4)$} &
{\footnotesize s:$x_{0}(2)$}{\footnotesize \par}

{\footnotesize $\oplus x_{1}(0)$} &
{\footnotesize r:$x_{0}(2)$}{\footnotesize \par}

{\footnotesize $\oplus x_{1}(0)$}{\footnotesize \par}

{\footnotesize d:$x_{0}(2)$} &
{\footnotesize r:$x_{1}(3)$} &
{\footnotesize s:$x_{1}(3)$}\tabularnewline
\hline 
\end{tabular}
\end{table}

\emph{Chord Ring Networks}

\begin{figure}
\centering \includegraphics[bb=10bp 10bp 141bp 130bp,clip,width=0.2\textwidth]{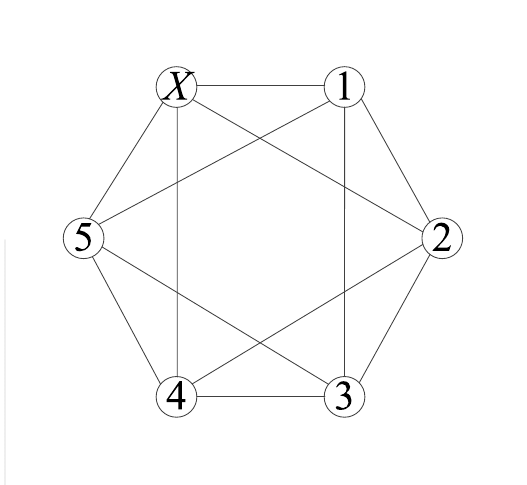}
\caption{A chord ring network}

\label{fig:chord-ring-network}
\end{figure}

Fig. \ref{fig:chord-ring-network} shows a chord ring network with
six nodes. By Theorem \ref{thm:max-throughput}, its broadcast throughput
upper bound is $4/5$. A transmission scheme for this network is described
in Appendix \ref{app:B-Transmission-scheme-for}.

\section{Color-based Scheduling}

Color-based scheduling can be used to achieve the throughput upper
bound in Theorem \ref{thm:max-throughput} in some networks.
\begin{defn}[\textbf{Vertex Coloring}]
Consider a graph $\mathcal{G}=(\mathcal{V},\mathcal{E})$. Let $\mathcal{C}$
be a set of colors. A \emph{vertex coloring} scheme $f:\mathcal{V}\to\mathcal{C}$
assigns to each vertex $v\in\mathcal{V}$ a color $c\in\mathcal{C}$.
\end{defn}

\begin{defn}[\textbf{Colored Graph}]
Consider a graph $\mathcal{G}=(\mathcal{V},\mathcal{E})$ and a particular
vertex coloring scheme $f:\mathcal{V}\to\mathcal{C}$ for it. The
resulting \emph{colored graph} is $\mathcal{G}_{f}=(\mathcal{V},\mathcal{E}_{f},f)$,
where $\mathcal{E}_{f}\subseteq\mathcal{E}$ such that an edge $e=\{k,k'\}\in\mathcal{E}$
is also an element in $\mathcal{E}_{f}$ if and only if $f(k)\neq f(k')$.
In other words, besides assigning colors to the vertices, we also
remove edges between vertices of like color to obtain the colored
graph $\mathcal{G}_{f}$. 
\end{defn}

\begin{defn}[\textbf{Color Group}]
Consider a graph $\mathcal{G}=(\mathcal{V},\mathcal{E})$ and an
associated colored graph $\mathcal{G}_{f}=(\mathcal{V},\mathcal{E}_{f},f)$.
Consider a qualified cut $C$ on the colored graph. The definitions
of $\mathcal{V}_{1}$, $\mathcal{V}_{2}$, $\mathcal{V}_{1}'$, $\mathcal{V}_{2}'$
are the same as in Definition \ref{def:cut-size} and Fig. \ref{fig:cut-partitioning}
(note that given the same $\mathcal{V}_{1}$ and $\mathcal{V}_{2}$
in the qualified cut, the $\mathcal{V}_{1}'$ and $\mathcal{V}_{2}'$
in $\mathcal{G}_{f}$ and the $\mathcal{V}_{1}'$ and $\mathcal{V}_{2}'$
in $\mathcal{G}$ may be different because $\mathcal{E}_{f}\subseteq\mathcal{E}$).
We divide the vertices in $\mathcal{V}_{1}'$ into different sets
according to the vertex color. Each of these vertex sets is called
a \emph{color group}. A \emph{color-$c$ group }is a color group wherein
the vertices have color $c\in\mathcal{C}$.
\end{defn}

\begin{defn}[\textbf{Color-group Cut Size}]
With the same definitions of $\mathcal{G}$, $\mathcal{G}_{f}$,
$C$, $\mathcal{V}_{1}$, $\mathcal{V}_{2}$, $\mathcal{V}_{1}'$,
$\mathcal{V}_{2}'$ as above. Let $\mathcal{U}_{c}$ be a color-$c$
group in $\mathcal{V}_{1}'$. We form a matrix where rows represent
vertices in $\mathcal{U}_{c}$, and columns represent vertices in
$\mathcal{V}_{2}'$. Consider $u\in\mathcal{U}_{c},v\in\mathcal{V}_{2}'$.
The $(u,v)$-th entry is 1 if and only if $(u,v)\in\mathcal{E}$;
it is 0 otherwise. The \emph{color-group cut size} of $\mathcal{U}_{c}$
is the rank of this matrix.
\end{defn}

\begin{defn}[\textbf{Color-cut Size}]
With the same definitions of $\mathcal{G}$, $\mathcal{G}_{f}$,
$C$, $\mathcal{V}_{1}$, $\mathcal{V}_{2}$, $\mathcal{V}_{1}'$,
$\mathcal{V}_{2}'$, $\mathcal{U}_{c}$ as above. Let $G_{c}$ be
the color-group cut size of $\mathcal{U}_{c}$. The color-cut size
of $C$ is the sum of the color-group cut size of all color groups,
$\sum_{c\in\mathcal{C}}G_{c}$.
\end{defn}

\begin{figure}
\centering \includegraphics[bb=10bp 10bp 166bp 149bp,clip,width=0.35\textwidth]{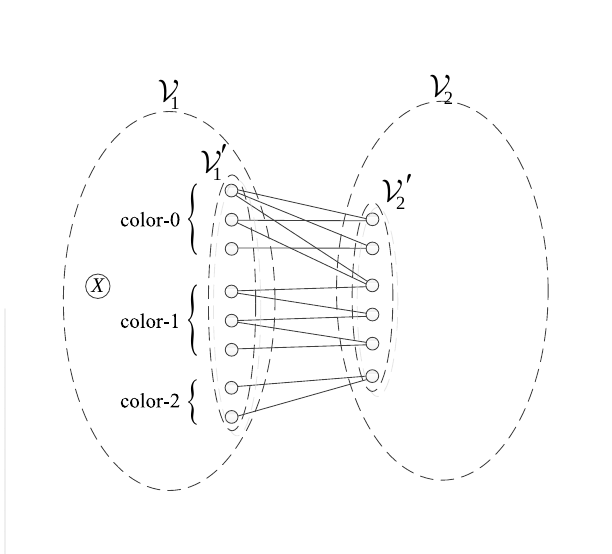}
\caption{A qualified cut with color-cut size 6}

\label{fig:color-cut-size-example}
\end{figure}

For illustration, Fig. \ref{fig:color-cut-size-example} shows the
partitioning of a colored graph with a qualified cut. There are three
colors $\{0,1,2\}$. The color-0 group has a color-group cut size
of 2, as
\[
\mbox{rank}\left[\begin{array}{cccccc}
1 & 1 & 1 & 0 & 0 & 0\\
1 & 0 & 1 & 0 & 0 & 0\\
0 & 1 & 0 & 0 & 0 & 0
\end{array}\right]=2.
\]
The color-1 group has a color-group cut size of 3, as
\[
\mbox{rank}\left[\begin{array}{cccccc}
0 & 0 & 1 & 1 & 0 & 0\\
0 & 0 & 0 & 1 & 1 & 0\\
0 & 0 & 0 & 0 & 1 & 0
\end{array}\right]=3.
\]
The color-2 group has a color-group cut size of 1, as
\[
\mbox{rank}\left[\begin{array}{cccccc}
0 & 0 & 0 & 0 & 0 & 1\\
0 & 0 & 0 & 0 & 0 & 1
\end{array}\right]=1.
\]
Thus, the color-cut size is $2+3+1=6$.

\begin{defn}[\textbf{\label{def:Color-based-Scheduling}Color-based Scheduling}]
Consider a network and an associated colored graph. A \emph{color-based
scheduling} is a schedule such that nodes with the same color transmit
in the same time slots, and nodes with different colors transmit in
different time slots.
\end{defn}
With a color-based scheduling, the color-group cut size represents
the maximum number of linearly independent equations the color group
can deliver across the cut in a time slot.
\begin{thm}
\label{thm:throughput-color-based}Consider a non-trivial network
$\mathcal{G}$ and an associated colored graph $\mathcal{G}_{f}$.
Suppose that the minimum color-cut size among all qualified cuts is
$n_{f}$. Then $\rho\leq n_{f}/(n_{f}+1)$ with any color-based scheduling.\end{thm}
\begin{IEEEproof}[Proof of Theorem \ref{thm:throughput-color-based}]
With respect to Definition \ref{def:Qualified-Cut-min-cut}, let
$C$ be a qualified cut of $\mathcal{G}_{f}$ with minimum color-cut
size $n_{f}$. The number of distinct colors in $\mathcal{V}_{1}'$
is $|f(\mathcal{V}_{1}')|$. We divide the nodes in $\mathcal{V}_{1}'$
into $|f(\mathcal{V}_{1}')|$ color groups. Let $G_{c}$ be the color-group
cut size of color-$c$ group, $c\in f(\mathcal{V}_{1}')\subseteq\mathcal{C}$.
We have
\begin{equation}
n_{f}=\sum_{c\in|f(\mathcal{V}_{1}')|}G_{c}.\label{eq:n_f-sum-of-Gc}
\end{equation}
Let $T_{c}$ be the number of time slots during which nodes in color-$c$
group transmits within the $W_{D}$ time slots (see Definition \ref{def:Throughput}
for the definition of $W_{D}$); let $R_{c}$ be the number of time
slots during which nodes in color-$c$ group receives within the $W_{D}$
time slots. We have
\begin{equation}
W_{D}=T_{c}+R_{c}\;\enskip\forall c\in f(\mathcal{V}_{1}').
\end{equation}
Furthermore, since each node must receive at least $D$ linearly independent
equations to decode the $D$ native packets, we must have 
\begin{equation}
R_{c}\geq D\;\enskip\forall c\in f(\mathcal{V}_{1}').
\end{equation}
In addition, at least $D$ linearly independent equations must be
delivered to nodes in $\mathcal{V}{}_{2}'$ from nodes in $\mathcal{V}{}_{1}'$,
meaning 
\begin{equation}
\sum_{c\in f(\mathcal{V}_{1}')}T_{c}G_{c}\geq D.\label{eq:sum-Tc>=00003DD}
\end{equation}
The LHS of (\ref{eq:sum-Tc>=00003DD}) is from the fact that a color
group can deliver at most $G_{c}$ linearly independent equations
across the cut per time slot. Notice the difference between (\ref{eq:sum-Tc>=00003DD})
and (\ref{eq:T-summation>=00003DD}). Theorem \ref{thm:max-throughput}
gives the general upper bound. Theorem \ref{thm:throughput-color-based}
here considers the upper bound assuming the adoption of color-based
scheduling. Since in color-based scheduling nodes of the same color
transmit in the same time slots, the number of equations crossing
from $\mathcal{V}_{1}'$ to $\mathcal{V}_{2}'$ in the colored graph
is $\sum_{c\in f(\mathcal{V}_{1}')}T_{c}G_{c}$, of which $D$ must
be linearly independent. The network throughput is
\begin{equation}
\rho=\lim_{D\rightarrow\infty}\frac{D}{W_{D}}=\lim_{D\rightarrow\infty}\frac{D}{T_{c}+R_{c}}\;\enskip\forall c\in f(\mathcal{V}_{1}').
\end{equation}
Thus,
\begin{equation}
\rho\leq\min_{c\in f(\mathcal{V}_{1}')}\lim_{D\rightarrow\infty}\frac{D}{T_{c}+D}.
\end{equation}
To obtain an upper bound that is valid for all coloring schemes, we
solve the following optimization problem
\begin{equation}
\max\min_{c\in f(\mathcal{V}_{1}')}\lim_{D\rightarrow\infty}\frac{D}{T_{c}+D}
\end{equation}
such that $\sum_{c\in f(\mathcal{V}_{1}')}T_{c}G_{c}\geq D$. The
minimum $\lim_{D\rightarrow\infty}D/(T_{c}+D)$ is obtained at the
maximum $T_{c}$; therefore, to maximize $\min_{c\in f(\mathcal{V}_{1}')}\lim_{D\rightarrow\infty}D/(T_{c}+D)$,
we need to minimize $\max T_{c}$. Suppose that $T_{c_{1}}=\max T_{c}$
for some $c_{1}\in f(\mathcal{V}_{1}')$. If there exists $T_{c_{2}}$,
$c_{2}\in f(\mathcal{V}_{1}')$, such that 
\begin{equation}
T_{c_{1}}>T_{c_{2}}
\end{equation}
then $T_{c_{1}}=\max T_{c}$ can be made smaller by allocating part
of the time slots from $T_{c_{1}}$ to $T_{c_{2}}$. As $D\rightarrow\infty$,
we require that
\begin{equation}
T_{c_{1}}=T_{c_{2}}=...=\max T_{c}\;\enskip c_{1},c_{2},...\in f(\mathcal{V}_{1}')
\end{equation}
in order that $\max T_{c}$ is minimized. Let $T$ be such $\max T_{c}$.
The solution to the above problem is given by 
\begin{align}
 & \sum_{c\in f(\mathcal{V}_{1}')}T_{c}G_{c}\geq D\Rightarrow T\sum_{c\in f(\mathcal{V}_{1}')}G_{c}\geq D\Rightarrow Tn_{f}\geq D\\
\Rightarrow & T\geq\frac{D}{n_{f}}
\end{align}
An upper bound of $\rho$ is therefore given by 
\begin{equation}
\rho\leq\lim_{D\rightarrow\infty}\frac{D}{\frac{D}{n_{f}}+D}=\frac{n_{f}}{n_{f}+1}
\end{equation}

\end{IEEEproof}
With Theorems \ref{thm:max-throughput} and \ref{thm:throughput-color-based},
we now have
\begin{equation}
\rho\leq\min(\frac{n}{n+1},\frac{n_{f}}{n_{f}+1}),
\end{equation}
where $n$ is the minimum qualified cut in $\mathcal{G}$ and $n_{f}$
is the minimum color-cut size among all qualified cuts in $\mathcal{G}_{f}$.
In general, it is obvious that $n\geq n_{f}$. Thus, if color-based
scheduling is used, $n_{f}/(n_{f}+1)$ is a tighter bound than $n/(n+1)$,
which is obvious since colored-based scheduling is only a subset of
possible scheduling schemes. In order that the general upper bound
$n/(n+1)$ can be achieved with color-based scheduling, we must color
$\mathcal{G}$ in such a way that the resulting colored graph $\mathcal{G}_{f}$
has $n_{f}=n$. We will show in Section \ref{sec:Broadcast-Schemes-grid}
that a specific color-based scheduling scheme can be constructed for
grid networks to achieve the broadcast throughput upper bound $n/(n+1)=2/3$.
In this scheme, the associated colored graph $\mathcal{G}_{f}$ for
the grid has $n_{f}=n=2$.

\section{Broadcast Scheme for Grid Networks\label{sec:Broadcast-Schemes-grid}}

\subsection{Overview}

In grid networks, $n=2$ and thus $\rho\leq n/(n+1)=2/3$. The goal
of this section is to prove that the throughput upper bound is achievable.
As the proof is quite involved, requiring the introduction of a number
of new concepts, we first give an overview of our approach here.

Our proof is a constructive proof. Specifically, we adopt a color-based
scheduling scheme with three colors. 

In Section \ref{sub:Coloring-of-networks}, we first review the concept
of an embedded Hamiltonian cycle within a grid network. The nodes
in the grid are colored according to their position in the embedded
Hamiltonian cycle, with the three colors assigned to successive nodes
in a repetitive manner, as in color 0, color 1, color 2, color 0,
color 1, color 2, and so on. In the resulting colored graph, each
node is guaranteed to have two neighbors of two different colors.
That is, the node and these two neighbors cover the three available
colors. Note that if the neighbors of a node were all of the same
color, then according to Theorem \ref{thm:throughput-color-based},
the throughput upper bound would be $1/2$ (because $n_{f}=1$), and
our target of achieving throughput of $2/3$ would not be possible.In
that light, the Hamiltonian-cycle coloring scheme here is designed
to ensure a necessary condition is not violated. 

In color-based scheduling, nodes with the same color transmit in the
same time slots. In Section \ref{sub:Ternary-transmission-schedule},
we specify when a node should transmit and what it should transmit
during its transmission time slots. The key concept is that we divide
the time slots into rounds, with each round consisting of three time
slots corresponding to the three colors. Thus, each node gets to transmit
once in each round in the time slot associated with its color. Also,
each node receives in two time slots in each round. If the information
received during its reception time slots are independent, then throughput
of $2/3$ is possible. In our scheme, the information transmitted
by a node in a round is basically the sum of the two packets it received
in the last round multiplied by a coefficient. We refer to this coefficient
as the \emph{transmit coefficient}.

The transmit coefficients used by the nodes in the network determine
whether each node can receive two independent linear equations in
each round. In Section \ref{sub:Random-Coefficients}, we describe
a scheme in which random transmit coefficients are used by the node.
Specifically, the transmit coefficient used by a node is drawn from
$GF(2^{s})\backslash\{0\}$ uniform-randomly in each round. This is
an i.i.d. time-varying transmit coefficient assignment scheme, since
the transmit coefficient of a node changes from round to round. We
choose to use this scheme mainly to simplify the proof. We argue that
provided the field size $2^{s}$ is large, with high probability the
linear equations received in all time slots are linearly independent.
Hence, throughput of $2/3$ is achievable.

\subsection{\label{sub:Coloring-of-networks}Coloring of networks by constructing
Hamiltonian cycles}

\begin{figure}
\centering\subfloat[\label{fig:grid-mxn-Hamiltonian-cycle}A Hamiltonian cycle in an $M\times N$
network ($M$ is even).]{\includegraphics[bb=252bp 245bp 380bp 380bp,clip,width=0.23\textwidth]{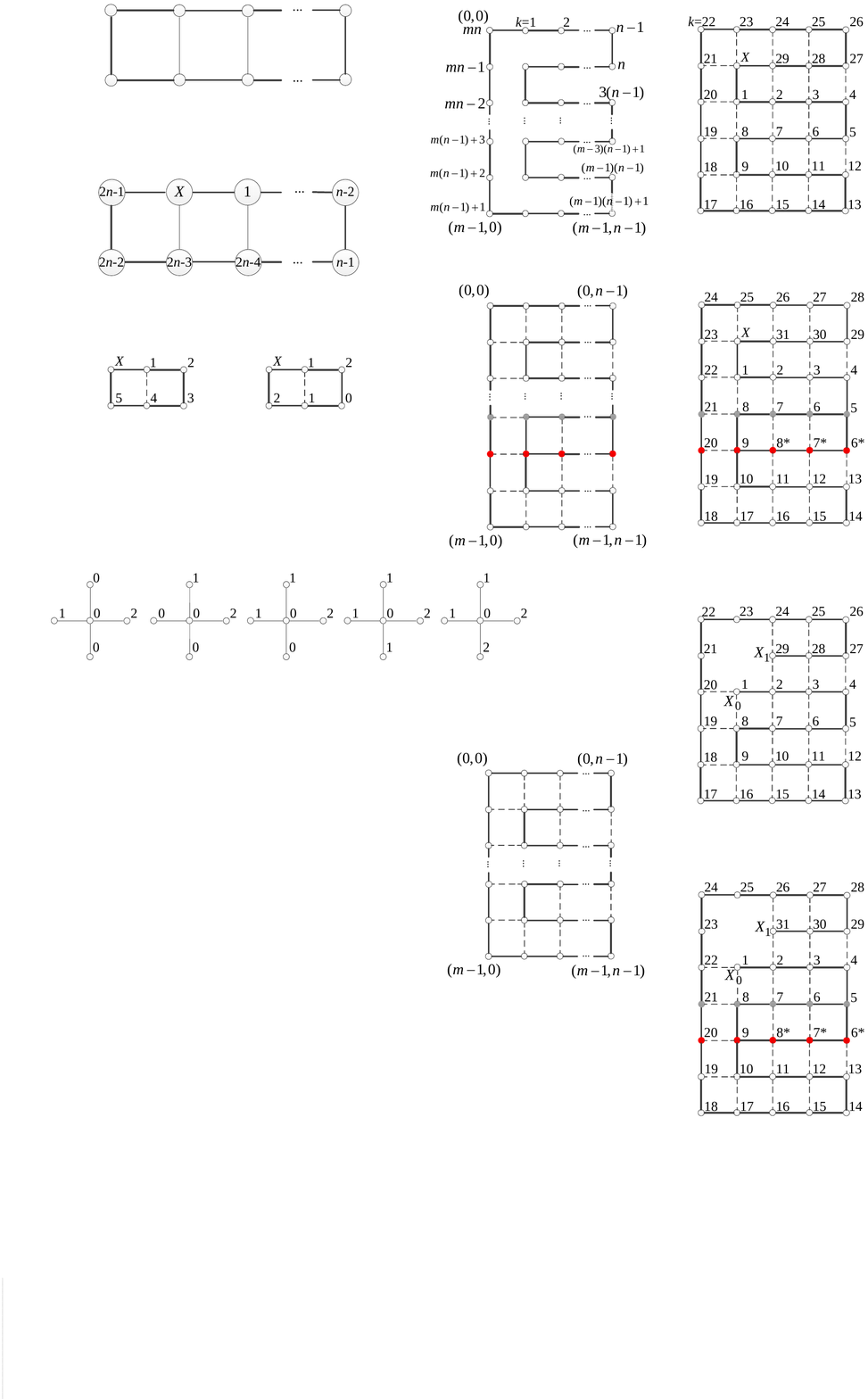}

}\qquad{} \subfloat[\label{fig:grid-6x5-numbering}Numbering of nodes in a $6\times5$
network after renumbering.]{\includegraphics[bb=395bp 678bp 515bp 813bp,clip,width=0.2\textwidth]{6E__Dropbox_PG_Sem6_Thesis_figures_ISIT_Hamiltonian.pdf}

} \caption{$M\times N$ networks with at least one of $M,N$ being even}
\end{figure}

A Hamiltonian cycle is a path that visits each node in a graph exactly
once and ends at its starting point. First, for any $M\times N$ grid
graph with at least one of $M,N$ being even, there is a Hamiltonian
cycle in the graph \cite{bollobas1979graph}. For example, we can
construct a comb-shaped Hamiltonian cycle as shown in Fig. \ref{fig:grid-mxn-Hamiltonian-cycle}.
Starting from a neighbor of $X$ on the cycle, we number along the
path by 1-2-3-...-({\small $MN$}-1). Depicted in Fig. \ref{fig:grid-6x5-numbering}
is an example of this numbering scheme in a $6\times5$ grid with
$X=(1,1)$.

\begin{figure}
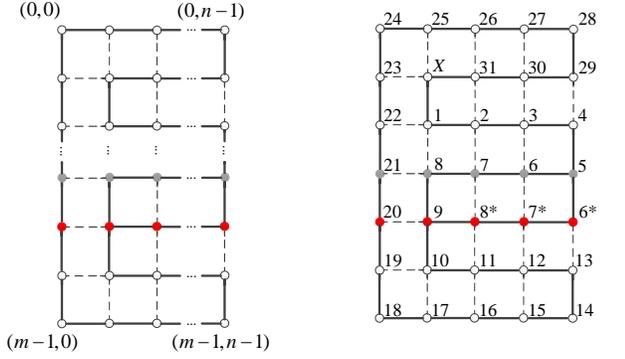

\centering\subfloat[\label{fig:grid-mxn-Hamiltonian-cycle-odd}A Hamiltonian cycle in
an $M\times N$ network ($M$ and $N$ are both odd). Part of the
cycle contains two parallel paths.]{\includegraphics[bb=260bp 495bp 380bp 655bp,clip,width=0.2\textwidth]{6E__Dropbox_PG_Sem6_Thesis_figures_ISIT_Hamiltonian.pdf}

}\qquad{} \subfloat[\label{fig:grid-7x5-numbering}Numbering of nodes in a $7\times5$
network.]{\includegraphics[bb=395bp 495bp 515bp 665bp,clip,width=0.2\textwidth]{6E__Dropbox_PG_Sem6_Thesis_figures_ISIT_Hamiltonian.pdf}

} \caption{$M\times N$ networks with both $M,N$ being odd}
\end{figure}

Next, if $M$ and $N$ are both odd, then it is not possible to construct
a Hamiltonian cycle in the grid \cite{bollobas1979graph}. Instead,
we construct a pseudo Hamiltonian cycle called the ``Split-Merge
Hamiltonian cycle'', as illustrated in Fig. \ref{fig:grid-mxn-Hamiltonian-cycle-odd}
and \ref{fig:grid-7x5-numbering}. As shown in Fig. \ref{fig:grid-7x5-numbering},
after visiting node 5, the visits split into two paths. Nodes 6 and
6{*} are visited in parallel next; node 7 and 7{*} after that; and
then node 8 and 8{*}; finally the parallel visits merge back to node
9. By splitting and merging as such, a result is the insertion of
a new row (as indicated red in Fig. \ref{fig:grid-mxn-Hamiltonian-cycle-odd}
and \ref{fig:grid-7x5-numbering}) into an $(M-1)\times N$ network
which already has a Hamiltonian cycle constructed (because $M-1$
is even, we can construct such a cycle).

\textbf{Hamiltonian Node Coloring\label{Hamiltonian-Node-Coloring}:}
Now that we have a node numbering scheme for all grid networks, we
\emph{partition the non-source nodes into nodes of three distinct
colors} $c\in\{0,1,2\}$ with a vertex coloring scheme given by $f(k)=k\bmod3$.
That is, we assign node $k$ the color $c=k\bmod3$. We also let $f(k^{*}):=f(k)$,
i.e., let the node $k^{*}$ in the split paths have the same color
as node $k$.

\begin{figure}
\centering\subfloat[\label{fig:grid-6x5-colored}$6\times5$]{\includegraphics[bb=23bp 6bp 128bp 149bp,clip,width=0.2\textwidth]{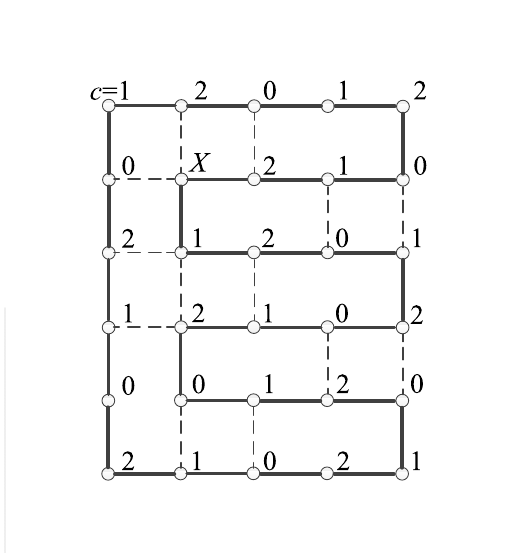}

}\qquad{} \subfloat[\label{fig:grid-7x5-colored}$7\times5$]{\includegraphics[bb=15bp 18bp 120bp 161bp,clip,width=0.2\textwidth]{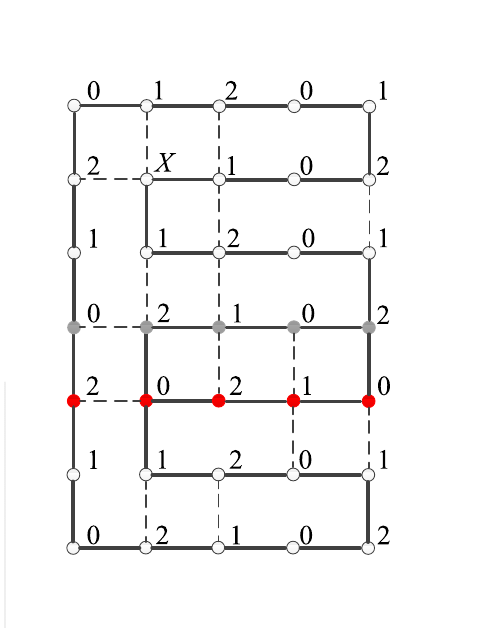}

} \caption{\label{fig:Colored-graphs}Colored graphs of $6\times5$ and $7\times5$
grid networks.}
\end{figure}

Fig. \ref{fig:Colored-graphs} shows the colored graphs produced after
applying the Hamiltonian coloring to networks in Fig. \ref{fig:grid-6x5-numbering}
and Fig. \ref{fig:grid-7x5-numbering}. Once we have colored the nodes,
we then remove links between nodes of the same color. The result is
a colored graph embedded in the original grid graph. The minimum qualified
cut size is $n=2$ in grid networks. One can verify that the minimum
color-cut size is $n_{f}=2$ in grid networks with the coloring scheme
above. Therefore by Theorem \ref{thm:throughput-color-based} the
broadcast throughput upper bound is still $2/3$, thus is not decreased
by the Hamiltonian coloring scheme.

The motivation for the above Hamiltonian coloring is as follows. A
node $k$ that is two or more hops away from $X$ can only receive
the broadcast information from $X$ through its neighbors $\mathcal{N}(k)$.
In our time-slotted scheme, the nodes with the same color transmit
at the same time. The Hamiltonian coloring scheme ensures that each
node $k$ has at least two neighbors assigned with the two colors
different from the color of node $k$. These two other colors correspond
to the time slots in which node $k$ receives. Thus, each node $k$
receives in at least two time slots out of every three time slots.
Section \ref{sub:Ternary-transmission-schedule} specifies this transmission
scheme more exactly.

\subsection{\label{sub:Ternary-transmission-schedule}Ternary transmission schedule}

Our basic idea is to let nodes with color $c$ transmit in time slots
$3t+c,t=0,1,2...$ As a consequence, every node transmits once and
receives twice in every round\emph{ }$t,t=0,1,2,...$, a set of three
time slots $\{3t,3t+1,3t+2\}$.

If we could ensure that in each round, every node receives two packets
that contain new information, then the broadcast throughput would
be $2/3$, which is the upper bound given by Theorem \ref{thm:max-throughput}.
Toward that end, we propose a schedule with the following four rules
(also summarized in Table \ref{tab:Transmission-schedule}):

\begin{table*}
\caption{\label{tab:Transmission-schedule}Ternary transmission schedule}
\centering\setlength\tabcolsep{2pt}%
\begin{tabular}{|c|>{\centering}m{1.1cm}|>{\centering}m{1.6cm}|>{\centering}m{5cm}|>{\centering}m{1.8cm}|>{\centering}m{1.8cm}|>{\centering}m{1.8cm}|}
\hline 
\emph{ts } &
{\footnotesize Node $X$} &
{\footnotesize $X_{0}$ (Node 1)} &
{\footnotesize $X_{1}$ (Node $MN-1$)} &
{\footnotesize Color-0 nodes} &
{\footnotesize Color-1 nodes} &
{\footnotesize Color-2 nodes}\tabularnewline
\hline 
\hline 
{\footnotesize $3t$} &
{\footnotesize $x_{0}(t)$} &
{\footnotesize -} &
{\footnotesize $x_{1}(t-1)$ if $(MN-1)\bmod3=0$; nothing otherwise} &
{\footnotesize $\alpha_{k}(z_{0}^{k}(t)\oplus z_{1}^{k}(t))$} &
{\footnotesize -} &
{\footnotesize -}\tabularnewline
\hline 
{\footnotesize $3t+1$} &
{\footnotesize $x_{1}(t)$} &
{\footnotesize $x_{0}(t-1)$} &
{\footnotesize $x_{1}(t-1)$ if $(MN-1)\bmod3=1$; nothing otherwise} &
{\footnotesize -} &
{\footnotesize $\alpha_{k}(z_{0}^{k}(t)\oplus z_{1}^{k}(t))$} &
{\footnotesize -}\tabularnewline
\hline 
{\footnotesize $3t+2$} &
{\footnotesize $x_{0}(t)\oplus x_{1}(t)$} &
{\footnotesize -} &
{\footnotesize $x_{1}(t-1)$ if $(MN-1)\bmod3=2$; nothing otherwise} &
{\footnotesize -} &
{\footnotesize -} &
{\footnotesize $\alpha_{k}(z_{0}^{k}(t)\oplus z_{1}^{k}(t))$}\tabularnewline
\hline 
\end{tabular}
\end{table*}

\emph{Rule 1)} \label{enu:rule-source}\textbf{Transmissions by source
$\bm{X}$}: Let the sequence of native packets to be broadcast by
$X$ be $\{x_{0}(t)\}_{t=0,1,2,..}$ and $\{x_{1}(t)\}_{t=0,1,2,..}$.
In time slot $3t$, $X$ transmits $x_{0}(t)$; in time slot $3t+1$,
$X$ transmits $x_{1}(t)$; in time slot $3t+2$, $X$ transmits $x_{0}(t)\oplus x_{1}(t)$.

\emph{Rule 2)} \label{enu:rule-non-source}\textbf{Transmissions by
nodes not adjacent to $\bm{X}$}: In time slot $3t+c,c\in\{0,1,2\}$,
node $k$, $k\notin\mathcal{N}(X)$, with color $c$ transmits\vspace{-10pt}
 
\[
y^{k}(t)=\alpha_{k}(z_{0}^{k}(t)\oplus z_{1}^{k}(t)),
\]
where $\alpha_{k}\in GF(2^{s})\backslash\{0\}$ is a transmit coefficient%
\footnote{We will treat the case of time-varying transmit coefficients later.
Here, we omit the dependency of the coefficients on time for simple
presentation.%
}, and{\small 
\[
z_{0}^{k}(t)=r_{0}^{k}(t-1)
\]
\[
z_{1}^{k}(t)=r_{1}^{k}(t-1)
\]
}In the above, $r_{0}^{k}(t-1),r_{1}^{k}(t-1)$ are the two packets
node $k$ received from its neighbors in the previous round in time
slots $3t-3+((c-1)\bmod3)$ and $3t-3+((c+1)\bmod3)$.

\emph{Rule 3)} \label{enu:rule-virtual-source}\textbf{Transmissions
by node 1 and node (}\textbf{\emph{\small MN}}\textbf{-1)}: Node $1$
and node ({\small $MN$}-1) only transmit native packets times a transmit
coefficient. Specifically, node 1 transmits $x_{0}(t-1)$ in time
slot $3t+1$. Node ({\small $MN$}-1) transmits $x_{1}(t-1)$ in time
slot $3t+c$, where {\small $c=(MN-1)\bmod3$} is its color.\\
We refer to these two nodes as ``\emph{virtual sources}''. Node
$1$ and node ({\small $MN$}-1) are responsible for forwarding $\{x_{0}(t)\}_{t=0,1,2,..}$
and $\{x_{1}(t)\}_{t=0,1,2,..}$, respectively. 

The role of nodes $1$ and ({\small $MN$}-1) is similar to that of
nodes $1$ and $5$ in the ring example in Section \ref{sec:Broadcast-Schemes-line-ring}.
In fact, our scheduling strategy for the grid network here is inspired
by the strategy for the ring network. If all links other than those
in the embedded Hamiltonian cycle are removed from the grid, we will
then have a ring, and the simple ring scheduling will work to give
a throughput of $2/3$. Unfortunately, because of the interference
from other links, the situation in the grid network is a bit more
complicated. Henceforth, let node 1 be denoted by $X_{0}$ and node
({\small $MN$}-1) be denoted by $X_{1}$. Being adjacent to $X$,
they can both derive $x_{0}(t-1)$ and $x_{1}(t-1)$ by the end of
round $t-1$, as explained below.

In the three time slots in round $t-1$, source $X$ transmits $x_{0}(t-1)$,
$x_{1}(t-1)$, and $x_{0}(t-1)\oplus x_{1}(t-1)$ respectively. Since
each neighbor of $X$ is colored with one color only, it is in the
receive mode in two of the three time slots. Both $x_{0}(t-1)$ and
$x_{1}(t-1)$ can be derived by any neighbor of $X$ (including $X_{0}$
and $X_{1}$) based on the receptions in these two time slots. 

\emph{Rule 4)} \label{enu:rule-adjacent-nodes-of-X}\textbf{Transmissions
by neighbors of $\bm{X}$ who are not $\bm{X}_{0}$ or $\bm{X}_{1}$}:
By this rule, we ensure that only the two virtual sources can send
out the newest native packets of $X$. An adjacent node $k$ of $X$,
who is not $X_{0}$ or $X_{1}$, can also derive $x_{0}(t-1)$ and
$x_{1}(t-1)$ by the end of round $t-1$. In round $t$, node $k$
transmits 
\[
y^{k}(t)=\alpha_{k}(z_{0}^{k}(t)\oplus z_{1}^{k}(t)),
\]
where $\alpha_{k}\in GF(2^{s})\backslash\{0\}$ is a transmit coefficient,
and 
\[
z_{0}^{k}(t)=r_{0}^{k}(t-1)-x_{0}'(t-1)
\]
\[
z_{1}^{k}(t)=r_{1}^{k}(t-1)-x_{1}'(t-1).
\]
In the above, $r_{0}^{k}(t-1),r_{1}^{k}(t-1)$ are the two packets
node $k$ received from its neighbors in the previous round in time
slots $3t-3+((c-1)\bmod3)$ and $3t-3+((c+1)\bmod3)$, respectively.
$x_{0}'(t-1)$ and $x_{1}'(t-1)$ are the two packets sent by $X$
and received by this node in time slots $3t-3+((c-1)\bmod3)$ and
$3t-3+((c+1)\bmod3)$, respectively. For example, if a node $k$ has
color 0, then $x_{0}'(t-1)$ and $x_{1}'(t-1)$ are the two packets
sent by $X$ in time slots $3t-1$ and $3t-2$, respectively; i.e.,
$x_{0}'(t-1)=x_{0}(t-1)\oplus x_{1}(t-1)$ and $x_{1}'(t-1)=x_{1}(t-1)$.

\begin{figure}
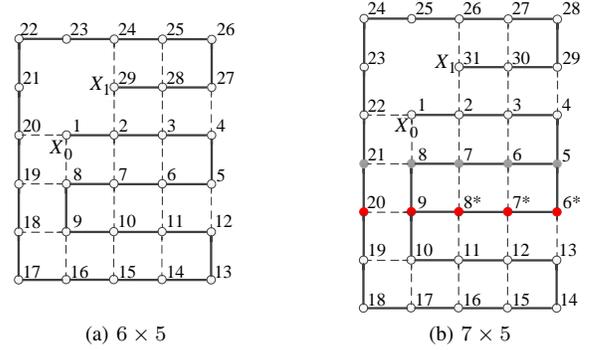

\centering\subfloat[\label{fig:grid-6x5-numbering-equivalent}$6\times5$]{\includegraphics[bb=395bp 335bp 515bp 475bp,clip,width=0.2\textwidth]{6E__Dropbox_PG_Sem6_Thesis_figures_ISIT_Hamiltonian.pdf}

}\qquad{} \subfloat[\label{fig:grid-7x5-numbering-equivalent}$7\times5$]{\includegraphics[bb=395bp 165bp 515bp 305bp,clip,width=0.2\textwidth]{6E__Dropbox_PG_Sem6_Thesis_figures_ISIT_Hamiltonian.pdf}

} \caption{\label{fig:grid-equivalent-networks}Equivalent networks after removing
$X$}
\end{figure}

\begin{figure}
\centering\subfloat[\label{fig:colored-graph-6x5-equivalent}$6\times5$]{\includegraphics[bb=12bp 7bp 120bp 149bp,clip,width=0.2\textwidth]{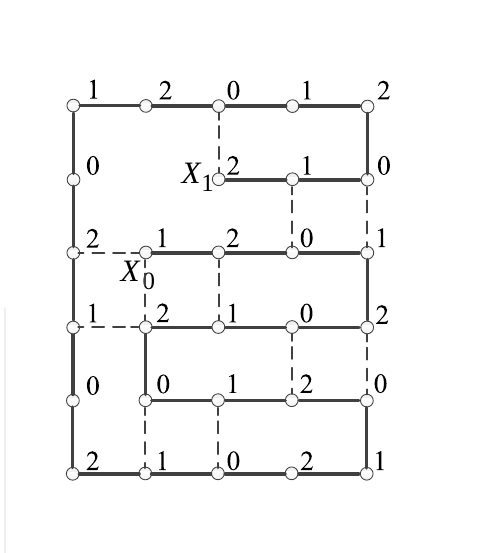}

}\qquad{} \subfloat[\label{fig:colored-graph-7x5-equivalent}$7\times5$]{\includegraphics[bb=12bp 20bp 120bp 162bp,clip,width=0.2\textwidth]{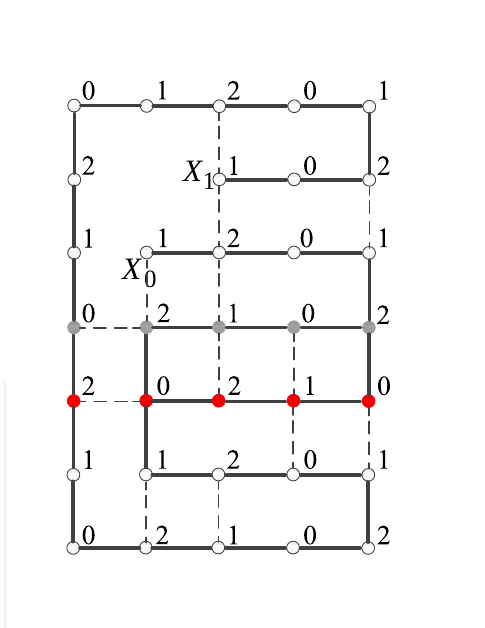}

} \caption{\label{fig:colored-graph-equivalent}Colored graph of the equivalent
networks}
\end{figure}

\textbf{Transformation to Two-Source Broadcast Problem:} With the
above rules, the virtual sources $X_{0}$ and $X_{1}$ can be considered
as the origins of the newest information. The single-source networks
in Fig. \ref{fig:grid-6x5-numbering} and Fig. \ref{fig:grid-7x5-numbering}
can then be transformed to two-source networks in Fig. \ref{fig:grid-equivalent-networks}.
In Fig. \ref{fig:grid-6x5-numbering-equivalent}, nodes 1 and 29 are
$X_{0}$ and $X_{1}$, respectively; in Fig. \ref{fig:grid-7x5-numbering-equivalent},
nodes 1 and 31 are $X_{0}$ and $X_{1}$, respectively. Fig. \ref{fig:colored-graph-equivalent}
shows the colored graph of the networks in Fig. \ref{fig:grid-equivalent-networks}.

\begin{figure}
\centering\subfloat[\label{fig:2x3-numbering}Numbering]{\includegraphics[bb=44bp 576bp 134bp 611bp,clip,width=0.2\textwidth]{6E__Dropbox_PG_Sem6_Thesis_figures_ISIT_Hamiltonian.pdf}

}\qquad{} \subfloat[\label{fig:2x3-color}Colored Graph]{\includegraphics[bb=2bp 18bp 92bp 53bp,clip,width=0.2\textwidth]{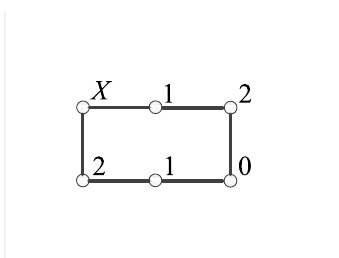}

} \caption{A $2\times3$ network example}
\end{figure}

\medskip{}

\emph{Example}

We first illustrate what happens when applying this schedule to a
simple $2\times3$ grid network. The source is located at (0,0). Fig.
\ref{fig:2x3-numbering} shows the numbering of this network and Fig.
\ref{fig:2x3-color} shows the corresponding colored graph. Although
node 1 and node 4 can overhear each other, they are of the same color
and thus transmit at the same time. Hence they will not interfere
each other. The network turns into a ring after coloring. In this
example, we set all transmit coefficients $\alpha_{k}$ to $1$. The
transmissions of our ternary schedule are the same as those shown
in Table \ref{tab:Transmission-schedule-ring} for the six-node ring
example. 

It can be observed that the two virtual sources node 1 and node 5
always have $x_{0}(0),...,x_{0}(t)$ and $x_{1}(0),...,x_{1}(t)$
by the end of round $t$. Therefore they can always derive $x_{0}(t+1)$
and $x_{1}(t+1)$ by the end of round $t+1$ even if they receive
PNC packets during round $t+1$, because they know all but one of
the unknowns (native packets).

In this example, each node can obtain two native packets in a round.
In a general grid network where there is interference among nodes,
obtaining native packets as such cannot be guaranteed. However, as
will be shown, we could ensure every non-source node still obtains
two linearly independent equations in each round.

\medskip{}

In a general grid network, depending on its position in the grid,
a node can have up to four neighbors. With the Hamiltonian Node Coloring,
it is possible for a node to have two or three neighbors of the same
color, one of which is an adjacent node on the Hamiltonian cycle (see
Section \ref{Hamiltonian-Node-Coloring} on Hamiltonian Node Coloring).
When multiple neighbors of the same color transmit simultaneously,
the node receives a PNC packet, for which the XOR of the simultaneous
transmissions of the neighbors of the same color is received. For
example, in Fig. \ref{fig:grid-7x5-numbering}, node 9 (color-0) has
four neighbors: node 8 (color-2), node 8{*} (color-2), node 10 (color-1)
and node 20 (color-2). As a consequence, in each round node 9 receives
from three nodes simultaneously in the color-2 time slot, which yields
a PNC packet; it receives from only one neighbor in the color-1 time
slot. For both packets, we need to make sure:
\begin{enumerate}
\item the packet received is linearly independent with all packets previously
received;
\item the packet received is not null.
\end{enumerate}

\subsection{\label{sub:Random-Coefficients}Random Transmit Coefficients}

In the previous simple $2\times3$ example, the transmit coefficients
$\alpha_{k}$ for all non-source nodes were set to $1$. In a general
grid, this scheme may not work. Henceforth, we consider a time-varying
random transmit coefficient scheme. Specifically, the transmit coefficient
$\alpha_{k}(t)$ of node $k$ in round $t$ is chosen uniform-randomly
from the non-zero elements of $GF(2^{s})$, and $\alpha_{k}(t)$ for
different $k$ and $t$ are i.i.d. The sequences of packets $\{x_{0}(t)\}_{t=0,1,2,...}$
and $\{x_{1}(t)\}_{t=0,1,2,...}$ transmitted by the virtual sources
$X_{0}$ and $X_{1}$ remain the same, and their transmit coefficients
can be considered as 1 throughout the process.

In grid networks, the newest information (i.e., signals embedded with
the latest native packets) come through a shortest path from the the
virtual sources $X_{0}$ and $X_{1}$, and in general this shortest
path may not be along the Hamiltonian cycle.
\begin{defn}[\textbf{Shortest Path}]
A \emph{shortest path} from a virtual source to a node is a shortest
sequence of adjacent nodes leading from the virtual source to the
node in the colored graph of the grid network.
\end{defn}
Note that in general there could be multiple shortest paths of the
same length leading from a virtual source to a node, and some of them
may share some common intermediate nodes.

\begin{defn}[\textbf{Coefficient Product/Path Coefficient}]
\textbf{\label{def:The-coefficient-product}}The \emph{coefficient
product} of $p=k_{1}-k_{2}-...-k_{l}-k$, a path from node $k_{1}$
to node $k$, in round $t$ is 
\begin{align}
g_{p}(t)= & \alpha_{k_{1}}(t-(l-1))\alpha_{k_{2}}(t-(l-2))\nonumber \\
 & ...\alpha_{k_{l-1}}(t-1)\alpha_{k_{l}}(t)\\
= & \prod_{r=1}^{l}\alpha_{k_{r}}(t-(l-r)),
\end{align}
where $\alpha_{k_{1}}(t-(l-1)),\alpha_{k_{2}}(t-(l-2),...,\alpha_{k_{l-1}}(t-1),\alpha_{k_{l}}(t)$
are the transmit coefficients of nodes $k_{1},k_{2},...,k_{l-1},k_{l}$
in rounds $t-(l-1),t-(l-2),...,t-1,t$, respectively. We will use
the terms ``coefficient product'' and ``path coefficient'' interchangeably
in this paper.
\end{defn}
If a native packet $x$ goes through a path $p$, then its coefficient
when it arrives at the last node in round $t$ will be $g_{p}(t)$,
the coefficient product (path coefficient) of this path. 
\begin{defn}[\textbf{Aggregated Path Coefficient}]
If a packet $x$ begins its journey from a node $j$ with an initial
transmit coefficient $\alpha_{j}$, splits and travels over multiple
paths $p_{1},p_{2},...$ of the same length, and then arrives at the
same node $k$ in the same time slot in round $t$, then the coefficient
of the packet $x$ when it is received at node $k$ is the sum of
coefficient products $\alpha_{j}g_{p_{1}}(t)\oplus\alpha_{j}g_{p_{2}}(t)\oplus...$,
thanks to PNC. This sum of coefficient products will be referred to
as the aggregated path coefficient.
\end{defn}
\medskip{}

\emph{Example of received packets}

For example in Fig. \ref{fig:grid-7x5-numbering}, in time slot $3t$,
node 10 receives from node 9. The shortest path from $X_{0}$ to node
$10$ for this time slot is $p_{0}=1-8-9-10$. However, this is the
shortest path for $\{x_{0}(t)\}_{t=0,1,...}$ only, because node $1$
only transmits $\{\alpha_{1}(t)x_{0}(t)\}_{t=0,1,...}$. The shortest
paths from $X_{1}$ to node $10$ for $\{x_{1}(t)\}_{t=0,1,...}$
are $p_{1}=31-2-1-8-9-10$, $p_{2}=31-2-7-8-9-10$ and $p_{3}=31-2-7-8^{*}-9-10$.
A native packet in $\{x_{0}(t)\}_{t=0,1,...}$ will be multiplied
by the transmit coefficient of the sender when being sent. The packet
node $10$ receives from node $9$ in time slot $3t$ is 
\begin{align}
r_{0}^{10}(t)= & g_{p_{0}}(t)x_{0}(t-3)\nonumber \\
 & \oplus(g_{p_{1}}(t)\oplus g_{p_{2}}(t)\oplus g_{p_{3}}(t))x_{1}(t-5)\nonumber \\
 & \oplus\epsilon(x_{0}(t-4),x_{0}(t-5),...,x_{0}(0))\nonumber \\
 & \oplus\epsilon(x_{1}(t-6),x_{1}(t-7),...,x_{1}(0))\\
= & \alpha_{1}(t-2)\alpha_{8}(t-1)\alpha_{9}(t)x_{0}(t-3)\nonumber \\
 & \oplus(\alpha_{31}(t-4)\alpha_{2}(t-3)\alpha_{1}(t-2)\alpha_{8}(t-1)\alpha_{9}(t)\nonumber \\
 & \oplus\alpha_{31}(t-4)\alpha_{2}(t-3)\alpha_{7}(t-2)\alpha_{8}(t-1)\alpha_{9}(t)\nonumber \\
 & \oplus\alpha_{31}(t-4)\alpha_{2}(t-3)\alpha_{7}(t-2)\alpha_{8^{*}}(t-1)\alpha_{9}(t))\nonumber \\
 & x_{1}(t-5)\nonumber \\
 & \oplus\epsilon(x_{0}(t-4),x_{0}(t-5),...,x_{0}(0))\nonumber \\
 & \oplus\epsilon(x_{1}(t-6),x_{1}(t-7),...,x_{1}(0))\label{eq:y^10_0(t+1)}
\end{align}
where $\epsilon(\cdot)$ is a linear combination of the arguments.
We see that the coefficient associated with a native packet embedded
in a reception is in general an aggregated path coefficient. 

The time indexes of the newest native packets in Eqn. (\ref{eq:y^10_0(t+1)})
escalate over time. Therefore each new $r_{0}^{10}(t)$ is linearly
independent of all $r_{0}^{10}(t'),t'<t$.

Packet $r_{1}^{10}(t)$, the packet node 10 receives in time slot
$3t+2$, will be a PNC packet from nodes 11 and 17. A shortest path
in this time slot for $\{x_{0}(t)\}_{t=0,1,...}$ is $1-8-21-20-19-18-17-10$;
and a shortest path for $\{x_{1}(t)\}_{t=0,1,2,...}$ is $31-2-7-8^{*}-7^{*}-12-11-10$.
Therefore the newest native packets in it would be $x_{0}(t-7)$ and
$x_{1}(t-7)$, each multiplied by some $\bigoplus_{p\in S}g_{p}(t)$,
where $S$ is the set of all shortest paths and $p$ is a shortest
path for the native packet. Note that for time slot $3t+2$, an even
shorter path exists for $\{x_{0}(t)\}_{t=0,1,...}$: 1-8-9-10-17-10.
However, this path has a loop and in general we do not consider paths
with loops because they can be eliminated in our computation for the
solution. For this path, a packet is sent from node 10 and cycled
back from node 17. The cycled back packet is $\alpha_{17}(t)r_{0}^{10}(t-1)+...$,
in which $\alpha_{17}(t)r_{0}^{10}(t-1)$ is the coefficient of node
17 times a packet already received by node 10, thus can be removed
easily. In fact, any cycled back information can be removed with the
knowledge of nodes on the cycle and the packets previously received.
Therefore, we only consider shortest paths with no cycles in this
work.

\medskip{}

\emph{Packets received by a general node}

We now give a general expression for received packets. Consider a
general node $k,k\neq1\mbox{ or }MN-1$, in the network. Focus on
one of its two receiving time slots in round $t$. Let $S_{0},S_{1}$
be the sets of shortest paths for $\{x_{0}(t)\}_{t=0,1,...}$ and
$\{x_{1}(t)\}_{t=0,1,...}$, respectively, in this time slot; and
$S_{0}^{+q},S_{1}^{+q}$ be the sets of paths that are $q$ hops longer
than $S_{0}$ and $S_{1}$ for $\{x_{0}(t)\}_{t=0,1,...}$ and $\{x_{1}(t)\}_{t=0,1,...}$,
respectively. Depending on $q$, each of $S_{0}^{+q}$ and $S_{1}^{+q}$
may or may not be empty. Node $k$ receives 
\begin{align}
r_{0}^{k}(t)= & a_{S_{0}}(t)x_{0}(t-i_{0})\oplus a_{S_{0}^{+1}}(t)x_{0}(t-i_{0}-1)\nonumber \\
 & \oplus...\oplus a_{S_{0}^{+t-i_{0}}}(t)x_{0}(0)\nonumber \\
 & \oplus a_{S_{1}}(t)x_{1}(t-i_{1})\oplus a_{S_{1}^{+1}}(t)x_{1}(t-i_{1}-1)\nonumber \\
 & \oplus...\oplus a_{S_{1}^{+t-i_{1}}}(t)x_{1}(0),\label{eq:z_0(t)}
\end{align}
where $i_{0}$ and $i_{1}$ are the lengths of paths in $S_{0}$ and
$S_{1}$, respectively; and where the aggregated path coefficients
are given by 

\begin{equation}
a_{S_{0}}(t)=\bigoplus_{p\in S_{0}}g_{p}(t),\; a_{S_{1}}(t)=\bigoplus_{p\in S_{1}}g_{p}(t),\label{eq:a_S0-a_S1}
\end{equation}
\begin{equation}
a_{S_{0}^{+q}}(t)=\bigoplus_{p\in S_{0}^{+q}}g_{p}(t),\; a_{S_{1}^{+q}}(t)=\bigoplus_{p\in S_{1}^{+q}}g_{p}(t).\label{eq:a_S0+q}
\end{equation}
Grouping the packets received in this colored slot in all rounds,
we have a linear equation system $\{r_{0}^{k}(t)\}_{t\geq\min(i_{0},i_{1})}$.

Now consider the other receiving time slot of node $k$ in round $t$.
Let $T_{0},T_{1}$ be the sets of shortest paths for $\{x_{0}(t)\}_{t=0,1,...}$
and $\{x_{1}(t)\}_{t=0,1,...}$, respectively, in this time slot;
and $T_{0}^{+q},T_{1}^{+q}$ be the sets of paths that are $q$ hops
longer than $T_{0}$ and $T_{1}$ for $\{x_{0}(t)\}_{t=0,1,...}$
and $\{x_{1}(t)\}_{t=0,1,...}$, respectively. Node $k$ receives
\begin{align}
r_{1}^{k}(t)= & a_{T_{0}}(t)x_{0}(t-j_{0})\oplus a_{T_{0}^{+1}}(t)x_{0}(t-j_{0}-1)\nonumber \\
 & \oplus...\oplus a_{T_{0}^{+t-j_{0}}}(t)x_{0}(0)\nonumber \\
 & \oplus a_{T_{1}}(t)x_{1}(t-j_{1})\oplus a_{T_{1}^{+1}}(t)x_{1}(t-j_{1}-1)\nonumber \\
 & \oplus...\oplus a_{T_{1}^{+t-j_{1}}}(t)x_{1}(0),\label{eq:z_1(t)}
\end{align}
where $j_{0}$ and $j_{1}$ are the lengths of paths in $T_{0}$ and
$T_{1}$, respectively; and where the aggregated path coefficients
are given by 

\begin{equation}
a_{T_{0}}(t)=\bigoplus_{p\in T_{0}}g_{p}(t),\; a_{T_{1}}(t)=\bigoplus_{p\in T_{1}}g_{p}(t),\label{eq:a_T0-a_T1}
\end{equation}
\begin{equation}
a_{T_{0}^{+q}}(t)=\bigoplus_{p\in T_{0}^{+q}}g_{p}(t),\; a_{T_{1}^{+q}}(t)=\bigoplus_{p\in T_{1}^{+q}}g_{p}(t).\label{eq:a_T0+q}
\end{equation}
Grouping the packets received in this colored slot in all rounds,
we have another linear equation system $\{r_{1}^{k}(t)\}_{t\geq\min(j_{0},j_{1})}$.

Note that we do not consider paths with cycles, because they can be
eliminated in our computation for the solution. For example, if a
packet $y^{k}(t)$ sent by node $k$ in round $t$ is cycled back
via a path $k-k_{1}-k_{2}-...-k_{l}-k$, then the resulting component
in a packet received in round $t+l$ will be $\alpha_{k_{1}}(t+1)\alpha_{k_{2}}(t+2)...\alpha_{k_{l}}(t+l)y^{k}(t)$,
which can be eliminated from the packet, as node $k$ already knows
$y^{k}(t)$. In other words, cycle-back information does not contain
anything new.

Each of the virtual sources, $X_{0}$ or $X_{1}$, broadcasts $D/2$
native packets (assuming $D$ is even for simplicity). That is, $\{x_{0}(t)\}_{t=0,1,...}=\{x_{0}(t)\}_{t=0,1,...,\frac{D}{2}-1}$
and $\{x_{1}(t)\}_{t=0,1,...}=\{x_{1}(t)\}_{t=0,1,...,\frac{D}{2}-1}$.
After all native packets have been sent by the source, we allow $MN$
more time slots for them to circulate in the network. During these
$MN$ time slots, the source and the virtual sources can be considered
as transmitting null packets, which do not increase the number of
unknowns in the network. 

We select $\{r_{0}^{k}(t_{0})\}_{\min(i_{0},i_{1})\leq t_{0}\leq\max(i_{0},i_{1})+\frac{D}{2}-1}$,
a subset of $\{r_{0}^{k}(t)\}_{t\geq\min(i_{0},i_{1})}$, and $\{r_{1}^{k}(t_{1})\}_{\min(j_{0},j_{1})\leq t_{1}\leq\max(j_{0},j_{1})+\frac{D}{2}-1}$,
a subset of $\{r_{1}^{k}(t)\}_{t\geq\min(j_{0},j_{1})}$, and group
them together as a single linear equation system

\begin{equation}
\{r_{0}^{k}(t_{0}),r_{1}^{k}(t_{1})\}_{\substack{\min(i_{0},i_{1})\leq t_{0}\leq\max(i_{0},i_{1})+\frac{D}{2}-1\\
\min(j_{0},j_{1})\leq t_{1}\leq\max(j_{0},j_{1})+\frac{D}{2}-1
}
}.\label{eq:grouped-linear-equation-system}
\end{equation}

\medskip{}

Before we discuss the relationship between the equations, we introduce
some additional definitions. In the following, we will use $|p|$
to represent the length of a path $p$, i.e., the number of hops in
$p$.

\begin{defn}[\textbf{Equal-hop Path Set}]
An equal-hop path set $S$ is a set of paths of the same length.
For example $S=\{p_{1},p_{2},...\}$, where $|p_{1}|=|p_{2}|=...$,
is an equal-hop path set.
\end{defn}

\begin{defn}[\textbf{$\bm{h}$-hop Path Set}]
An equal-hop path set $S$ is called an $h$-hop path set if all
paths $p\in S$ have the same length $|p|=h$.\end{defn}
\begin{conjecture}
\label{conj:two-disjoint-shortest-paths}\textup{\emph{Consider a
node $k$ that is not the source or one of the virtual sources. Let
}}\textup{$i_{0}$, $i_{1}$, $j_{0}$, $j_{1}$}\textup{\emph{ be
as defined in Eqn. (\ref{eq:z_0(t)}) and (\ref{eq:z_1(t)}). Suppose
that $c_{0}$ and $c_{1}$ are the two colors different from the color
of $k$. Also suppose that }}\textup{$i_{1}-i_{0}=j_{1}-j_{0}$. }\textup{\emph{Then
node $k$ always has two disjoint shortest paths $p_{0}$ and $p_{1}$
such that: 1) $p_{0}$ is from $X_{0}$ through the color-$c_{0}$
neighbors, and $p_{1}$ is from $X_{1}$ through the color-$c_{1}$
neighbors, to node $k$; or 2) $p_{0}$ is from $X_{0}$ through the
color-$c_{1}$ neighbors, and $p_{1}$ is from $X_{1}$ through the
color-$c_{0}$ neighbors, to node $k$.}}
\end{conjecture}
The verification of the conjecture is shown in Appendix \ref{app:Two-disjoint-shortest-paths}.
Although we cannot prove the conjecture, we verify with a computer
program that the conjecture is true for a large number of networks.
\begin{cor}[of Conjecture \ref{conj:two-disjoint-shortest-paths}]
\label{cor:not-equivalent-to-zero}\textup{\emph{Let $a_{S_{0}}(t)$,
$a_{S_{1}}(t)$, $a_{T_{0}}(t)$, $a_{T_{1}}(t)$ be as defined in
Eqn. (\ref{eq:a_S0-a_S1}) and (\ref{eq:a_T0-a_T1}), and let }}\textup{$i_{0}$,
$i_{1}$, $j_{0}$, $j_{1}$}\textup{\emph{ be defined as in Eqn.
(\ref{eq:z_0(t)}) and (\ref{eq:z_1(t)}). Suppose that }}$i_{0}<i_{1}$,
$j_{0}<j_{1}$ and $i_{1}-i_{0}=j_{1}-j_{0}$, then $a_{S_{0}}(i_{1})a_{T_{1}}(j_{1})\oplus a_{S_{1}}(i_{1})a_{T_{0}}(j_{1})\not\equiv0$.
\end{cor}
The proof of the corollary is in Appendix \ref{app:No-equivalent-to-zero}.

Remark: Note that in Corollary \ref{cor:not-equivalent-to-zero} and
the subsequent discussion, we use the equivalent sign $\equiv$ to
mean the equivalence of the two expressions on the LHS and the RHS.
In Corollary \ref{cor:not-equivalent-to-zero}, $a_{S_{0}}(i_{1})a_{T_{1}}(j_{1})\oplus a_{S_{1}}(i_{1})a_{T_{0}}(j_{1})\not\equiv0$
means the variables inside the expression on the LHS does not cancel
out to zero. In our scheme, transmit coefficients are i.i.d. uniform
random variables with values drawn from $GF(2^{s})\backslash\{0\}$.
It is entirely possible that a given set of realizations for the transmit
coefficients causes the above the expression to be 0. However, not
all realizations do so if the expression is not identically equal
to 0. 
\begin{cor}[of Conjecture \ref{conj:two-disjoint-shortest-paths}]
\label{cor:A-is-nonsingular}\textup{\emph{We can derive all native
packets from }}(\ref{eq:grouped-linear-equation-system})\textup{\emph{
with probability greater than }}\textup{$(1-2MN/(2^{s}-1))^{D/2}$}\textup{\emph{.}}\end{cor}
\begin{IEEEproof}[Proof of Corollary \ref{cor:A-is-nonsingular}]
We first note that we can solve for $\{x_{0}(t)\}_{0\leq t<i_{1}-i_{0}}$
from 
\[
\{r_{0}^{k}(t_{0})\}_{\min(i_{0},i_{1})\leq t_{0}<\max(i_{0},i_{1})}
\]
alone without considering $r_{1}^{k}(t_{1})$ when $i_{0}<i_{1}$
as long as $a_{S_{0}}(t_{0})\neq0$ for all $\min(i_{0},i_{1})\leq t_{0}<\max(i_{0},i_{1})$,
or $\{x_{1}(t)\}_{0\leq t<i_{0}-i_{1}}$ from the above when $i_{0}>i_{1}$
as long as $a_{S_{1}}(t_{0})\neq0$ for all $\min(i_{0},i_{1})\leq t_{0}<\max(i_{0},i_{1})$.
Similarly we can solve for $\{x_{0}(t)\}_{0\leq t<j_{1}-j_{0}}$ from
\[
\{r_{1}^{k}(t_{1})\}_{\min(j_{0},j_{1})\leq t_{1}<\max(j_{0},j_{1})}
\]
alone without considering $r_{0}^{k}(t_{0})$ when $j_{0}<j_{1}$
as long as $a_{T_{0}}(t_{1})\neq0$ for all $\min(j_{0},j_{1})\leq t_{1}<\max(j_{0},j_{1})$,
or $\{x_{1}(t)\}_{0\leq t<j_{0}-j_{1}}$ from the above when $j_{0}>j_{1}$
as long as $a_{T_{1}}(t_{1})\neq0$ for all $\min(j_{0},j_{1})\leq t_{1}<\max(j_{0},j_{1})$.
We only need to consider $r_{0}^{k}(t)$ and $r_{1}^{k}(t)$ together
when solving $x_{0}(t)$ or $x_{1}(t)$ for large $t$.

There are several cases depending on $i_{0}$, $i_{1}$, $j_{0}$
and $j_{1}$:

1. $i_{0}<i_{1}$ and $j_{0}>j_{1}$: 

We can solve for $\{x_{0}(t)\}_{0\leq t<i_{1}-i_{0}}$ from $\{r_{0}^{k}(t_{0})\}_{\min(i_{0},i_{1})\leq t_{0}<\max(i_{0},i_{1})}$
as long as $a_{S_{0}}(t_{0})\neq0$ for all $\min(i_{0},i_{1})\leq t_{0}<\max(i_{0},i_{1})$,
and $\{x_{1}(t)\}_{0\leq t<j_{0}-j_{1}}$ from $\{r_{1}^{k}(t_{1})\}_{\min(j_{0},j_{1})\leq t_{1}<\max(j_{0},j_{1})}$
as long as $a_{S_{1}}(t_{1})\neq0$ for all $\min(j_{0},j_{1})\leq t_{1}<\max(j_{0},j_{1})$. 

We express $a_{S_{0}}(t_{0})$ by 
\begin{equation}
a_{S_{0}}(t_{0})=P_{1}\oplus P_{2}\oplus...\oplus P_{L}\label{eq:RHS-of-det-2}
\end{equation}
where each of $\{P_{j}\}_{j=1,2,...,L}$ is a path coefficient, i.e.,
\[
P_{j}=g_{p}(t_{0})
\]
for some $p\in S_{0}$. Each of $\{P_{j}\}_{j=1,2,...,L}$ is a product
of $R=i_{0}$ transmit coefficients, i.e.,
\[
P_{j}=g_{p}(t_{0})=\prod_{u\mbox{ is in }p}\alpha_{u}(t_{u})
\]
where $t_{u}$ is the round node $u$ is visited. Each $P_{j}$ contains
$R=i_{0}$ factors. A path cannot have more than $MN$ hops, as $MN$
is the number of nodes in the network. Thus, $R=i_{0}\leq MN$. Without
loss of generality, suppose $P_{1}$ contains a set of distinct factors
that is not exactly the same as that of any $P_{j},2\leq j\leq L$.
Such $P_{1}$ exists because no two shortest paths contain the same
set of nodes. By Lemma \ref{lem:Not-equivalent} in Appendix \ref{app:No-equivalent-to-zero},
$a_{S_{0}}(t_{0})\not\equiv0$; by Lemma \ref{lem:not-equal-to-0}
in Appendix \ref{app:No-equivalent-to-zero}, 
\begin{align*}
\Pr\left(a_{S_{0}}(t_{0})=0\right) & =\Pr\left(P_{1}\oplus P_{2}\oplus...\oplus P_{L}=0\right)\\
 & \leq\frac{R}{2^{s}-1}\leq\frac{MN}{2^{s}-1},
\end{align*}
i.e.,
\[
\Pr\left(a_{S_{0}}(t_{0})\neq0\right)>1-MN/(2^{s}-1).
\]
By similar argument, we have $a_{T_{1}}(t_{1})\not\equiv0$ and 
\[
\Pr\left(a_{T_{1}}(t_{1})\neq0\right)>1-MN/(2^{s}-1).
\]

Thus, we can solve for $\{x_{0}(t)\}_{0\leq t<i_{1}-i_{0}}$ from
$\{r_{0}^{k}(t_{0})\}_{\min(i_{0},i_{1})\leq t_{0}<\max(i_{0},i_{1})}$
with probability greater than $(1-MN/(2^{s}-1))^{i_{1}-i_{0}}$, and
$\{x_{1}(t)\}_{0\leq t<j_{0}-j_{1}}$ from $\{r_{1}^{k}(t_{1})\}_{\min(j_{0},j_{1})\leq t_{1}<\max(j_{0},j_{1})}$
with probability $(1-MN/(2^{s}-1))^{j_{0}-j_{1}}$.

The first pair of equations that cannot be solved from its own series
is
\begin{align*}
r_{0}^{k}(i_{1})= & a_{S_{0}}(i_{1})x_{0}(i_{1}-i_{0})\oplus a_{S_{0}^{+1}}(i_{1})x_{0}(i_{1}-i_{0}-1)\\
 & \oplus...\oplus a_{S_{0}^{+i_{1}-i_{0}}}(i_{1})x_{0}(0)\\
 & \oplus a_{S_{1}}(i_{1})x_{1}(0)\\
r_{1}^{k}(j_{0})= & a_{T_{0}}(j_{0})x_{0}(0)\oplus\\
 & \oplus a_{T_{1}}(j_{0})x_{1}(j_{0}-j_{1})\oplus a_{T_{1}^{+1}}(j_{0})x_{1}(j_{0}-j_{1}-1)\\
 & \oplus...\oplus a_{T_{1}^{+j_{0}-j_{1}}}(j_{0})x_{1}(0).
\end{align*}
$x_{0}(i_{1}-i_{0})$ can be solved from $r_{0}^{k}(i_{1})$ as long
as $a_{S_{0}}(i_{1})\neq0$ because $x_{0}(0),...,x_{0}(i_{1}-i_{0}-1)$
and $x_{1}(0)$ are already known; and $x_{1}(j_{0}-j_{1})$ can be
solved from $r_{1}^{k}(j_{0})$ as long as $a_{T_{1}}(j_{0})\neq0$
because $x_{0}(0)$ and $x_{1}(0),...,x_{1}(j_{0}-j_{1}-1)$ are already
known. 

Since $\Pr\left(a_{S_{0}}(i_{1})\neq0\right)>1-MN/(2^{s}-1)$ and
$\Pr\left(a_{T_{1}}(j_{0})\neq0\right)>1-MN/(2^{s}-1)$, we can solve
for $x_{0}(i_{1}-i_{0})$ and $x_{1}(j_{0}-j_{1})$ from $r_{0}^{k}(i_{1})$
and $r_{1}^{k}(j_{0})$ with probability greater than $(1-MN/(2^{s}-1))^{2}\geq1-2MN/(2^{s}-1)$
conditioning on that $\{x_{0}(t)\}_{0\leq t<i_{1}-i_{0}}$ and $\{x_{1}(t)\}_{0\leq t<j_{0}-j_{1}}$
are already known. Similarly $x_{0}(i_{1}-i_{0}+1)$ and $x_{1}(j_{0}-j_{1}+1)$
can be solved from $r_{0}^{k}(i_{1}+1)$ and $r_{1}^{k}(j_{0}+1)$
with probability greater than $1-2MN/(2^{s}-1)$, conditioning on
that $\{x_{0}(t)\}_{0\leq t<i_{1}-i_{0}+1}$ and $\{x_{1}(t)\}_{0\leq t<j_{0}-j_{1}+1}$
are already known. By induction all native packets can be solved from
the equations in (\ref{eq:grouped-linear-equation-system}) with probability
greater than $(1-2MN/(2^{s}-1))^{D/2}$, as there are $D/2$ native
packets in both $\{x_{0}(t)\}_{t=0,...,\frac{D}{2}-1}$ and $\{x_{1}(t)\}_{t=0,...,\frac{D}{2}-1}$.

2. $i_{0}>i_{1}$ and $j_{0}<j_{1}$: 

The proof of this case is similar to the previous case, thus will
not be discussed here.

3. $i_{0}<i_{1}$, $j_{0}<j_{1}$, and $i_{1}-i_{0}\neq j_{1}-j_{0}$: 

We can solve for $\{x_{0}(t)\}_{0\leq t<i_{1}-i_{0}}$ from $\{r_{0}^{k}(t_{0})\}_{\min(i_{0},i_{1})\leq t_{0}<\max(i_{0},i_{1})}$,
and $\{x_{0}(t)\}_{0\leq t<j_{1}-j_{0}}$ from $\{r_{1}^{k}(t_{1})\}_{\min(j_{0},j_{1})\leq t_{1}<\max(j_{0},j_{1})}$.
Without loss of generality, assume $i_{1}-i_{0}>j_{1}-j_{0}$. The
first pair of equations that cannot be solved from its own series
is 
\begin{align*}
r_{0}^{k}(i_{1})= & a_{S_{0}}(i_{1})x_{0}(i_{1}-i_{0})\oplus a_{S_{0}^{+1}}(i_{1})x_{0}(i_{1}-i_{0}-1)\\
 & \oplus...\oplus a_{S_{0}^{+i_{1}-i_{0}}}(i_{1})x_{0}(0)\\
 & \oplus a_{S_{1}}(i_{1})x_{1}(0)\\
r_{1}^{k}(j_{1})= & a_{T_{0}}(j_{1})x_{0}(j_{1}-j_{0})\oplus a_{T_{0}^{+1}}(j_{1})x_{0}(j_{1}-j_{0}-1)\\
 & \oplus...\oplus a_{T_{0}^{+j_{1}-j_{0}}}(j_{1})x_{0}(0)\\
 & \oplus a_{T_{1}}(j_{1})x_{1}(0).
\end{align*}
$x_{1}(0)$ can be solved from $r_{1}^{k}(j_{1})$ as long as $a_{T_{1}}(j_{1})\neq0$
because $x_{0}(0),...,x_{0}(j_{1}-j_{0}),...,x_{0}(i_{1}-i_{0}-1)$
are already known; then $x_{0}(i_{1}-i_{0})$ can be solved from $r_{0}^{k}(i_{1})$
as long as $a_{S_{0}}(i_{1})\neq0$ because $x_{0}(0),...,x_{0}(i_{1}-i_{0}-1)$
and $x_{1}(0)$ are already known. With similar argument as in Case
1, all native packets can be solved from the equations in (\ref{eq:grouped-linear-equation-system})
with probability greater than $(1-2MN/(2^{s}-1))^{D/2}$.

4. $i_{0}>i_{1}$, $j_{0}>j_{1}$, and $i_{1}-i_{0}\neq j_{1}-j_{0}$: 

The proof of this case is similar to the previous case, thus will
not be discussed here.

5. $i_{0}<i_{1}$, $j_{0}<j_{1}$, and $i_{1}-i_{0}=j_{1}-j_{0}$: 

Let $\delta=i_{1}-i_{0}=j_{1}-j_{0}$. We can solve for $\{x_{0}(t)\}_{0\leq t<\delta}$
from either $\{r_{0}^{k}(t_{0})\}_{\min(i_{0},i_{1})\leq t_{0}<\max(i_{0},i_{1})}$
or $\{r_{1}^{k}(t_{1})\}_{\min(j_{0},j_{1})\leq t_{1}<\max(j_{0},j_{1})}$.
The first pair of equation that cannot be solved from its own series
is
\begin{align*}
r_{0}^{k}(i_{1})= & a_{S_{0}}(i_{1})x_{0}(\delta)\oplus a_{S_{0}^{+1}}(i_{1})x_{0}(\delta-1)\\
 & \oplus...\oplus a_{S_{0}^{+\delta}}(i_{1})x_{0}(0)\\
 & \oplus a_{S_{1}}(i_{1})x_{1}(0)\\
r_{1}^{k}(j_{1})= & a_{T_{0}}(j_{1})x_{0}(\delta)\oplus a_{T_{0}^{+1}}(j_{1})x_{0}(\delta-1)\\
 & \oplus...\oplus a_{T_{0}^{+\delta}}(j_{1})x_{0}(0)\\
 & \oplus a_{T_{1}}(j_{1})x_{1}(0).
\end{align*}
Putting the unknown part on the LHS and known part on the RHS yields
\begin{align*}
a_{S_{0}}(i_{1})x_{0}(\delta)\oplus a_{S_{1}}(i_{1})x_{1}(0)= & r_{0}^{k}(i_{1})\oplus a_{S_{0}^{+1}}(i_{1})x_{0}(\delta-1)\\
 & \oplus...\oplus a_{S_{0}^{+\delta}}(i_{1})x_{0}(0)\\
a_{T_{0}}(j_{1})x_{0}(\delta)\oplus a_{T_{1}}(j_{1})x_{1}(0)= & r_{1}^{k}(j_{1})\oplus a_{T_{0}^{+1}}(j_{1})x_{0}(\delta-1)\\
 & \oplus...\oplus a_{T_{0}^{+\delta}}(j_{1})x_{0}(0).
\end{align*}
As long as the matrix
\[
\left(\begin{array}{cc}
a_{S_{0}}(i_{1}) & a_{S_{1}}(i_{1})\\
a_{T_{0}}(j_{1}) & a_{T_{1}}(j_{1})
\end{array}\right)
\]
has full rank, we can solve for $x_{0}(\delta)$ and $x_{1}(0)$ from
the two equations. The determinant of the matrix is
\begin{align}
A & =\det\left(\begin{array}{cc}
a_{S_{0}}(i_{1}) & a_{S_{1}}(i_{1})\\
a_{T_{0}}(j_{1}) & a_{T_{1}}(j_{1})
\end{array}\right)\label{eq:determinant_2x2}\\
 & =a_{S_{0}}(i_{1})a_{T_{1}}(j_{1})\oplus a_{S_{1}}(i_{1})a_{T_{0}}(j_{1}).
\end{align}
$A\not\equiv0$ (i.e., $A$ is not always equal to zero) by Appendix
\ref{app:No-equivalent-to-zero}.

We express $A$ by 
\begin{equation}
A=Q_{1}\oplus Q_{2}\oplus...\oplus Q_{L}\label{eq:RHS-of-det}
\end{equation}
where each of $\{Q_{j}\}_{j=1,2,...,L}$ is a product of two path
coefficients, i.e., 
\[
Q_{j}=g_{p}(i_{1})g_{p'}(j_{1})
\]
for some $p\in S_{0}$ and $p'\in T_{1}$, or $p\in S_{1}$ and $p'\in T_{0}$.
Each of $\{Q_{j}\}_{j=1,2,...,L}$ is a product of $R=i_{1}+j_{0}=i_{0}+j_{1}$
transmit coefficients, i.e.,
\[
Q_{j}=g_{p}(i_{1})g_{p'}(j_{1})=\prod_{u\mbox{ is in }p}\alpha_{u}(t_{u})\prod_{v\mbox{ is in }p'}\alpha_{v}(t_{v})
\]
where $t_{u}$ and $t_{v}$ are the rounds node $u$ and $v$ are
visited, respectively. Each $Q_{j}$ contains $R$ factors. A path
cannot have more than $MN$ hops, as $MN$ is the number of nodes
in the network. Thus, $R=i_{1}+j_{0}=i_{0}+j_{1}\leq2MN$. Without
loss of generality, suppose $Q_{1}$ contains a set of distinct factors
that is not exactly the same as that of any $Q_{j},2\leq j\leq L$.
Such $Q_{1}$ is showed to exist in Appendix \ref{app:No-equivalent-to-zero}.
Also, without loss of generality, we assume if $Q_{j}=Q_{k}$ for
$j,k\neq1$, then they will be removed from $Q_{1}\oplus Q_{2}\oplus...\oplus Q_{L}$
so that $Q_{j}\neq Q_{k}$ in the remaining $Q_{1}\oplus Q_{2}\oplus...\oplus Q_{L}$.
We write 
\[
Q_{1}=\prod_{m=1}^{R}r_{m},
\]
where $r_{m}$ is $\alpha_{u}(t_{u})$ for some $u$ in $p$ or $\alpha_{v}(t_{v})$
for some $v$ in $p'$, and $r_{m_{1}}\neq r_{m_{2}}$ for $m_{1}\neq m_{2},1\leq m_{1},m_{2}\leq R$.
By Lemma \ref{lem:not-equal-to-0}, $\Pr\left(A=0\right)=\Pr\left(Q_{1}\oplus Q_{2}\oplus...\oplus Q_{L}=0\right)\leq R/(2^{s}-1)\leq2MN/(2^{s}-1)$. 

Therefore, we can solve for $x_{0}(\delta)$ and $x_{1}(0)$ from
$r_{0}^{k}(i_{1})$ and $r_{1}^{k}(j_{1})$ with probability greater
than $1-2MN/(2^{s}-1)$. Similarly, we can solve for $x_{0}(\delta+1)$
and $x_{1}(2)$ from 
\begin{align*}
r_{0}^{k}(i_{1}+1)= & a_{S_{0}}(i_{1}+1)x_{0}(\delta+1)\oplus a_{S_{0}^{+1}}(i_{1}+1)x_{0}(\delta)\\
 & \oplus...\oplus a_{S_{0}^{+\delta+1}}(i_{1}+1)x_{0}(0)\\
 & \oplus a_{S_{1}}(i_{1}+1)x_{1}(1)\oplus a_{S_{1}^{+1}}(i_{1}+1)x_{1}(0)\\
r_{1}^{k}(j_{1}+1)= & a_{T_{0}}(j_{1}+1)x_{0}(\delta+1)\oplus a_{T_{0}^{+1}}(j_{1}+1)x_{0}(\delta)\\
 & \oplus...\oplus a_{T_{0}^{+\delta+1}}(j_{1}+1)x_{0}(0)\\
 & \oplus a_{T_{1}}(j_{1}+1)x_{1}(1)\oplus a_{T_{1}^{+1}}(j_{1}+1)x_{1}(0)
\end{align*}
with probability greater than $1-2MN/(2^{s}-1)$, conditioning on
that $\{x_{0}(t)\}_{0\leq t<\delta+1}$ and $x_{1}(0)$ are already
known. By induction all native packets can be solved from the equations
in (\ref{eq:grouped-linear-equation-system}) with probability greater
than $(1-2MN/(2^{s}-1))^{D/2}$, as there are $D/2$ native packets
in both $\{x_{0}(t)\}_{t=0,...,\frac{D}{2}-1}$ and $\{x_{1}(t)\}_{t=0,...,\frac{D}{2}-1}$.

6. $i_{0}>i_{1}$, $j_{0}>j_{1}$, and $i_{0}-i_{1}=j_{0}-j_{1}$: 

The proof of this case is similar to the previous case, thus will
not be discussed here.

In conclusion, all native packets can be solved from (\ref{eq:grouped-linear-equation-system})
with probability greater than $(1-2MN/(2^{s}-1))^{D/2}$.
\end{IEEEproof}
With the above lemmas, we can show that the broadcast throughput upper
bound is achievable in grid networks with high probability. This result
is presented in the following theorem.
\begin{cor}[of Conjecture \ref{conj:two-disjoint-shortest-paths}]
\label{thm:throughput-reachable}Suppose that $M$ and $N$ are fixed.
The broadcast throughput in grid networks reaches $2/3$ with high
probability when $s$ is of order larger than $\log D$.\end{cor}
\begin{IEEEproof}[Proof of Corollary \ref{thm:throughput-reachable}]
By Corollary \ref{cor:A-is-nonsingular}, all native packets can
be solved from (\ref{eq:grouped-linear-equation-system}) with probability
greater than 
\begin{equation}
(1-\frac{2MN}{2^{s}-1})^{\frac{D}{2}}.
\end{equation}
For large $s$, $(1-\frac{2MN}{2^{s}-1})^{\frac{D}{2}}$ can be approximated
by 
\begin{equation}
\exp(-\frac{MND}{2^{s}-1}).
\end{equation}
Thus, if $s$ is of order larger than $\log D$ (e.g., $s=\log D^{e}$),
where $e>1$, the limit of the above probability as $D\to\infty$
is 
\begin{equation}
\lim_{D\to\infty}(1-\frac{2MN}{2^{s}-1})^{\frac{D}{2}}=1.
\end{equation}

Therefore if $D$ is large, at the end of round $(D/2+\max(i_{0},i_{1})-1)$,
node $k$ can derive all native packets from $X$ with a high probability.
At the end of round $D/2+MN-2$, all nodes can derive all native packets
from $X$ with a high probability. The throughput is
\begin{equation}
\rho=\lim_{D\to\infty}\frac{D}{3(\frac{D}{2}+MN-2)}=\frac{2}{3}.
\end{equation}

\end{IEEEproof}
As a consequence, the broadcast throughput upper bound is achievable
with high probability when the field size is of order larger than
the logarithm of the number of packets.

\section{\label{sec:Conclusions}Conclusions}

In this work, we have investigated the broadcast throughput of half-duplex
wireless networks. We show that the theoretical throughput upper bound
is $n/(n+1)$ for single-source broadcast, where $n$ is the minimum
vertex-cut size of the network. This upper bound is not always achievable
in general, but is achievable in many networks, including line, ring,
chord ring, and grid networks.

\appendix

\subsection{\label{app:A-Broadcast-throughput-of}Broadcast throughput of the
network in Fig. \ref{fig:example-not-reaching-upper-bound}}

In this appendix, we argue that the broadcast throughput in the network
in Fig. \ref{fig:example-not-reaching-upper-bound} cannot reach the
upper bound given by Theorem \ref{thm:max-throughput}\emph{.} Let
$\mathcal{T}_{i}$ be the set of time slots during which node $i$
transmits within the $W_{D}$ time slots; let $\mathcal{R}_{i}$ be
the set of time slots during which node $i$ receives within the $W_{D}$
time slots.

To achieve the the throughput upper bound $2/3$, we need
\[
|\mathcal{R}_{i}|\geq\frac{2}{3}W_{D},\forall i\in\{X_{0},X_{1},1,2,3\}.
\]
Since a node is either in the transmission mode or the receiving mode,
we have
\begin{align}
 & |\mathcal{T}_{i}|=W_{D}-|\mathcal{R}_{i}|\leq\frac{1}{3}W_{D},\forall i\in\{X_{0},X_{1},1,2,3\}.\label{eq:T_i-leq-W_D/3}
\end{align}
Consider node 1. Its throughput is upper-bounded as follows:
\begin{align*}
\rho_{1} & \leq|\mathcal{T}_{X_{0}}\cup\mathcal{T}_{X_{1}}|=|\mathcal{T}_{X_{0}}|+|\mathcal{T}_{X_{1}}|-|\mathcal{T}_{X_{0}}\cap\mathcal{T}_{X_{1}}|\\
 & \leq\frac{2}{3}W_{D}-|\mathcal{T}_{X_{0}}\cap\mathcal{T}_{X_{1}}|
\end{align*}
In order that $\rho_{1}\geq\frac{2}{3}W_{D}$, we need
\begin{equation}
|\mathcal{T}_{X_{0}}\cap\mathcal{T}_{X_{1}}|=0.\label{eq:T_X0-inter-T_X1=00003D0}
\end{equation}
Applying the same argument on nodes 2 and 3 gives
\begin{equation}
|\mathcal{T}_{X_{0}}\cap\mathcal{T}_{3}|=0\label{eq:T_X0-inter-T_3=00003D0}
\end{equation}
\begin{equation}
|\mathcal{T}_{X_{1}}\cap\mathcal{T}_{2}|=0\label{eq:T_X1-inter-T_2=00003D0}
\end{equation}

The throughput of node 2 is
\begin{align}
\rho_{2} & \leq|\mathcal{T}_{X_{0}}|+|\mathcal{T}_{3}|-|\mathcal{T}_{X_{0}}\cap\mathcal{T}_{2}|\label{eq:appendix-node-2}\\
 & \leq\frac{2}{3}W_{D}-|\mathcal{T}_{X_{0}}\cap\mathcal{T}_{2}|\label{eq:appendix-node-2-b}\\
\Rightarrow & |\mathcal{T}_{X_{0}}\cap\mathcal{T}_{2}|=0\label{eq:T_X0-inter-T_2=00003D0}
\end{align}
In the above, (\ref{eq:appendix-node-2}) is from the half-duplex
constraint that when node 2 transmits during the $T_{2}$ slots, if
$X_{0}$ transmits at the same time, no information can be received.
(\ref{eq:appendix-node-2-b}) is derived from (\ref{eq:T_i-leq-W_D/3}).
Similarly, we can argue that
\begin{align}
|\mathcal{T}_{X_{1}}\cap\mathcal{T}_{3}| & =0\label{eq:T_X1-inter-T_3=00003D0}\\
|\mathcal{T}_{2}\cap\mathcal{T}_{3}| & =0\label{eq:T_2-inter-T_3=00003D0}
\end{align}
Now, as $W_{D}$ is the total number of time slots under consideration,
we have
\begin{align*}
 & |\mathcal{T}_{X_{0}}\cup\mathcal{T}_{3}\cup\mathcal{T}_{X_{1}}\cup\mathcal{T}_{2}|\leq W_{D}\\
\Rightarrow & |\mathcal{T}_{X_{0}}\cup\mathcal{T}_{3}|+|\mathcal{T}_{X_{1}}\cup\mathcal{T}_{2}|-|(\mathcal{T}_{X_{0}}\cup\mathcal{T}_{3})\cap(\mathcal{T}_{X_{1}}\cup\mathcal{T}_{2})|\leq W_{D}
\end{align*}
In order that $\rho_{2}=\frac{2}{3}W_{D}$ and $\rho_{3}=\frac{2}{3}W_{D}$,
we need 
\[
|\mathcal{T}_{X_{0}}\cup\mathcal{T}_{3}|\geq\frac{2}{3}W_{D}
\]
\[
|\mathcal{T}_{X_{1}}\cup\mathcal{T}_{2}|\geq\frac{2}{3}W_{D}
\]
This is because a node can receive information from the transmissions
of its neighbors only. Therefore, 
\begin{align}
 & |(\mathcal{T}_{X_{0}}\cup\mathcal{T}_{3})\cap(\mathcal{T}_{X_{1}}\cup\mathcal{T}_{2})|\geq\frac{1}{3}W_{D}\nonumber \\
\Rightarrow & |(\mathcal{T}_{X_{0}}\cap\mathcal{T}_{X_{1}})\cup(\mathcal{T}_{X_{0}}\cap\mathcal{T}_{2})\cup(\mathcal{T}_{3}\cap\mathcal{T}_{X_{1}})\cup(\mathcal{T}_{3}\cup\mathcal{T}_{2})|\geq\frac{1}{3}W_{D}\nonumber \\
\Rightarrow & |\mathcal{T}_{X_{0}}\cap\mathcal{T}_{X_{1}}|+|\mathcal{T}_{X_{0}}\cap\mathcal{T}_{2}|+|\mathcal{T}_{3}\cap\mathcal{T}_{X_{1}}|+|\mathcal{T}_{3}\cup\mathcal{T}_{2}|\geq\frac{1}{3}W_{D}\nonumber \\
\Rightarrow & |\mathcal{T}_{X_{0}}\cap\mathcal{T}_{2}|+|\mathcal{T}_{3}\cap\mathcal{T}_{X_{1}}|+|\mathcal{T}_{3}\cup\mathcal{T}_{2}|\geq\frac{1}{3}W_{D}\label{eq:sum-geq-W/3}
\end{align}
(\ref{eq:sum-geq-W/3}) is derived from (\ref{eq:T_X0-inter-T_X1=00003D0}).
However, (\ref{eq:sum-geq-W/3}) contradicts (\ref{eq:T_X0-inter-T_2=00003D0}),
(\ref{eq:T_X1-inter-T_3=00003D0}) and (\ref{eq:T_2-inter-T_3=00003D0}).
Therefore, the throughput upper bound cannot be achieved.

\subsection{\label{app:B-Transmission-scheme-for}Transmission scheme for the
chord ring network in Fig. \ref{fig:chord-ring-network}}

In this appendix we show a transmission scheme for the chord ring
network in Fig. \ref{fig:chord-ring-network}. The scheduling of nodes
is shown in Table \ref{tab:Transmission-schedule-chord-ring}.

Here we denote the sequence of native packets to be broadcast by $X$
by $\{x_{0}(t)\}_{t=0,1,...}$, $\{x_{1}(t)\}_{t=0,1,...}$, $\{x_{2}(t)\}_{t=0,1,...}$
and $\{x_{3}(t)\}_{t=0,1,...}$. We define a round $t,t=0,1,...$
to be a set of five time slots \{$5t,5t+1,5t+2,5t+3,5t+4$\}. The
source $X$ transmits $x_{0}(t)$, $x_{1}(t)$, $x_{2}(t)$, $x_{3}(t)$
and $x_{0}(t)\oplus x_{1}(t)\oplus x_{2}(t)\oplus x_{3}(t)$ in time
slots $5t,5t+1,5t+2,5t+3$ and $5t+4$, respectively. Node $1$ transmits
$x_{0}(t-1)$ in time slot $5t$; node $2$ transmits $x_{1}(t-1)$
in time slot $5t+1$; node $4$ transmits $x_{2}(t-1)$ in time slot
$5t+2$; and node $5$ transmits $x_{3}(t-1)$ in time slot $5t+3$.

From round 1 onwards, every non-source node receives sufficient information
for it to derive four new native packets in each round. Thus, the
broadcast throughput is $4/5$.

\begin{table}
\caption{\label{tab:Transmission-schedule-chord-ring}Broadcast schedule for
the chord ring network in Fig. \ref{fig:chord-ring-network}. ``s'',
``r'' and ``d'' indicate ``send'', ``receive'' and ``derive'',
respectively.}
\centering\setlength\tabcolsep{3pt}{\footnotesize }%
\begin{tabular}{|l|c|>{\raggedright}p{1.1cm}|>{\raggedright}p{1cm}|>{\raggedright}p{1cm}|>{\raggedright}p{1cm}|>{\raggedright}p{1cm}|>{\raggedright}p{1cm}|}
\hline 
\emph{\footnotesize t} &
\emph{\footnotesize ts} &
{\footnotesize Node $X$} &
{\footnotesize Node 1} &
{\footnotesize Node 2} &
{\footnotesize Node 3} &
{\footnotesize Node 4} &
{\footnotesize Node 5}\tabularnewline
\hline 
\hline 
\multirow{5}{*}{0} &
{\footnotesize 0} &
{\footnotesize s:$x_{0}(0)$} &
{\footnotesize r:$x_{0}(0)$} &
{\footnotesize r:$x_{0}(0)$} &
{\footnotesize -} &
{\footnotesize r:$x_{0}(0)$} &
{\footnotesize r:$x_{0}(0)$}\tabularnewline
\cline{2-8} 
 & {\footnotesize 1} &
{\footnotesize s:$x_{1}(0)$} &
{\footnotesize r:$x_{1}(0)$} &
{\footnotesize r:$x_{1}(0)$} &
{\footnotesize -} &
{\footnotesize r:$x_{1}(0)$} &
{\footnotesize r:$x_{1}(0)$}\tabularnewline
\cline{2-8} 
 & {\footnotesize 2} &
{\footnotesize s:$x_{2}(0)$} &
{\footnotesize r:$x_{2}(0)$} &
{\footnotesize r:$x_{2}(0)$} &
{\footnotesize -} &
{\footnotesize r:$x_{2}(0)$} &
{\footnotesize r:$x_{2}(0)$}\tabularnewline
\cline{2-8} 
 & {\footnotesize 3} &
{\footnotesize s:$x_{3}(0)$} &
{\footnotesize r:$x_{3}(0)$} &
{\footnotesize r:$x_{3}(0)$} &
{\footnotesize -} &
{\footnotesize r:$x_{3}(0)$} &
{\footnotesize r:$x_{3}(0)$}\tabularnewline
\cline{2-8} 
 & {\footnotesize 4} &
{\footnotesize s:$x_{0}(0)$}{\footnotesize \par}

{\footnotesize $\oplus x_{1}(0)$}{\footnotesize \par}

{\footnotesize $\oplus x_{2}(0)$}{\footnotesize \par}

{\footnotesize $\oplus x_{3}(0)$} &
{\footnotesize -} &
{\footnotesize -} &
{\footnotesize -} &
{\footnotesize -} &
{\footnotesize -}\tabularnewline
\hline 
\multirow{5}{*}{1} &
{\footnotesize 5} &
{\footnotesize s:$x_{0}(1)$} &
{\footnotesize s:$x_{0}(0)$} &
{\footnotesize r:$x_{0}(0)$}{\footnotesize \par}

{\footnotesize $\oplus x_{0}(1)$}{\footnotesize \par}

{\footnotesize d:$x_{0}(1)$} &
{\footnotesize r:$x_{0}(0)$} &
{\footnotesize r:$x_{0}(1)$} &
{\footnotesize r:$x_{0}(0)$}{\footnotesize \par}

{\footnotesize $\oplus x_{0}(1)$}{\footnotesize \par}

{\footnotesize d:$x_{1}(1)$}\tabularnewline
\cline{2-8} 
 & {\footnotesize 6} &
{\footnotesize s:$x_{1}(1)$} &
{\footnotesize r:$x_{1}(0)$}{\footnotesize \par}

{\footnotesize $\oplus x_{1}(1)$}{\footnotesize \par}

{\footnotesize d:$x_{1}(1)$} &
{\footnotesize s:$x_{1}(0)$} &
{\footnotesize r:$x_{1}(0)$} &
{\footnotesize r:$x_{1}(0)$}{\footnotesize \par}

{\footnotesize $\oplus x_{1}(1)$}{\footnotesize \par}

{\footnotesize d:$x_{1}(1)$} &
{\footnotesize r:$x_{1}(1)$}\tabularnewline
\cline{2-8} 
 & {\footnotesize 7} &
{\footnotesize s:$x_{2}(1)$} &
{\footnotesize r:$x_{2}(1)$} &
{\footnotesize r:$x_{2}(0)$}{\footnotesize \par}

{\footnotesize $\oplus x_{2}(1)$}{\footnotesize \par}

{\footnotesize d:$x_{2}(1)$} &
{\footnotesize r:$x_{2}(0)$} &
{\footnotesize s:$x_{2}(0)$} &
{\footnotesize r:$x_{2}(0)$}{\footnotesize \par}

{\footnotesize $\oplus x_{2}(1)$}{\footnotesize \par}

{\footnotesize d:$x_{2}(1)$}\tabularnewline
\cline{2-8} 
 & {\footnotesize 8} &
{\footnotesize s:$x_{3}(1)$} &
{\footnotesize r:$x_{3}(0)$}{\footnotesize \par}

{\footnotesize $\oplus x_{3}(1)$}{\footnotesize \par}

{\footnotesize d:$x_{3}(1)$} &
{\footnotesize r:$x_{3}(1)$} &
{\footnotesize r:$x_{3}(0)$} &
{\footnotesize r:$x_{3}(0)$}{\footnotesize \par}

{\footnotesize $\oplus x_{3}(1)$}{\footnotesize \par}

{\footnotesize d:$x_{3}(1)$} &
{\footnotesize s:$x_{3}(0)$}\tabularnewline
\cline{2-8} 
 & {\footnotesize 9} &
{\footnotesize s:$x_{0}(1)$}{\footnotesize \par}

{\footnotesize $\oplus x_{1}(1)$}{\footnotesize \par}

{\footnotesize $\oplus x_{2}(1)$}{\footnotesize \par}

{\footnotesize $\oplus x_{3}(1)$} &
{\footnotesize r:$x_{0}(1)$}{\footnotesize \par}

{\footnotesize $\oplus x_{1}(1)$}{\footnotesize \par}

{\footnotesize $\oplus x_{2}(1)$}{\footnotesize \par}

{\footnotesize $\oplus x_{3}(1)$}{\footnotesize \par}

{\footnotesize d:$x_{0}(1)$} &
{\footnotesize r:$x_{0}(1)$}{\footnotesize \par}

{\footnotesize $\oplus x_{1}(1)$}{\footnotesize \par}

{\footnotesize $\oplus x_{2}(1)$}{\footnotesize \par}

{\footnotesize $\oplus x_{3}(1)$}{\footnotesize \par}

{\footnotesize d:$x_{1}(1)$} &
{\footnotesize -} &
{\footnotesize r:$x_{0}(1)$}{\footnotesize \par}

{\footnotesize $\oplus x_{1}(1)$}{\footnotesize \par}

{\footnotesize $\oplus x_{2}(1)$}{\footnotesize \par}

{\footnotesize $\oplus x_{3}(1)$}{\footnotesize \par}

{\footnotesize d:$x_{2}(1)$} &
{\footnotesize r:$x_{0}(1)$}{\footnotesize \par}

{\footnotesize $\oplus x_{1}(1)$}{\footnotesize \par}

{\footnotesize $\oplus x_{2}(1)$}{\footnotesize \par}

{\footnotesize $\oplus x_{3}(1)$}{\footnotesize \par}

{\footnotesize d:$x_{3}(1)$}\tabularnewline
\hline 
\multirow{5}{*}{{\footnotesize 2}} &
{\footnotesize 10} &
{\footnotesize s:$x_{0}(2)$} &
{\footnotesize s:$x_{0}(1)$} &
{\footnotesize r:$x_{0}(1)$}{\footnotesize \par}

{\footnotesize $\oplus x_{0}(2)$}{\footnotesize \par}

{\footnotesize d:$x_{0}(2)$} &
{\footnotesize r:$x_{0}(1)$} &
{\footnotesize r:$x_{0}(2)$} &
{\footnotesize r:$x_{0}(1)$}{\footnotesize \par}

{\footnotesize $\oplus x_{0}(2)$}{\footnotesize \par}

{\footnotesize d:$x_{1}(2)$}\tabularnewline
\cline{2-8} 
 & {\footnotesize 11} &
{\footnotesize s:$x_{1}(2)$} &
{\footnotesize r:$x_{1}(1)$}{\footnotesize \par}

{\footnotesize $\oplus x_{1}(2)$}{\footnotesize \par}

{\footnotesize d:$x_{1}(2)$} &
{\footnotesize s:$x_{1}(1)$} &
{\footnotesize r:$x_{1}(1)$} &
{\footnotesize r:$x_{1}(1)$}{\footnotesize \par}

{\footnotesize $\oplus x_{1}(2)$}{\footnotesize \par}

{\footnotesize d:$x_{1}(2)$} &
{\footnotesize r:$x_{1}(2)$}\tabularnewline
\cline{2-8} 
 & {\footnotesize 12} &
{\footnotesize s:$x_{2}(2)$} &
{\footnotesize r:$x_{2}(2)$} &
{\footnotesize r:$x_{2}(1)$}{\footnotesize \par}

{\footnotesize $\oplus x_{2}(2)$}{\footnotesize \par}

{\footnotesize d:$x_{2}(2)$} &
{\footnotesize r:$x_{2}(1)$} &
{\footnotesize s:$x_{2}(1)$} &
{\footnotesize r:$x_{2}(1)$}{\footnotesize \par}

{\footnotesize $\oplus x_{2}(2)$}{\footnotesize \par}

{\footnotesize d:$x_{2}(2)$}\tabularnewline
\cline{2-8} 
 & {\footnotesize 13} &
{\footnotesize s:$x_{3}(2)$} &
{\footnotesize r:$x_{3}(1)$}{\footnotesize \par}

{\footnotesize $\oplus x_{3}(2)$}{\footnotesize \par}

{\footnotesize d:$x_{3}(2)$} &
{\footnotesize r:$x_{3}(2)$} &
{\footnotesize r:$x_{3}(1)$} &
{\footnotesize r:$x_{3}(1)$}{\footnotesize \par}

{\footnotesize $\oplus x_{3}(2)$}{\footnotesize \par}

{\footnotesize d:$x_{3}(2)$} &
{\footnotesize s:$x_{3}(1)$}\tabularnewline
\cline{2-8} 
 & {\footnotesize 14} &
{\footnotesize s:$x_{0}(2)$}{\footnotesize \par}

{\footnotesize $\oplus x_{1}(2)$}{\footnotesize \par}

{\footnotesize $\oplus x_{2}(2)$}{\footnotesize \par}

{\footnotesize $\oplus x_{3}(2)$} &
{\footnotesize r:$x_{0}(2)$}{\footnotesize \par}

{\footnotesize $\oplus x_{1}(2)$}{\footnotesize \par}

{\footnotesize $\oplus x_{2}(2)$}{\footnotesize \par}

{\footnotesize $\oplus x_{3}(2)$}{\footnotesize \par}

{\footnotesize d:$x_{0}(2)$} &
{\footnotesize r:$x_{0}(2)$}{\footnotesize \par}

{\footnotesize $\oplus x_{1}(2)$}{\footnotesize \par}

{\footnotesize $\oplus x_{2}(2)$}{\footnotesize \par}

{\footnotesize $\oplus x_{3}(2)$}{\footnotesize \par}

{\footnotesize d:$x_{1}(2)$} &
{\footnotesize -} &
{\footnotesize r:$x_{0}(2)$}{\footnotesize \par}

{\footnotesize $\oplus x_{1}(2)$}{\footnotesize \par}

{\footnotesize $\oplus x_{2}(2)$}{\footnotesize \par}

{\footnotesize $\oplus x_{3}(2)$}{\footnotesize \par}

{\footnotesize d:$x_{2}(2)$} &
{\footnotesize r:$x_{0}(2)$}{\footnotesize \par}

{\footnotesize $\oplus x_{1}(2)$}{\footnotesize \par}

{\footnotesize $\oplus x_{2}(2)$}{\footnotesize \par}

{\footnotesize $\oplus x_{3}(2)$}{\footnotesize \par}

{\footnotesize d:$x_{3}(2)$}\tabularnewline
\hline 
\end{tabular}
\end{table}

\subsection{\label{app:No-equivalent-to-zero}Proof of Corollary \ref{cor:not-equivalent-to-zero}:}
\begin{IEEEproof}
Suppose the colors of the receiving node -- node $k$'s neighbors
are $c_{0}=(k-1)\bmod3$ and $c_{1}=(k+1)\bmod3$. $a_{S_{0}}(i_{1})$
and $a_{S_{1}}(i_{1})$ are the aggregated path coefficients associated
with the shortest paths from $X_{0}$ and $X_{1}$, respectively,
to the color-$c_{0}$ neighbors; and $a_{T_{0}}(j_{1})$ and $a_{T_{1}}(j_{1})$
are the aggregated path coefficients associated with the shortest
paths from $X_{0}$ and $X_{1}$, respectively, to the color-$c_{1}$
neighbors. The summands in $a_{S_{0}}(i_{1})$, $a_{S_{1}}(i_{1})$,
$a_{T_{0}}(j_{1})$ and $a_{T_{1}}(j_{1})$ are products of $i_{0}$,
$i_{1}$, $j_{0}$ and $j_{1}$ transmit coefficients, respectively.
Thus, the summands in $a_{S_{0}}(i_{1})a_{T_{1}}(j_{1})$ and $a_{S_{1}}(i_{1})a_{T_{0}}(j_{1})$
are products of $i_{0}+j_{1}$ and $i_{1}+j_{0}$ transmit coefficients,
respectively. Note that $i_{0}+j_{1}=i_{1}+j_{0}$ by the statement
of the corollary.

According to our observation in Appendix \ref{app:Two-disjoint-shortest-paths},
we conjecture (in Conjecture \ref{conj:two-disjoint-shortest-paths}
in the main body of this report) that node $k$ always has two disjoint
shortest paths $p_{0}$ and $p_{1}$ such that $p_{0}$ is from a
virtual source through the color-$c_{0}$ neighbors, and $p_{1}$
is from the other virtual source through the color-$c_{1}$ neighbors,
to node $k$. By disjoint we mean there is no node that is in both
of the two shortest paths. 

Without loss of generality, suppose $p_{0}\in S_{0}$ and $p_{1}\in T_{1}$.
We first show that there does not exist two different shortest paths
$\tilde{p}_{0}$ and $\tilde{p}_{1}$ such that $g_{p_{0}}(i_{1})g_{p_{1}}(j_{1})\equiv g_{\tilde{p}_{0}}(i_{1})g_{\tilde{p}_{1}}(j_{1})$.

Suppose on the contrary that there are two different shortest paths
$\tilde{p}_{0}$ and $\tilde{p}_{1}$ such that $g_{p_{0}}(i_{1})g_{p_{1}}(j_{1})\equiv g_{\tilde{p}_{0}}(i_{1})g_{\tilde{p}_{1}}(j_{1})$.
Let $p_{0}=u_{1}-u_{2}-...-u_{i_{0}}-k$ and $p_{1}=v_{1}-v_{2}-...-v_{j_{1}}-k$,
where $u_{1}=X_{0}$ and $v_{1}=X_{1}$. Then
\begin{multline}
g_{p_{0}}(i_{1})g_{p_{1}}(j_{1})=\left(\prod_{r=1}^{i_{0}}\alpha_{u_{r}}(i_{1}-(i_{0}-r))\right)\left(\prod_{l=1}^{j_{1}}\alpha_{v_{l}}(l)\right)\\
=\left(\alpha_{u{}_{1}}(i_{1}-i_{0}+1)\alpha_{u{}_{2}}(i_{1}-i_{0}+2)...\alpha_{u_{i_{0}}}(i_{1})\right)\\
\cdot\left(\alpha_{v_{1}}(1)\alpha_{v_{2}}(2)...\alpha_{v_{j_{1}}}(j_{1})\right)\label{eq:appendix-assumption-1}
\end{multline}

Consider any node $u_{r},1\leq r\leq i_{0}$, it must be visited in
round $i_{1}-i_{0}+r$ because $\alpha_{u_{r}}(i_{1}-i_{0}+r)$, its
transmit coefficient in round $i_{1}-i_{0}+r$, appears on the RHS
of (\ref{eq:appendix-assumption-1}). Similarly any node $v_{l},1\leq l\leq j_{1}$,
must be visited in round $l$. Therefore, for a given round, $\tilde{p}_{0}$
and $\tilde{p}_{1}$ must contain exactly the same one or two nodes
as $p_{0}$ and $p_{1}$. Without loss of generality, suppose $\tilde{p}_{0}$
starts from $X_{0}$ and $\tilde{p}_{1}$ starts from $X_{1}$. $\tilde{p}_{0}$
and $\tilde{p}_{1}$ can and only can be generated from $p_{0}$ and
$p_{1}$ by \emph{switching} nodes that are visited in the same rounds,
because two nodes cannot be visited at the same time in a single path. 

Suppose that only a pair of nodes are switched: 
\[
\tilde{p}_{0}=u_{1}-u_{2}-...-u_{r}-v_{l+1}-u_{r+2}-...-u_{i_{0}}-k
\]
\[
\tilde{p}_{1}=v_{1}-v_{2}-...-v_{l}-u_{r+1}-v_{l+2}-...v_{j_{1}}-k
\]
$1\leq r<i_{0}$, $1\leq l<j_{1}$. $u_{r}$ and $v_{l}$ have two
common neighbors, $u_{r+1}$ and $v_{l+1}$, that are both the next
node in $p_{0}$. Fig. \ref{fig:two-nodes-common-neighbors} shows
a possible condition for $u_{r}$ and $v_{l}$ in $p_{0}$ and $p_{1}$
(other conditions are similar). $p_{0}$ and $p_{1}$ merge at node
$k$ at the end, forming a closed area $A$, as shown in Fig. \ref{fig:p0-p1-form-closed-area-1}.
One of the virtual sources must be in area $A$. For example, $X_{1}$
is in area $A$ in Fig. \ref{fig:p0-p1-form-closed-area-1} (this
is because the two neighbors of $v_{l}$ are inside the area $A$,
and $p_{0}$ and $p_{1}$ being disjoint shortest paths means that
the same node cannot appear twice within the union of the nodes of
$p_{0}$ and $p_{1}$), and $X_{0}$ is in area $A$ in Fig. \ref{fig:p0-p1-form-closed-area-2}.
The other virtual source is in area $B$. Thus, there is a path $u_{r}-...-k-...-v_{l}$
separating $X_{0}$ and $X_{1}$. The possible relative positions
for $X$, $X_{0}$ and $X_{1}$ are shown in Fig. \ref{fig:source-case}.
$u_{r}$ and $v_{l}$ cannot be switched if $X$, $X_{0}$ and $X_{1}$
are positioned as in Fig. \ref{fig:source-case-1}, because it is
impossible for a path to separate $X_{0}$ and $X_{1}$ in this case.
If $X$, $X_{0}$ and $X_{1}$ are positioned as in Fig. \ref{fig:source-case-2},
the middle of $X_{0}$ and $X_{1}$ is $X$, which cannot be in the
middle of any path. Both possible positions contradicts our supposition
that $u_{r}$ and $v_{l}$ can be switched to generate two new shortest
paths $\tilde{p}_{0}$ and $\tilde{p}_{1}$ from $p_{0}$ and $p_{1}$
such that $g_{p_{0}}(i_{1})g_{p_{1}}(j_{1})\equiv g_{\tilde{p}_{0}}(i_{1})g_{\tilde{p}_{1}}(j_{1})$. 

The cases that more than a pair of nodes are switched are also impossible,
because the first pair already cannot be switched by our argument
above. As a consequence, the supposition that that there are two different
shortest paths $\tilde{p}_{0}$ and $\tilde{p}_{1}$ such that $g_{p_{0}}(i_{1})g_{p_{1}}(j_{1})\equiv g_{\tilde{p}_{0}}(i_{1})g_{\tilde{p}_{1}}(j_{1})$
is impossible.

\begin{figure}
\centering\includegraphics[bb=5bp 10bp 61bp 56bp,clip,width=0.15\textwidth]{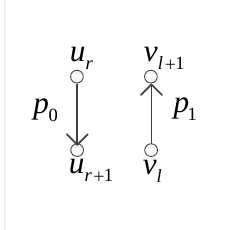}
\caption{\label{fig:two-nodes-common-neighbors}Two nodes $u_{r}$ and $v_{l}$
with two neighbors that are both the next node on the path}
\end{figure}
\begin{figure}
\centering\subfloat[\label{fig:p0-p1-form-closed-area-1}]{\includegraphics[bb=17bp 17bp 200bp 150bp,clip,width=0.34\textwidth]{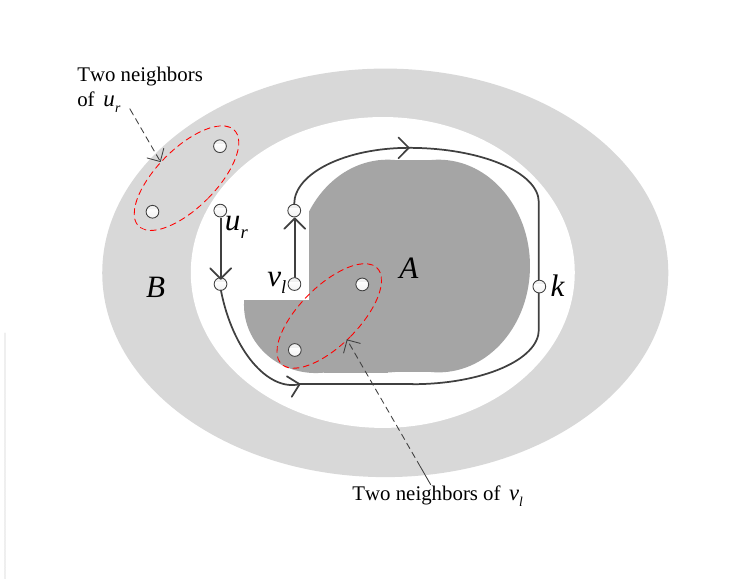}

}

\subfloat[\label{fig:p0-p1-form-closed-area-2}]{\includegraphics[bb=17bp 17bp 200bp 149bp,clip,width=0.34\textwidth]{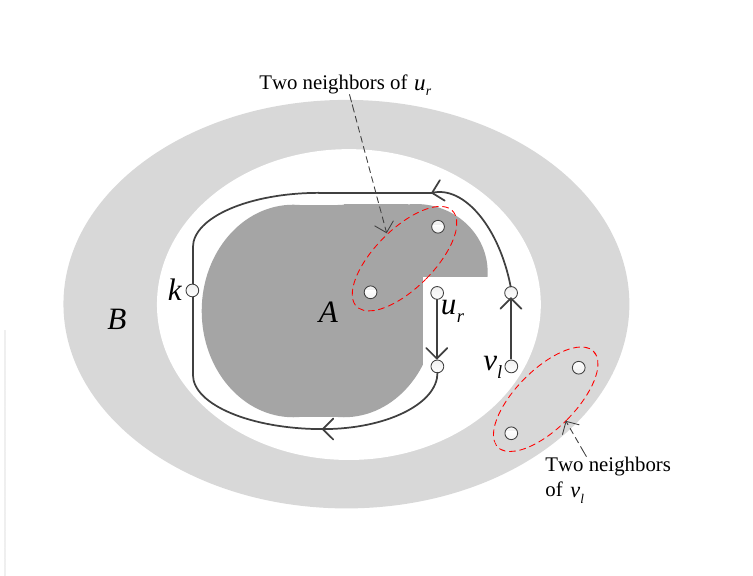}

} \caption{\label{fig:p0-p1-form-closed-area}$p_{0}$ and $p_{1}$ merge at
node $k$ to form a closed area $A$.}
\end{figure}
\begin{figure}
\centering\subfloat[\label{fig:source-case-1}]{\includegraphics[bb=10bp 18bp 75bp 58bp,clip,width=0.15\textwidth]{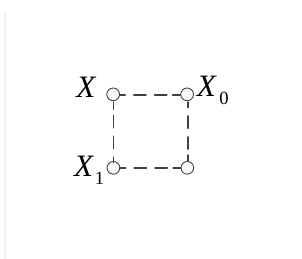}

}\qquad{}\qquad{}\subfloat[\label{fig:source-case-2}]{\includegraphics[bb=15bp 6bp 80bp 46bp,clip,width=0.15\textwidth]{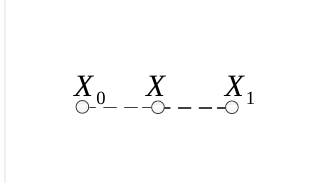}

} \caption{\label{fig:source-case}Possible relative positions for $X$, $X_{0}$
and $X_{1}$.}
\end{figure}

As a consequence, $g_{p_{0}}(i_{1})g_{p_{1}}(j_{1})$ is a term in
$a_{S_{0}}(i_{1})a_{T_{1}}(j_{1})\oplus a_{S_{1}}(i_{1})a_{T_{0}}(j_{1})$
that is not equivalent to any other term. Thus, $g_{p_{0}}(i_{1})g_{p_{1}}(j_{1})$
contains a set of factors that is different from any other term.

We can represent $a_{S_{0}}(i_{1})a_{T_{1}}(j_{1})\oplus a_{S_{1}}(i_{1})a_{T_{0}}(j_{1})$
by 
\begin{equation}
Q_{1}\oplus Q_{2}\oplus...\oplus Q_{L}\label{eq:RHS-of-det-1}
\end{equation}
where each element in $\{Q_{j}\}_{j=1,2,...,L}$ is a product of two
path coefficients, i.e., 
\[
Q_{j}=g_{p}(i_{1})g_{p'}(j_{1})
\]
for some $p\in S_{0}$ and $p'\in T_{1}$, or $p\in S_{1}$ and $p'\in T_{0}$.
An element in $\{Q_{j}\}_{j=1,2,...,L}$ is 
\[
Q_{j}=g_{p}(i_{1})g_{p'}(j_{1})=\prod_{u\mbox{ is in }p}\alpha_{u}(t_{u})\prod_{v\mbox{ is in }p'}\alpha_{v}(t_{v})
\]
where $t_{u}$ and $t_{v}$ are the rounds node $u$ and $v$ are
visited, respectively. Thus, $Q_{j}$ is a product of $R=i_{1}+j_{0}=i_{0}+j_{1}$
transmit coefficients. Without loss of generality, let $Q_{1}=g_{p_{0}}(i_{1})g_{p_{1}}(j_{1})$.
Then $Q_{1}$ contains a set of distinct factors that is not exactly
the same as that of any $Q_{j},2\leq j\leq L$. By Lemma \ref{lem:Not-equivalent},
$a_{S_{0}}(i_{1})a_{T_{1}}(j_{1})\oplus a_{S_{1}}(i_{1})a_{T_{0}}(j_{1})\not\equiv0$. 

Therefore $a_{S_{0}}(i_{1})a_{T_{1}}(j_{1})\oplus a_{S_{1}}(i_{1})a_{T_{0}}(j_{1})\not\equiv0$
when $i_{0}<i_{1}$, $j_{0}<j_{1}$ and $i_{1}-i_{0}=j_{1}-j_{0}$.\end{IEEEproof}
\begin{lem}
\label{lem:Not-equivalent}Consider a degree-$R$ multivariable polynomial
$Q_{1}\oplus Q_{2}\oplus...\oplus Q_{L}$, where each of $Q_{i},1\leq i\leq L,$
is a product of $R\geq1$ factors (variables), wherein the same factor
can appear at most twice in each $Q_{i}$. Suppose that there is a
$Q_{j},1\leq j\leq L,$ whose factors are all distinct and whose factors
are not exactly the same as the factors of $Q_{i},1\leq i\leq L,i\neq j$
(i.e., there must be at least one factor in $Q_{j}$ that is not in
$Q_{i}$). Then $Q_{1}\oplus Q_{2}\oplus...\oplus Q_{L}\not\equiv0$.
\end{lem}
Remark: This is an intuitively trivial lemma. It basically says that
if there is a $Q_{j}$ that is pairwise distinct from any other $Q_{i},i\neq j$,
then the overall polynomial cannot cancel out to zero algebraically.
Here, we have not considered substituting the variables in the polynomial
with specific values. Later in Lemma \ref{lem:not-equal-to-0}, we
will consider assigning i.i.d. random values to each of the variables
in the polynomial in the context of our random transmit coefficients. 
\begin{IEEEproof}
Without loss of generality, suppose that $j=1$ (i.e., $Q_{j}$ is
$Q_{1}$). Also, without loss of generality, we assume if $Q_{i}=Q_{l}$
for $i,l\neq1,i\neq l$, then they will be removed from $Q_{1}\oplus Q_{2}\oplus...\oplus Q_{L}$
so that $Q_{i}\neq Q_{l}$ in the remaining $Q_{1}\oplus Q_{2}\oplus...\oplus Q_{L}$.
We write 
\[
Q_{1}=\prod_{m=1}^{R}r_{m},
\]
where $r_{m}\neq r_{m'}$ for $m\neq m',1\leq m,m'\leq R$.

First, consider the case of $R=1$. If $L=1$, then $Q_{1}\oplus Q_{2}\oplus...\oplus Q_{L}=Q_{1}=r_{1}\not\equiv0$.
For $L>1$, since $Q_{i}\neq Q_{l}$ for $i\neq l$ it is also clear
that $Q_{1}\oplus Q_{2}\oplus...\oplus Q_{L}\not\equiv0$, since all
of $Q_{i}$ consist of one distinct variable.

Now, suppose that the lemma is true for $R=k$ for some $k\geq1$,
we will show that it is also true for $R=k+1$. We write 
\begin{align*}
Q_{1}\oplus Q_{2}\oplus...\oplus Q_{L} & =r_{1}(r_{2}...r_{k+1}\oplus\sum_{n}s_{n}^{(k)})\oplus\sum_{n}s_{n}^{(k+1)}
\end{align*}
where $s_{n}^{(k)}$ and $s_{n}^{(k+1)}$ are products of $k$ and
$k+1$ factors, respectively. Specifically, each of $s_{n}^{(k)}$
represents a $Q_{i}$ with factor $r_{1}$, and each of $s_{n}^{(k+1)}$
represents a $Q_{i}$ that does not have factor $r_{1}$. We know
that $\nexists s_{n}^{(k)}$ such that $s_{n}^{(k)}=r_{2}r_{3}...r_{k+1}$,
otherwise $\exists Q_{i}=r_{1}s_{n}^{(k)}$ such that $Q_{i}=Q_{1}$,
contradicting our supposition that $Q_{1}$ contains a set of factors
that is not the same as any of $Q_{i},2\leq i\leq L$. If $\sum_{n}s_{n}^{(k+1)}\equiv0$,
then
\begin{align*}
 & r_{1}(r_{2}...r_{k+1}\oplus\sum_{n}s_{n}^{(k)})\oplus\sum_{n}s_{n}^{(k+1)}\equiv0\\
\Leftrightarrow & r_{2}...r_{k+1}\oplus\sum_{n}s_{n}^{(k)}\equiv0
\end{align*}
which is impossible by the supposition that this lemma is true for
$R=k$. If $\sum_{n}s_{n}^{(k+1)}\not\equiv0$, it is trivially true
that $r_{1}(r_{2}...r_{k+1}\oplus\sum_{n}s_{n}^{(k)})\not\equiv\sum_{n}s_{n}^{(k+1)}$,
because $r_{1}$ does not appear on the RHS while it does on the LHS.

In conclusion, the lemma is true for any $R\geq1$.\end{IEEEproof}
\begin{lem}
\label{lem:not-equal-to-0}Consider a degree-$R$ multivariable polynomial
$Q_{1}\oplus Q_{2}\oplus...\oplus Q_{L}$, where each of $Q_{i},1\leq i\leq L,$
is a product of $R\geq1$ factors (variables), wherein the same factor
can appear at most twice in each $Q_{i}$. Suppose that there is a
$Q_{j},1\leq j\leq L,$ whose factors are all distinct and whose factors
are not exactly the same as the factors of $Q_{i},1\leq i\leq L,i\neq j$
(i.e., there must be at least one factor in $Q_{j}$ that is not in
$Q_{i}$). Further suppose that the factors (variables) in the polynomial
$Q_{1}\oplus Q_{2}\oplus...\oplus Q_{L}$ are i.i.d. uniform random
variables with values drawn from $GF(2^{s})\backslash\{0\}$. Then
$\Pr\left(Q_{1}\oplus Q_{2}\oplus...\oplus Q_{L}=0\right)\leq R/(2^{s}-1)$.\end{lem}
\begin{IEEEproof}
$Q_{1}\oplus Q_{2}\oplus...\oplus Q_{L}\not\equiv0$ by Lemma \ref{lem:Not-equivalent}.
Without loss of generality, suppose that $j=1$ (i.e., $Q_{j}$ is
$Q_{1}$). Also, without loss of generality, we assume if $Q_{i}=Q_{l}$
for $i,l\neq1,i\neq l$, then they will be removed from $Q_{1}\oplus Q_{2}\oplus...\oplus Q_{L}$
so that $Q_{i}\neq Q_{l}$ in the remaining $Q_{1}\oplus Q_{2}\oplus...\oplus Q_{L}$.
We write 
\[
Q_{1}=\prod_{m=1}^{R}r_{m},
\]
where $r_{m}\neq r_{m'}$ for $m\neq m',1\leq m,m'\leq R$. 

First, consider the case of $R=1$. 
\begin{align*}
\Pr\left(Q_{1}\oplus Q_{2}\oplus...\oplus Q_{L}=0\right)= & \Pr\left(r_{1}=Q_{2}\oplus...\oplus Q_{L}\right)\\
\leq & \frac{1}{2^{s}-1}.
\end{align*}
Note that to arrive at the inequality above, if the realization of
$Q_{2}\oplus...\oplus Q_{L}=0$, then $\Pr\left(r_{1}=0\right)=0$;
if on the other hand $Q_{2}\oplus...\oplus Q_{L}\neq0$, then $\Pr\left(r_{1}=Q_{2}\oplus...\oplus Q_{L}\right)=1/(2^{s}-1)$,
regardless of what non-zero realization $Q_{2}\oplus...\oplus Q_{L}$
take.

Next, suppose that this lemma is true for $R=k$ for some $k\geq1$,
we want to show that it is also true for $R=k+1$. We write 
\begin{align*}
 & Q_{1}\oplus Q_{2}\oplus...\oplus Q_{L}\\
= & r_{1}...r_{k+1}\oplus Q_{2}\oplus...\oplus Q_{L}\\
= & r_{1}^{2}\sum_{n}s_{n}^{(k-1)}\oplus r_{1}(r_{2}...r_{k+1}\oplus\sum_{n}s_{n}^{(k)})\oplus\sum_{n}s_{n}^{(k+1)}
\end{align*}
where $s_{n}^{(k-1)}$, $s_{n}^{(k)}$ and $s_{n}^{(k+1)}$ are respectively
products of $k-1$, $k$ and $k+1$ factors, all of whom are not $r_{1}$.
We know that $\nexists s_{n}^{(k)}$ such that $s_{n}^{(k)}\equiv r_{2}r_{3}...r_{R}$,
otherwise $\exists Q_{i}=r_{1}s_{n}^{(k)}$ such that $Q_{i}=Q_{1}$,
contradicting our supposition that $Q_{1}$ contains a set of factors
that is not the same as any $Q_{i},2\leq i\leq L$. 

$r_{1}(r_{2}...r_{k+1}\oplus\sum_{n}s_{n}^{(k)})\not\equiv0$ because
$r_{2}...r_{k+1}\oplus\sum_{n}s_{n}^{(k)}\not\equiv0$ by Lemma \ref{lem:Not-equivalent}.
There are four cases to be considered as follows: 

1. $\sum_{n}s_{n}^{(k-1)}\equiv0,\sum_{n}s_{n}^{(k+1)}\equiv0$ (this
is the case where $r_{1}$ appears once in all of $Q_{i}$):
\begin{align*}
 & \Pr\left(Q_{1}\oplus Q_{2}\oplus...\oplus Q_{L}=0\right)\\
= & \Pr\left(r_{1}(r_{2}...r_{k+1}\oplus\sum_{n}s_{n}^{(k)})=0\right)\\
= & \Pr\left(r_{2}...r_{k+1}\oplus\sum_{n}s_{n}^{(k)}=0\right)\\
\leq & \frac{k}{2^{s}-1}\\
\leq & \frac{k+1}{2^{s}-1}
\end{align*}
by the supposition that this lemma is true for $R=k$,

2. $\sum_{n}s_{n}^{(k-1)}\equiv0,\sum_{n}s_{n}^{(k+1)}\not\equiv0$:
{\footnotesize 
\begin{align*}
 & \Pr\left(Q_{1}\oplus Q_{2}\oplus...\oplus Q_{L}=0\right)\\
= & \Pr\left(r_{1}(r_{2}...r_{k+1}\oplus\sum_{n}s_{n}^{(k)})\oplus\sum_{n}s_{n}^{(k+1)}=0\mid r_{2}...r_{k+1}\oplus\sum_{n}s_{n}^{(k)}=0\right)\\
 & \cdot\Pr\left(r_{2}...r_{k+1}\oplus\sum_{n}s_{n}^{(k)}=0\right)+\\
 & \Pr\left(r_{1}(r_{2}...r_{k+1}\oplus\sum_{n}s_{n}^{(k)})\oplus\sum_{n}s_{n}^{(k+1)}=0\mid r_{2}...r_{k+1}\oplus\sum_{n}s_{n}^{(k)}\neq0\right)\\
 & \cdot\Pr\left(r_{2}...r_{k+1}\oplus\sum_{n}s_{n}^{(k)}\neq0\right)\\
= & \Pr\left(\sum_{n}s_{n}^{(k+1)}=0\right)\Pr\left(r_{2}...r_{k+1}\oplus\sum_{n}s_{n}^{(k)}=0\right)+\\
 & \Pr\left(r_{1}=\frac{\sum_{n}s_{n}^{(k+1)}}{r_{2}...r_{k+1}\oplus\sum_{n}s_{n}^{(k)}}\mid r_{2}...r_{k+1}\oplus\sum_{n}s_{n}^{(k)}\neq0\right)\\
 & \cdot\Pr\left(r_{2}...r_{k+1}\oplus\sum_{n}s_{n}^{(k)}\neq0\right)\\
\leq & \Pr\left(r_{2}...r_{k+1}\oplus\sum_{n}s_{n}^{(k)}=0\right)+\\
 & \Pr\left(r_{1}=\frac{\sum_{n}s_{n}^{(k+1)}}{r_{2}...r_{k+1}\oplus\sum_{n}s_{n}^{(k)}}\mid r_{2}...r_{k+1}\oplus\sum_{n}s_{n}^{(k)}\neq0\right).
\end{align*}
}Now, 
\[
\Pr\left(r_{2}...r_{k+1}\oplus\sum_{n}s_{n}^{(k)}=0\right)\leq\frac{k}{2^{s}-1}
\]
by the supposition that this lemma is true for $R=k$. If $\sum_{n}s_{n}^{(k+1)}=0$,
then 
\begin{align*}
 & \Pr\left(r_{1}=\frac{\sum_{n}s_{n}^{(k+1)}}{r_{2}...r_{k+1}\oplus\sum_{n}s_{n}^{(k)}}\mid r_{2}...r_{k+1}\oplus\sum_{n}s_{n}^{(k)}\neq0\right)\\
= & \Pr\left(r_{1}=0\right)\\
= & 0.
\end{align*}
On the other hand, if $\sum_{n}s_{n}^{(k+1)}\neq0$, then 
\begin{align*}
 & \Pr\left(r_{1}=\frac{\sum_{n}s_{n}^{(k+1)}}{r_{2}...r_{k+1}\oplus\sum_{n}s_{n}^{(k)}}\mid r_{2}...r_{k+1}\oplus\sum_{n}s_{n}^{(k)}\neq0\right)\\
= & \frac{1}{2^{s}-1}
\end{align*}
as $r_{1},...,r_{k+1}$ are i.i.d. uniform random variables with values
drawn from $GF(2^{s})\backslash\{0\}$ and $r_{1}$ is not a factor
in either $s_{n}^{(k)},\forall n$ or $s_{n}^{(k+1)},\forall n$.
Thus, {\footnotesize 
\begin{align*}
\Pr\left(Q_{1}\oplus Q_{2}\oplus...\oplus Q_{L}=0\right)\leq & \frac{k}{2^{s}-1}+\frac{1}{2^{s}-1}\\
= & \frac{k+1}{2^{s}-1}.
\end{align*}
}{\footnotesize \par}

3. $\sum_{n}s_{n}^{(k-1)}\not\equiv0,\sum_{n}s_{n}^{(k+1)}\equiv0$:
{\footnotesize 
\begin{align*}
 & \Pr\left(Q_{1}\oplus Q_{2}\oplus...\oplus Q_{L}=0\right)\\
= & \Pr\left(r_{1}^{2}\sum_{n}s_{n}^{(k-1)}\oplus r_{1}(r_{2}...r_{k+1}\oplus\sum_{n}s_{n}^{(k)})=0\right)\\
= & \Pr\left(r_{1}\sum_{n}s_{n}^{(k-1)}\oplus(r_{2}...r_{k+1}\oplus\sum_{n}s_{n}^{(k)})=0\right)\\
= & \Pr\left(r_{1}\sum_{n}s_{n}^{(k-1)}=0\mid r_{2}...r_{k+1}\oplus\sum_{n}s_{n}^{(k)}=0\right)\\
 & \cdot\Pr\left(r_{2}...r_{k+1}\oplus\sum_{n}s_{n}^{(k)}=0\right)+\\
 & \Pr\left(r_{1}\sum_{n}s_{n}^{(k-1)}\oplus(r_{2}...r_{k+1}\oplus\sum_{n}s_{n}^{(k)})=0\right.\\
 & \left.\mid r_{2}...r_{k+1}\oplus\sum_{n}s_{n}^{(k)}\neq0\right)\\
 & \cdot\Pr\left(r_{2}...r_{k+1}\oplus\sum_{n}s_{n}^{(k)}\neq0\right)\\
\leq & \Pr\left(r_{2}...r_{k+1}\oplus\sum_{n}s_{n}^{(k)}=0\right)+\\
 & \Pr\left(r_{1}\sum_{n}s_{n}^{(k-1)}\oplus(r_{2}...r_{k+1}\oplus\sum_{n}s_{n}^{(k)})=0\right.\\
 & \left.\mid r_{2}...r_{k+1}\oplus\sum_{n}s_{n}^{(k)}\neq0\right)
\end{align*}
}where {\small 
\[
\Pr\left(r_{2}...r_{k+1}\oplus\sum_{n}s_{n}^{(k)}=0\right)\leq\frac{k}{2^{s}-1}
\]
}by the supposition that this lemma is true for $R=k$. If $\sum_{n}s_{n}^{(k-1)}=0$,
then{\small 
\begin{align*}
 & \Pr\left(r_{1}\sum_{n}s_{n}^{(k-1)}\oplus(r_{2}...r_{k+1}\oplus\sum_{n}s_{n}^{(k)})=0\right.\\
 & \left.\mid r_{2}...r_{k+1}\oplus\sum_{n}s_{n}^{(k)}\neq0\right)\\
= & \Pr\left(r_{2}...r_{k+1}\oplus\sum_{n}s_{n}^{(k)}=0\mid r_{2}...r_{k+1}\oplus\sum_{n}s_{n}^{(k)}\neq0\right)\\
= & 0.
\end{align*}
}On the other hand, if $\sum_{n}s_{n}^{(k-1)}\neq0$, then {\small 
\begin{align*}
 & \Pr\left(r_{1}\sum_{n}s_{n}^{(k-1)}\oplus(r_{2}...r_{k+1}\oplus\sum_{n}s_{n}^{(k)})=0\right.\\
 & \left.\mid r_{2}...r_{k+1}\oplus\sum_{n}s_{n}^{(k)}\neq0\right)\\
= & \Pr\left(r_{1}=\frac{r_{2}...r_{k+1}\oplus\sum_{n}s_{n}^{(k)}}{\sum_{n}s_{n}^{(k+1)}}\mid r_{2}...r_{k+1}\oplus\sum_{n}s_{n}^{(k)}\neq0\right)\\
= & \frac{1}{2^{s}-1}
\end{align*}
}as $r_{1},...,r_{k+1}$ are i.i.d. uniform random variables with
values drawn from $GF(2^{s})\backslash\{0\}$ and $r_{1}$ is not
a factor in either $s_{n}^{(k)},\forall n$ or $s_{n}^{(k+1)},\forall n$.
Thus,

{\footnotesize 
\begin{align*}
\Pr\left(Q_{1}\oplus Q_{2}\oplus...\oplus Q_{L}=0\right)\leq & \frac{k}{2^{s}-1}+\frac{1}{2^{s}-1}\\
= & \frac{k+1}{2^{s}-1}.
\end{align*}
}{\footnotesize \par}

4. $\sum_{n}s_{n}^{(k-1)}\not\equiv0,\sum_{n}s_{n}^{(k+1)}\not\equiv0$:
The equation
\[
r_{1}^{2}\sum_{n}s_{n}^{(k-1)}\oplus r_{1}(r_{2}...r_{k+1}\oplus\sum_{n}s_{n}^{(k)})\oplus\sum_{n}s_{n}^{(k+1)}=0
\]
is a second order polynomial with at most two solutions as far as
$r_{1}$ is concerned. The probability for $r_{1}$ to be one of the
two solutions is $2/(2^{s}-1)$. Thus, 
\[
\Pr\left(Q_{1}\oplus Q_{2}\oplus...\oplus Q_{L}=0\right)\leq\frac{2}{2^{s}-1}\leq\frac{k+1}{2^{s}-1}.
\]

In conclusion, the lemma is true for any $R\geq1$.
\end{IEEEproof}

\subsection{\label{app:Two-disjoint-shortest-paths}Verification of Conjecture
\ref{conj:two-disjoint-shortest-paths}.}

Consider the $7\times5$ grid network in Fig. \ref{fig:grid-7x5-numbering}.
It can be transformed to the network in Fig. \ref{fig:grid-7x5-numbering-equivalent},
whose colored graph is shown in Fig. \ref{fig:colored-graph-7x5-equivalent}.
Table \ref{tab:Two-disjoint-7x5} lists the two disjoint shortest
paths for all non-virtual-source nodes.

\begin{table*}
\caption{\label{tab:Two-disjoint-7x5}Two disjoint shortest paths for all non-virtual-source
nodes in Fig. \ref{fig:grid-7x5-numbering-equivalent}.}
\centering%
\begin{tabular}{|c|c|c|}
\hline 
Node \# &
Shortest path from $X_{0}$ &
Shortest path from $X_{1}$\tabularnewline
\hline 
\hline 
2 &
1-2 &
31-30-29-4-3-2\tabularnewline
\hline 
3 &
1-2-3 &
31-30-29-4-3\tabularnewline
\hline 
4 &
1-2-3-4 &
31-30-29-4\tabularnewline
\hline 
5 &
1-2-7-6-5 &
31-30-29-4-5\tabularnewline
\hline 
6 &
1-2-7-6 &
31-30-29-4-5-6\tabularnewline
\hline 
6{*} &
1-2-7-6-7{*}-6{*} &
31-30-29-4-5-6{*}\tabularnewline
\hline 
7 &
1-2-7 &
31-30-29-4-5-6-7\tabularnewline
\hline 
7{*} &
1-8-9-8{*}-7{*} &
31-30-29-4-5-6{*}-7{*}\tabularnewline
\hline 
8 &
1-8 &
31-2-7-8{*}-9-8\tabularnewline
\hline 
8{*} &
1-8-9-8{*} &
31-2-7-8{*}\tabularnewline
\hline 
9 &
1-8-9 &
31-2-7-8{*}-7{*}-12-11-10\tabularnewline
\hline 
10 &
1-8-9-10 &
31-2-7-8{*}-7{*}-12-11-10\tabularnewline
\hline 
11 &
1-8-9-10-11 &
31-2-7-8{*}-7{*}-12-11\tabularnewline
\hline 
12 &
1-8-9-10-11-12 &
31-2-7-8{*}-7{*}-12\tabularnewline
\hline 
13 &
1-8-9-10-11-16-15-14-13 &
31-2-7-8{*}-7{*}-12-13\tabularnewline
\hline 
14 &
1-8-9-10-11-16-15-14 &
31-2-7-8{*}-7{*}-12-13-14\tabularnewline
\hline 
15 &
1-8-9-10-11-16-15 &
31-2-7-8{*}-7{*}-12-13-14-15\tabularnewline
\hline 
16 &
1-8-9-10-11-16 &
31-2-7-8{*}-7{*}-12-13-14-15-16\tabularnewline
\hline 
17 &
1-8-21-20-19-18-17 &
31-2-7-8{*}-9-10-17\tabularnewline
\hline 
18 &
1-8-21-20-19-18 &
31-2-7-8{*}-9-10-17-18\tabularnewline
\hline 
19 &
1-8-21-20-19 &
31-2-7-8{*}-9-10-17-18-19\tabularnewline
\hline 
20 &
1-8-21-20 &
31-2-7-8{*}-9-10-17-18-19-20\tabularnewline
\hline 
21 &
1-8-21 &
31-26-25-24-23-22-21\tabularnewline
\hline 
22 &
1-8-21-22 &
31-26-25-24-23-22\tabularnewline
\hline 
23 &
1-8-21-22-23 &
31-26-25-24-23\tabularnewline
\hline 
24 &
1-8-21-22-23-24 &
31-26-25-24\tabularnewline
\hline 
25 &
1-8-21-22-23-24-25 &
31-26-25\tabularnewline
\hline 
26 &
1-2-3-4-29-28-27-26 &
31-26\tabularnewline
\hline 
27 &
1-2-3-4-29-28-27 &
31-26-27\tabularnewline
\hline 
28 &
1-2-3-4-29-28 &
31-26-27-28\tabularnewline
\hline 
29 &
1-2-3-4-29 &
31-30-29\tabularnewline
\hline 
30 &
1-2-3-4-29-30 &
31-30\tabularnewline
\hline 
\end{tabular}
\end{table*}

\begin{figure*}
\centering\subfloat[\label{fig:grid-10x9}Numbering]{\includegraphics[bb=10bp 18bp 224bp 226bp,clip,width=0.35\textwidth]{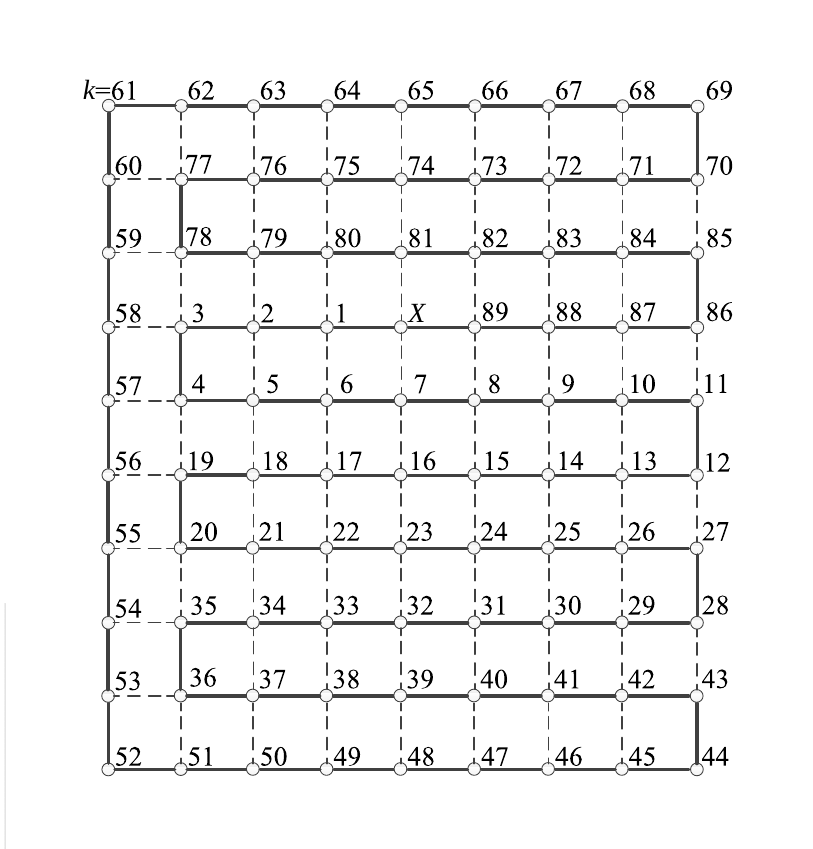}

}

\subfloat[\label{fig:colored-graph-10x9}Colored graph]{\includegraphics[bb=10bp 18bp 224bp 226bp,clip,width=0.35\textwidth]{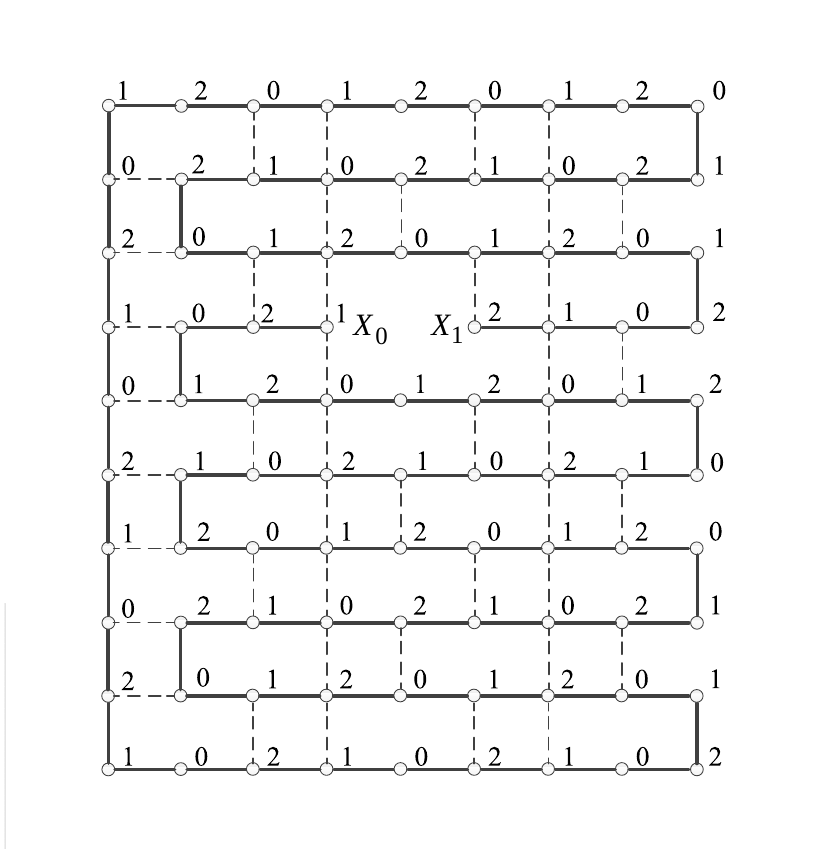}

}

\subfloat[\label{fig:colored-graph-10x9-equivalent}Equivalent network]{\includegraphics[bb=10bp 18bp 224bp 226bp,clip,width=0.35\textwidth]{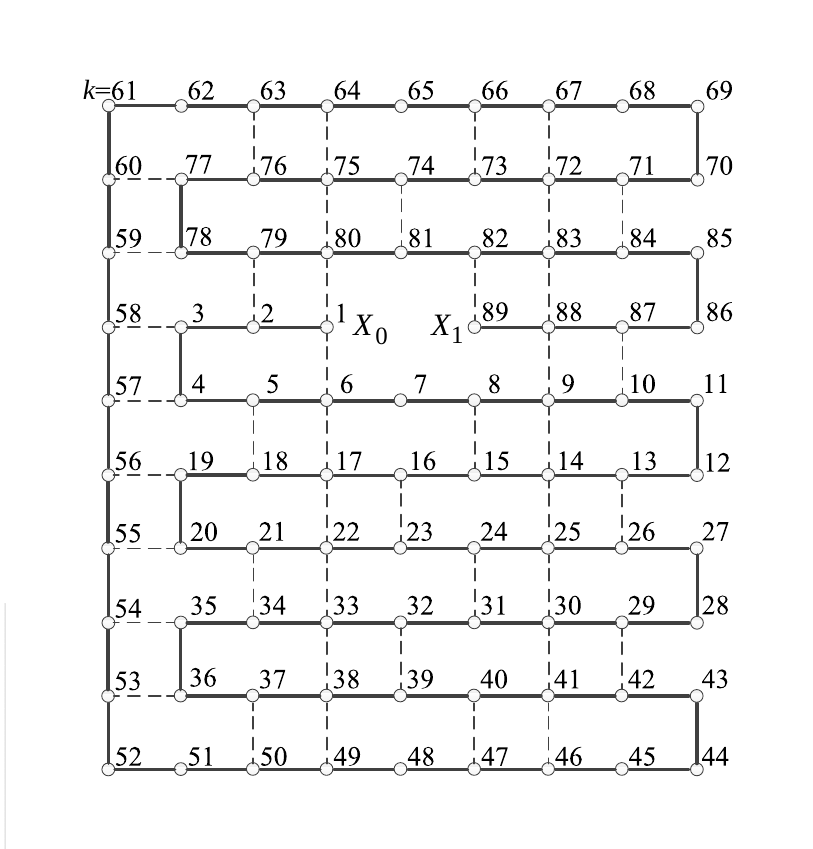}

} \caption{\label{fig:10x9-grid}A $10\times9$ network}
\end{figure*}

Consider the $10\times9$ grid network in Fig. \ref{fig:grid-10x9}.
It can be transformed to the network in Fig. \ref{fig:colored-graph-10x9-equivalent},
whose colored graph is shown in Fig. \ref{fig:colored-graph-10x9}.
Table \ref{tab:Two-disjoint-10x9-1} and \ref{tab:Two-disjoint-10x9-2}
list the two disjoint shortest paths for all non-virtual-source nodes.

To verify the conjecture, we used a computer program to enumerate
all shortest paths from a node to the two virtual sources. As the
time needed to enumerate all shortest paths in a grid grows exponentially
with the number of nodes, and due to the limit of time, we cannot
thoroughly investigate large grid networks. According to our result,
the conjecture at least applies to all grid networks not larger than
$10\times10$.

\begin{figure}
\centering\subfloat[\label{fig:flip-rotate-1}Original Hamiltonian cycle]{\includegraphics[bb=15bp 15bp 174bp 219bp,clip,width=0.2\textwidth]{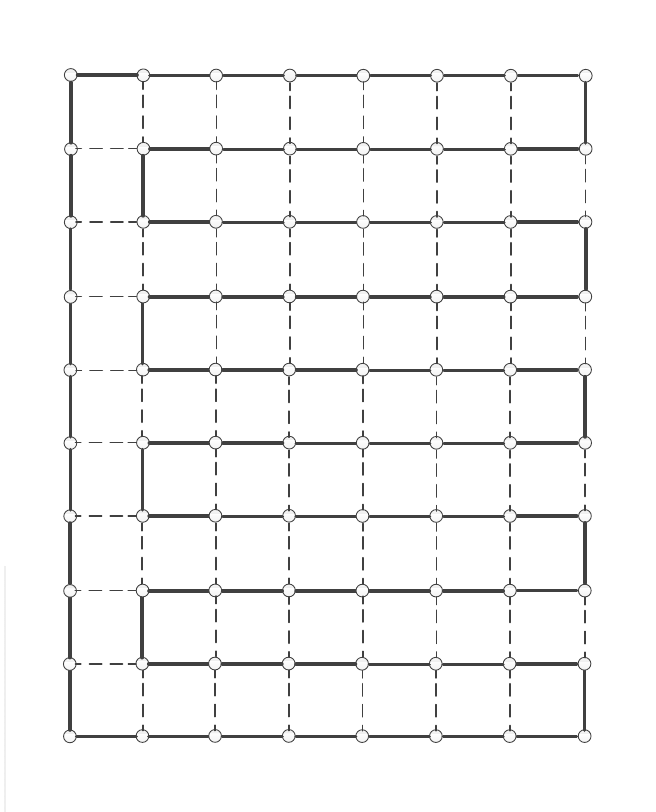}

}\qquad{} \subfloat[\label{fig:flip-rotate-2}Flipped Hamiltonian cycle]{\includegraphics[bb=15bp 15bp 174bp 219bp,clip,width=0.2\textwidth]{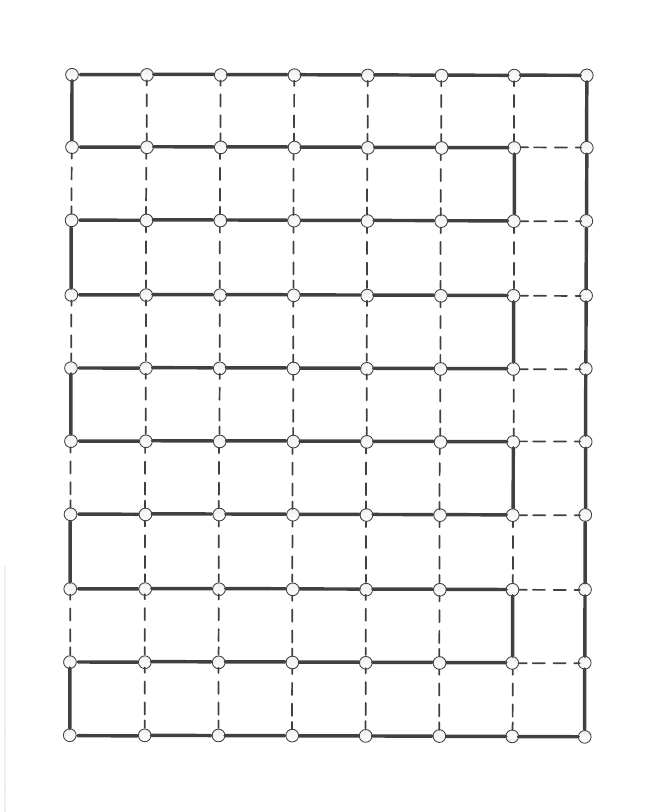}

} \caption{\label{fig:flip-rotate}A $10\times8$ grid network with the original
and flipped Hamiltonian cycles.}
\end{figure}

The conjecture does not apply directly to some grid networks with
special source positions. For example, the conjecture does not apply
to the $10\times8$ grid network when the source is at one of the
following positions: (1,1), (1,7), (2,1), (3,7), (4,1), (4,7), (5,1),
(5,7), (6,7), (7,1), (8,1), (8,7). However, we can flip the Hamiltonian
cycle horizontally (as shown in Fig. \ref{fig:flip-rotate}) to transform
these positions to (1,6), (1,0), (2,6), (3,0), (4,6), (4,0), (5,6),
(5,0), (6,0), (7,6), (8,6), (8,0), where the conjecture applies. We
have verified that the conjecture applies to all networks not larger
than $10\times10$ if we allow flipping the Hamiltonian cycle.

\begin{table*}
\caption{\label{tab:Two-disjoint-10x9-1}Two disjoint shortest paths for non-virtual-source
nodes 2-44 in Fig. \ref{fig:colored-graph-10x9-equivalent}.}
\centering%
\begin{tabular}{|c|c|c|}
\hline 
Node \# &
Shortest path from $X_{0}$ &
Shortest path from $X_{1}$\tabularnewline
\hline 
\hline 
2 &
1-2 &
89-88-9-8-7-6-5-4-3-8\tabularnewline
\hline 
3 &
1-2-3 &
89-88-9-8-7-6-5-4-3\tabularnewline
\hline 
4 &
1-2-3-4 &
89-88-9-8-7-6-5-4\tabularnewline
\hline 
5 &
1-2-3-4-5 &
89-88-9-8-7-6-5\tabularnewline
\hline 
6 &
1-2-3-4-5-6 &
89-88-9-8-7-6\tabularnewline
\hline 
7 &
1-6-7 &
89-88-9-8-7\tabularnewline
\hline 
8 &
1-6-7-8 &
89-88-9-8\tabularnewline
\hline 
9 &
1-6-7-8-9 &
89-88-9\tabularnewline
\hline 
10 &
1-6-7-8-9-14-13-12-11-10 &
89-88-87-10\tabularnewline
\hline 
11 &
1-6-7-8-9-14-13-12-11 &
89-88-87-10-11\tabularnewline
\hline 
12 &
1-6-7-8-9-14-13-12 &
89-88-87-10-11-12\tabularnewline
\hline 
13 &
1-6-7-8-9-14-13 &
89-88-87-10-11-12-13\tabularnewline
\hline 
14 &
1-6-7-8-9-14 &
89-88-87-10-11-12-13-14\tabularnewline
\hline 
15 &
1-6-17-16-15 &
89-88-9-14-15\tabularnewline
\hline 
16 &
1-6-17-16 &
89-88-9-14-15-16\tabularnewline
\hline 
17 &
1-6-17 &
89-88-9-14-15-16-17\tabularnewline
\hline 
18 &
1-2-3-4-57-56-19-18 &
89-88-9-8-7-6-17-18\tabularnewline
\hline 
19 &
1-2-3-4-57-56-19 &
89-88-9-8-7-6-17-18-19\tabularnewline
\hline 
20 &
1-6-17-18-19-20 &
89-88-9-8-15-16-23-22-21-20\tabularnewline
\hline 
21 &
1-6-17-18-19-20-21 &
89-88-9-8-15-16-23-22-21\tabularnewline
\hline 
22 &
1-6-17-18-19-20-21-22 &
89-88-9-8-15-16-23\tabularnewline
\hline 
23 &
1-6-17-22-23 &
89-88-9-14-25-24-23\tabularnewline
\hline 
24 &
1-6-17-22-23 &
89-88-9-14-25-24\tabularnewline
\hline 
25 &
1-6-17-22-23-24-25 &
89-88-9-14-25\tabularnewline
\hline 
26 &
1-6-17-22-23-24-25-30-29-28-27-26 &
89-88-9-14-13-26\tabularnewline
\hline 
27 &
1-6-17-22-23-24-25-30-29-28-27 &
89-88-9-14-13-26-27\tabularnewline
\hline 
28 &
1-6-17-22-23-24-25-30-29-28 &
89-88-9-14-13-26-27-28\tabularnewline
\hline 
29 &
1-6-17-22-23-24-25-30-29 &
89-88-9-14-13-26-27-28-29\tabularnewline
\hline 
30 &
1-6-17-22-23-24-25-30 &
89-88-9-14-13-26-27-28-29-30\tabularnewline
\hline 
31 &
1-6-17-22-33-32-31 &
89-88-9-14-25-30-31\tabularnewline
\hline 
32 &
1-6-17-22-33-32 &
89-88-9-14-25-30-31-32\tabularnewline
\hline 
33 &
1-6-17-22-33 &
89-88-9-14-25-30-31-32-33\tabularnewline
\hline 
34 &
1-2-3-58-57-56-55-54-35-34 &
89-88-9-8-7-6-17-22-33-34\tabularnewline
\hline 
35 &
1-2-3-58-57-56-55-54-35 &
89-88-9-8-7-6-17-22-33-34-35\tabularnewline
\hline 
36 &
1-6-17-22-33-34-35-36 &
89-88-9-14-25-30-41-40-39-38-37-36\tabularnewline
\hline 
37 &
1-6-17-22-33-34-35-37 &
89-88-9-14-25-30-41-40-39-38-37\tabularnewline
\hline 
38 &
1-6-17-22-33-34-35-37-38 &
89-88-9-14-25-30-41-40-39-38\tabularnewline
\hline 
39 &
1-6-17-22-33-38-39 &
89-88-9-14-25-30-41-40-39\tabularnewline
\hline 
40 &
1-6-17-22-33-38-39-40 &
89-88-9-14-25-30-41-40\tabularnewline
\hline 
41 &
1-6-17-22-33-38-39-40-41 &
89-88-9-14-25-30-41\tabularnewline
\hline 
42 &
1-6-17-22-33-38-39-40-41-46-45-44-43-42 &
89-88-9-14-25-30-29-42\tabularnewline
\hline 
43 &
1-6-17-22-33-38-39-40-41-46-45-44-43 &
89-88-9-14-25-30-29-42-43\tabularnewline
\hline 
44 &
1-6-17-22-33-38-39-40-41-46-45-44 &
89-88-9-14-25-30-29-42-43-44\tabularnewline
\hline 
\end{tabular}
\end{table*}
\begin{table*}
\caption{\label{tab:Two-disjoint-10x9-2}Two disjoint shortest paths for non-virtual-source
nodes 45-88 in Fig. \ref{fig:colored-graph-10x9-equivalent}.}

\centering%
\begin{tabular}{|c|c|c|}
\hline 
Node \# &
Shortest path from $X_{0}$ &
Shortest path from $X_{1}$\tabularnewline
\hline 
\hline 
45 &
1-6-17-22-33-38-39-40-41-46-45 &
89-88-9-14-25-30-29-42-43-44-45\tabularnewline
\hline 
46 &
1-6-17-22-33-38-39-40-41-46 &
89-88-9-14-25-30-29-42-43-44-45-46\tabularnewline
\hline 
47 &
1-6-17-22-33-38-49-48-47 &
89-88-9-14-25-30-41-46-47\tabularnewline
\hline 
48 &
1-6-17-22-33-38-49-48 &
89-88-9-14-25-30-41-46-47-48\tabularnewline
\hline 
49 &
1-6-17-22-33-38-49 &
89-88-9-14-25-30-41-46-47-48-49\tabularnewline
\hline 
50 &
1-2-3-58-57-56-55-54-53-52-51-50 &
89-88-9-14-25-30-41-46-47-48-49-50\tabularnewline
\hline 
51 &
1-2-3-58-57-56-55-54-53-52-51 &
89-88-9-14-25-30-41-46-47-48-49-50-51\tabularnewline
\hline 
52 &
1-2-3-58-57-56-55-54-53-52 &
89-88-9-14-25-30-41-46-47-48-49-50-51-52\tabularnewline
\hline 
53 &
1-2-3-58-57-56-55-54-53 &
89-88-9-14-25-30-41-46-47-48-49-50-51-52-53\tabularnewline
\hline 
54 &
1-2-3-58-57-56-55-54 &
89-88-9-14-25-30-31-32-34-35-54\tabularnewline
\hline 
55 &
1-2-3-58-57-56-55 &
89-88-9-14-25-30-31-32-34-35-54-55\tabularnewline
\hline 
56 &
1-2-3-58-57-56 &
89-88-9-14-15-16-17-18-19-56\tabularnewline
\hline 
57 &
1-2-3-58-57 &
89-88-9-14-15-16-17-18-19-56-57\tabularnewline
\hline 
58 &
1-2-3-58 &
89-82-81-80-79-78-59-58\tabularnewline
\hline 
59 &
1-2-3-58-59 &
89-82-81-80-79-78-59\tabularnewline
\hline 
60 &
1-2-3-58-59-60 &
89-82-81-80-75-76-63-62-61-60\tabularnewline
\hline 
61 &
1-2-3-58-59-60-61 &
89-82-81-80-75-76-63-62-61\tabularnewline
\hline 
62 &
1-2-3-58-59-60-61-62 &
89-82-81-80-75-76-63-62\tabularnewline
\hline 
63 &
1-2-3-58-59-60-61-62-63 &
89-82-81-80-75-76-63\tabularnewline
\hline 
64 &
1-80-75-64 &
89-82-83-72-73-66-65-64\tabularnewline
\hline 
65 &
1-80-75-64-65 &
89-82-83-72-73-66-65\tabularnewline
\hline 
66 &
1-80-75-64-65-66 &
89-82-83-72-73-66\tabularnewline
\hline 
67 &
1-80-75-64-65-66-67 &
89-82-83-72-71-70-69-68-67\tabularnewline
\hline 
68 &
1-80-75-64-65-66-67-68 &
89-82-83-72-71-70-69-68\tabularnewline
\hline 
69 &
1-80-75-64-65-66-67-68-69 &
89-82-83-72-71-70-69\tabularnewline
\hline 
70 &
1-80-75-64-65-66-67-68-69-70 &
89-82-83-72-71-70\tabularnewline
\hline 
71 &
1-80-75-64-65-66-67-68-69-70-71 &
89-82-83-72-71\tabularnewline
\hline 
72 &
1-80-75-74-73-72 &
89-82-83-72\tabularnewline
\hline 
73 &
1-80-75-74-73 &
89-82-83-72-73\tabularnewline
\hline 
74 &
1-80-75-74 &
89-82-83-72-73-74\tabularnewline
\hline 
75 &
1-80-79-78-77-76-75 &
89-82-81-74-75\tabularnewline
\hline 
76 &
1-80-79-78-77-76 &
89-82-81-74-75-76\tabularnewline
\hline 
77 &
1-80-79-78-77 &
89-82-81-74-75-76-77\tabularnewline
\hline 
78 &
1-80-79-78 &
89-82-81-74-75-76-77-78\tabularnewline
\hline 
79 &
1-80-79 &
89-82-81-74-75-76-77-78-79\tabularnewline
\hline 
80 &
1-80 &
89-82-81-80\tabularnewline
\hline 
81 &
1-80-81 &
89-82-81\tabularnewline
\hline 
82 &
1-80-81-82 &
89-82\tabularnewline
\hline 
83 &
1-80-81-82-83 &
89-88-87-86-85-84-83\tabularnewline
\hline 
84 &
1-80-81-82-83-84 &
89-88-87-86-85-84\tabularnewline
\hline 
85 &
1-80-81-82-83-84-85 &
89-88-87-86-85\tabularnewline
\hline 
86 &
1-80-81-82-83-84-85-86 &
89-88-87-86\tabularnewline
\hline 
87 &
1-80-81-82-83-84-85-86-87 &
89-88-87\tabularnewline
\hline 
88 &
1-6-7-8-9-10-87-88 &
89-88\tabularnewline
\hline 
\end{tabular}
\end{table*}

\bibliographystyle{23E__Dropbox_PG_Sem6_Thesis_IEEEtran}
\bibliography{22E__Dropbox_PG_Sem6_Thesis_reference}

\end{document}